\documentclass[11pt,a4paper]{article}
\usepackage[margin=1in]{geometry}
\usepackage{authblk}
\usepackage[backend=biber, style=numeric, sorting=none]{biblatex}
\usepackage{blindtext}
\usepackage[T1]{fontenc}
\usepackage{graphicx}
\usepackage{soul}
\usepackage{dcolumn}
\usepackage{bm}
\usepackage{mathtools}
\usepackage{amssymb}
\usepackage{amsmath}
\usepackage{orcidlink}
\usepackage{amsthm}
\usepackage{listings}
\usepackage{bbold}
\usepackage{amsfonts}
\usepackage{appendix}
\usepackage{bm}
\usepackage[caption=false,subrefformat=parens]{subfig}
\usepackage{balance}
\usepackage[dvipsnames]{xcolor}
\usepackage{calc}
\usepackage{dsfont}
\usepackage{orcidlink}
\usepackage{bigints}
\usepackage{ragged2e}
\usepackage{tablefootnote}
\usepackage{braket}
\usepackage{verbatim}
\usepackage{multirow}
\usepackage{booktabs}
\usepackage{tabularx}
\usepackage{makecell}

\usepackage{fix-cm}
\usepackage{hyperref}
\hypersetup{colorlinks=true, allcolors=blue}

\DeclareSourcemap{
  \maps[datatype=bibtex]{
    \map[overwrite]{
      \step[fieldsource=month, match=\regexp{Jan.*}, replace=1]
      \step[fieldsource=month, match=\regexp{Feb.*}, replace=2]
      \step[fieldsource=month, match=\regexp{Mar.*}, replace=3]
      \step[fieldsource=month, match=\regexp{Apr.*}, replace=4]
      \step[fieldsource=month, match=\regexp{May.*}, replace=5]
      \step[fieldsource=month, match=\regexp{Jun.*}, replace=6]
      \step[fieldsource=month, match=\regexp{Jul.*}, replace=7]
      \step[fieldsource=month, match=\regexp{Aug.*}, replace=8]
      \step[fieldsource=month, match=\regexp{Sep.*}, replace=9]
      \step[fieldsource=month, match=\regexp{Oct.*}, replace=10]
      \step[fieldsource=month, match=\regexp{Nov.*}, replace=11]
      \step[fieldsource=month, match=\regexp{Dec.*}, replace=12]
    }
  }
}

\newcommand{\jose}[1] {{\color{black}{#1}}\color{black}\normalsize}

\newcommand{\corr}[1] {{\color{black}{#1}}\color{black}\normalsize}

\newcommand{\saba}[1] {{\color{black}{#1}}\color{black}\normalsize}

\newcommand{\angstrom}{\text{\AA}}

\definecolor{reprogray}{gray}{0.85}
\newcommand{\rev}[1]{{\color{red}#1}}
\newcommand{\reprofig}[3]{%
  \setlength{\fboxsep}{0pt}%
  \fcolorbox{gray}{reprogray}{%
    \parbox[c][#2][c]{#1}{%
      \centering\color{black}\footnotesize
      \textbf{[Figure not shown in this version]}\\[2pt]
      Reproduced/adapted from the reference(s) below;\\
      see the published article for the original figure.\\[6pt]
      #3%
    }%
  }%
}
\newcommand{\reprobox}[3]{%
  \setlength{\fboxsep}{0pt}%
  \fcolorbox{gray}{reprogray}{%
    \parbox[c][#2][c]{#1}{\centering\color{black}\footnotesize #3}%
  }%
}

\begin{document}

\title{Theory of Electron Spin Resonance Scanning Tunneling Microscopy: The First Decade}

\author[1,2]{Saba Taherpour\,\orcidlink{0009-0006-6738-4397}}
\author[1,3]{Denis Jankovi\'{c}\,\orcidlink{0000-0002-9550-6412}}
\author[4,*]{Jose Reina-Galvez\,\orcidlink{0000-0002-1668-8984}}
\author[1,3,*]{Hoang-Anh Le\,\orcidlink{0000-0002-1668-8984}}
\author[1,3,*]{Christoph Wolf\,\orcidlink{0000-0002-9340-9782}}

\affil[1]{Center for Quantum Nanoscience, Institute for Basic Science, Seoul 03760, Republic of Korea}
\affil[2]{Department of Physics, Ewha Womans University, Seoul 03760, Republic of Korea}
\affil[3]{Ewha Womans University, Seoul 03760, Republic of Korea}
\affil[4]{Department of Physics, University of Konstanz, D-78457, Konstanz, Germany}
\affil[*]{Author to whom any correspondence should be addressed: \texttt{jose.reina-galvez@uni-konstanz.de}, \texttt{hoanganhle@qns.science}, \texttt{wolf.christoph@qns.science}}


\maketitle

\begin{abstract}

Electron spin resonance in scanning tunneling microscopy enabled the study of electronic transitions of magnetic impurities on surfaces at the atomic scale. This ESR-STM technique allows to spectroscopically probe and coherently manipulate spins using an all-electrical method without oscillating external magnetic driving fields. Here, we aim to review recent advancements in ESR-STM. We will discuss possible  fundamental mechanisms by which the electric field drives spin resonance based on Heisenberg exchange, Kondo scattering, and Anderson impurity models. We validate theoretical predictions against experimental observations, to understand how electronic correlations, spin exchange, and many-body effects manifest in ESR-STM signals. After reviewing coherent spin control in the STM junction, we discuss potential applications of the ESR-STM method for \corr{coherent} multi-spin control which enables multiple-qubit operations. \corr{Finally, we address recent developments in coupled electron–nuclear spin systems, including hyperfine-resolved ESR spectroscopy,  and the driving and polarization of nuclear spins in ESR-STM.}

\end{abstract}


\section{Introduction}

\corr{Electron spin resonance scanning tunneling microscopy (ESR-STM) is a novel technique that merges the atomic spatial resolution of the STM with the coherent spin manipulation capabilities of ESR. Scanning tunneling microscopy enabled the direct visualization of the atomic surface structure in the solid state~\cite{Binnig1982}. Since its invention in the 1980s the technique has seen rapid development of novel spectroscopic techniques~\cite{Bi2023review, Guo2026}, including, but not limited to, light-emission STM~\cite{Gimzewski1988, Gimzewski1993}, noise spectroscopy~\cite{Birk1995}, the combination with atomic force microscopy~\cite{Giessibl1998, Giessibl2000}, or tip-enhanced Raman spectroscopy~\cite{Stockle2000}. Ultrafast methods, relying on optical pump techniques combined STM with THz radiation~\cite{Cocker2013} or optical pump lasers~\cite{Terada2010}, have enabled the direct observation of physical processes at sub-picosecond time scales.

A parallel development has focused on the study of surface magnetism. This led to the development of spin-polarized STM~\cite{Wiesendanger1990, Wiesendanger1994SPM}, which showed that the STM readout can be sensitive to the spin direction of individual spins and demonstrated that spins can be incoherently manipulated using inelastic tunneling~\cite{Heinrich2004}. Early attempts to observe spin precession in the STM have been reported decades ago~\cite{Manassen1989}, but it took significant development until reproducible ESR was demonstrated in an STM setup~\cite{Baumann_Paul_science_2015}.

}

\corr{The power of ESR-STM lies in overcoming the spatial limitation of conventional ensemble ESR. Whilst ESR is a standard technique for probing  paramagnetic centers, providing information on electronic structures, spin dynamics, and magnetic interaction~\cite{Abragam1970}, it lacks spatial selectivity necessary to address individual spins in heterogeneous environments~\cite{Hall2016}. By combining the sub-atomic precision of STM with ESR, ESR-STM achieves this localization and has become a fundamental tool for quantum-coherent nanoscience~\cite{Heinrich2021}.}


\corr{ESR-STM experiments are generally performed in ultrahigh vacuum and at cryogenic temperatures ($<4$ K). Magnetic fields are used to lift the spin degeneracy via the Zeeman effect. The resulting energy gap defines the resonance frequency of the system. To drive transitions between these spin states a radio-frequency (RF, typically in the GHz range) is applied to the magnetic tip of the STM junction. This creates an effective oscillating magnetic field, which, when the RF frequency is on resonance with the Zeeman splitting, results in a coherent drive measurable as a change in the tunneling current. Unlike conventional STM spectroscopy, ESR-STM is not limited by thermal Fermi-Dirac broadening~\cite{Lambe1968VibrationSpectra, Song2010Highresolution, Ast2016quantumlimit} but only by the coherent properties of the system under study, allowing for energy resolution well below the thermal limit~\cite{Baumann_Paul_science_2015, DJ2025review}.}

\begin{figure*}[htbp!]
\centering
\includegraphics[width=1\columnwidth]{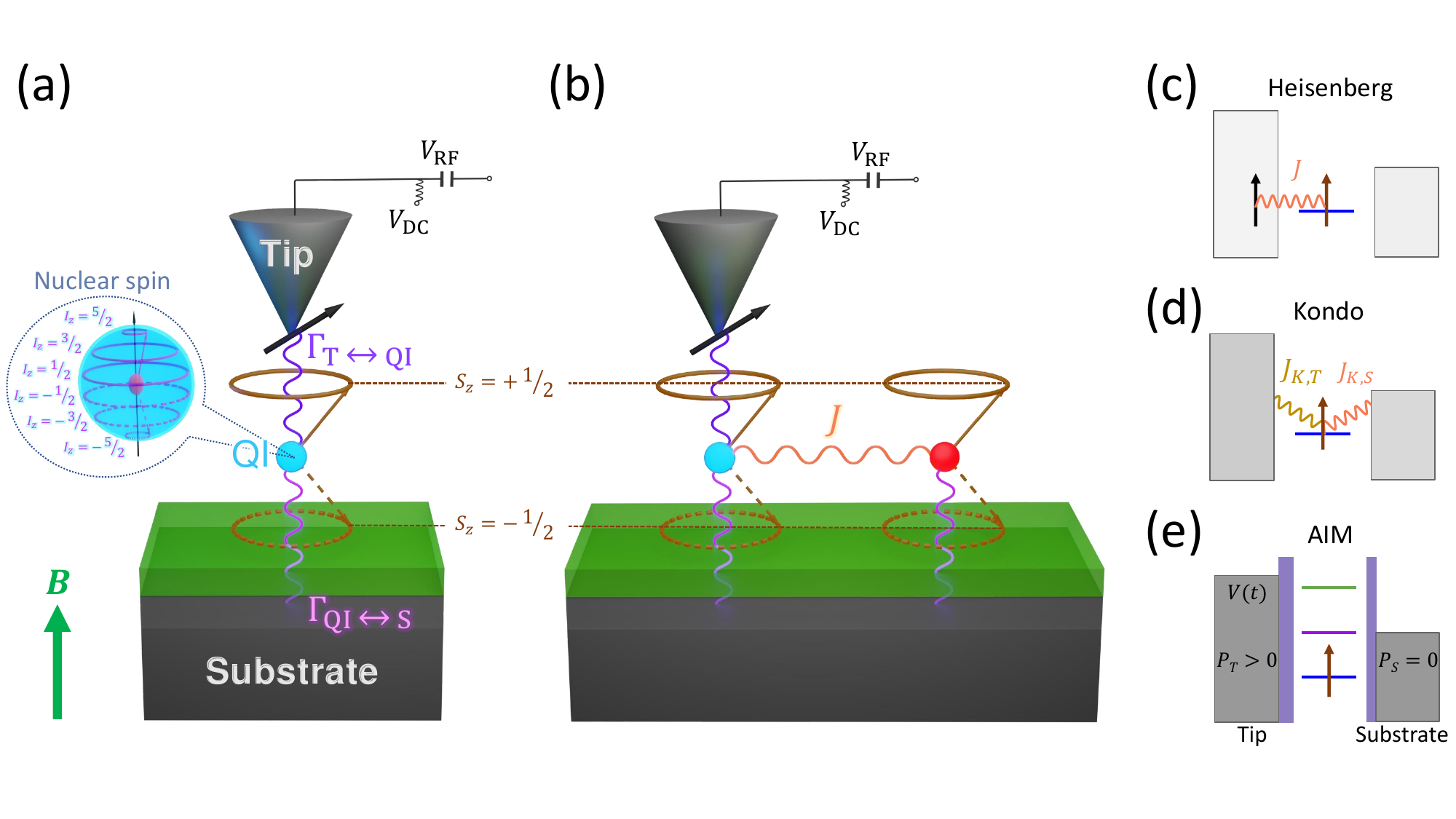}
\caption{\textbf{Schematic of the ESR--STM junction and effective models.}
(a) A quantum impurity (QI, blue; sensor spin) is positioned in the STM junction and interacts with the tip and substrate with tunneling rates $\Gamma_{\mathrm{T}\leftrightarrow \mathrm{QI}}$ and $\Gamma_{\mathrm{QI}\leftrightarrow \mathrm{S}}$, respectively. A DC bias $V_{\mathrm{DC}}$ and an RF modulation $V_{\mathrm{RF}}$ enable ESR driving. An external magnetic field $\mathbf{B}$ splits the electronic spin states ($S_z=\pm \tfrac12$). Hyperfine coupling produces a nuclear-spin manifold (illustrated for $I=5/2$) with projections $I_z=-5/2,\ldots,+5/2$, yielding a hyperfine-split electronic spectrum. (b) Extension to a coupled-spin system: the sensor spin (blue) is exchange-coupled ($J$) to a remote spin (red) adsorbed nearby on the surface. (c--e) Minimal descriptions of the spin-driving physics considered in this work: (c) Heisenberg exchange between localized spins, (d) Kondo coupling of the localized spin to tip and substrate electron baths with exchange constants $J_{K,\mathrm{T}}$ and $J_{K,\mathrm{S}}$, and (e) Anderson impurity model (AIM) with time-dependent bias voltage $V(t)$ and lead spin polarizations $P_\mathrm{T}$ (Tip) and $P_\mathrm{S}=0$ (substrate).}
\label{fig:fig1}
\end{figure*}

Understanding the physics behind ESR in the STM junction was a central question in the first decade of ESR-STM experiments and will be the focus of this \jose{review}. It is important to emphasize that ESR in the STM differs from conventional ESR in the sense that ESR-STM uses oscillating \textit{electric} fields instead of oscillating \textit{magnetic} fields. This is experimentally advantageous because it is relatively simple to generate fast electric signals by modulating the bias voltage \corr{however it is a non-trivial question \textit{how} spins can couple to time-varying electric fields}. A schematic view of an ESR-STM junction is given in Fig.~\ref{fig:fig1}(a). At the center is the magnetic impurity or quantum impurity (QI), which hosts one or more unpaired spins. This impurity is coupled to two \jose{fermionic} baths: the STM tip and substrate, which are both large metal objects. The respective couplings $\Gamma_{{\rm QI}\leftrightarrow\alpha}$ ($\alpha= \text{T, S}$) are generally weak, such that the QI, in good approximation, can be treated as a localized spin. The magnitude of the coupling can be controlled. The QI-tip coupling $\Gamma_{{\mathrm{ QI \leftrightarrow T}}}$ depends exponentially on the QI-tip distance. The QI-substrate coupling $\Gamma_{{\rm QI \leftrightarrow S}}$ is made sufficiently weak by growing thin insulating layers on top of the metal. Only spins located in the STM junction are directly subject to the tunneling current, which couples the QI to tip ($\Gamma_{\mathrm{QI\leftrightarrow T}}$) and the substrate ($\Gamma_{\mathrm{QI\leftrightarrow S}}$). In contrast, multi-spin control enables the manipulation of spins outside the junction, which only subject them to substrate scattering ($\Gamma_{\mathrm{QI\leftrightarrow T}}\approx 0$). \jose{In general, the couplings in the STM junction are asymmetric; the impurity couples more strongly to the substrate than to the STM tip, indicating that the QI is effectively not attached to the tip during an ESR measurement. Consequently, the DC bias drop is also asymmetric~\cite{Tu_X_double_barrier_2008}. Since $\Gamma_{\mathrm{QI\leftrightarrow T}} < \Gamma_{\mathrm{QI\leftrightarrow S}}$, the bias drop occurs predominantly at the tip side.}

In this review, we discuss recent advances in the field of ESR-STM. We want to put particular emphasis on the underlying mechanism and theoretical proposals. Hopefully, a broader overview that incorporates experiment and theory can serve as a guide for robust experimental designs and advance the field of quantum-coherent nanoscience at the atomic scale. This review is structured as follows. In Section~\ref{section:review_experiments}, we first review ESR-STM experiments in the past decade. We then discuss the theoretical models proposed to explain ESR driving mechanisms in ESR-STM. This discussion is divided into two parts: (i) Control of the spin in the STM junction (Fig.~\ref{fig:fig1}(a)), addressed in Section~\ref{sec:sensor_control}, and (ii) Control of spins outside the junction by utilizing the \textit{remote spin} approach (Figs.~\ref{fig:fig1}(b)) discussed in Section~\ref{sec:multi_control}. Finally, in Section~\ref{sec:electron-nuclear}, we extend the discussion to more complex systems, including coupled electron–electron and electron–nuclear spins.


\section{A brief review of ESR-STM experiments}
\label{section:review_experiments}

ESR-STM began with the first demonstration of spin resonance of a single Fe atom on MgO in 2015~\cite{Baumann_Paul_science_2015}. 
This experiment established the ability to detect and coherently drive the spin of an individual atom on surface using RF voltage. Shortly afterward, ESR-STM was extended to atomic-scale sensing. 
ESR frequency shifts of a single Fe atom were used to detect the dipolar field of a nearby Ho atom, enabling readout of its magnetic state~\cite{Natterer2017, Choi2017}. The technique was also applied to map magnetic dipolar fields with atomic resolution and to study the eigenstates of engineered coupled spin-$1/2$ atoms on surfaces~\cite{Yang2017Engineering}.

Following these initial demonstrations, ESR-STM experiments began to probe coherence and the coupling of surface spins to their surrounding environment. Measurements of relaxation and coherence times revealed two main decoherence mechanisms in the STM junction: tunneling current and thermally activated spin flips~\cite{Willke2018Probing}. Coherence was further investigated in exchange-coupled spin systems, where singlet–triplet transitions exhibited longer coherence times than isolated spins~\cite{Bae2018ScienceAdvance}. ESR-STM also resolved hyperfine interactions of individual Fe and Ti atoms, demonstrating strong sensitivity to the adsorption site and local coordination, and enabling atom-specific electron–nuclear spin spectroscopy~\cite{Willke2018Hyperfine}. In parallel, electrically driven nuclear polarization and single-atom nuclear magnetic resonance were demonstrated by pumping nuclear spins via repeated electron spin flips, establishing the STM junction as a source of dynamic nuclear polarization~\cite{Yang2018}.

A further milestone was the expansion of experimental control knobs, including vector magnetic fields, engineered tip stray fields, and tunable exchange interactions~\cite{Natterer2019}. These tools enabled tunable resonance frequency and amplitude, allowing efficient driving at low or zero external field~\cite{Willke2019Tuning}. Exchange interactions \corr{between the tip and surface spin were} tuned over a wide range from 1 mT to 10 T, demonstrated by combined ESR and inelastic tunneling measurements~\cite{Yang2019Tuning}. At the same time, ESR-based magnetic resonance imaging demonstrated spatially resolved measurements of spin interactions at the atomic scale~\cite{Willke2019MRI}. 
The introduction of pulsed ESR-STM further marked the transition from frequency-domain spectroscopy to time-domain measurements~\cite{KaiYang_science2019}.

ESR-STM was subsequently extended in both temperature range and material scope. 
Experiments performed at millikelvin temperatures enabled high-resolution spectroscopy of atomic adsorbates such as hydrogenated Ti on MgO~\cite{Weerdenburg2021,Steinbrecher2021}.
For the same adsorbate, vector-field ESR measurements revealed a strong dependence of resonance frequency and amplitude on field orientation, reflecting the roles of spin polarization, local fields, and spin-orbit coupling~\cite{JK2021vectormagneticfield}. 
The interpretations of these adsorbates were later clarified based on combined experimental and density functional theory studies~\cite{JK2022Hyperfine,Farinacci2022,phark2025Ti3ML}. 
ESR-STM was also extended to molecular systems, including single Iron Phthalocyanine (FePc) molecule on MgO/Ag~\cite{Zhang2022FePc} and molecular radical-ion states such as C$_{60}$ on graphene~\cite{HAZAN2023107377}. 
Pulsed ESR-STM experiments demonstrated coherent control of these molecular spins, with decoherence largely dominated by tunneling electrons~\cite{Willke2021ControlMolecules}. \corr{Beyond single-spin and molecular platforms, ESR-STM was applied to engineered spin-1/2 Ti arrays, where it resolved many-body energy spectra with atomic selectivity and enabled ESR transitions between spin multiplets, with peak shifts revealing the underlying exchange couplings~\cite{Yang2021RVB}.}

Time-resolved ESR-STM captured coherent flip-flop dynamics in exchange-coupled Ti atom pairs, revealing nanosecond oscillations driven by exchange-mediated spin transport~\cite{Veldman2021cohevolution}. 
Improved sensitivity also enabled detailed spectroscopy of local spin environments, including \corr{high-frequency measurements up to 100 GHz~\cite{Drost2022}, measurements at variable temperatures between 1 and 10 K~\cite{Hwang2022}, or the determination} of anisotropic hyperfine interactions for $^{47,49}$Ti atoms on MgO/Ag using vector-field ESR~\cite{JK2022Hyperfine,Farinacci2022}. ESR-STM was further applied to adsorbate assemblies, where alkali-metal dimers on MgO/Ag substrate exhibited a single ESR resonance associated with an unpaired electron in a bonding molecular orbital, confirmed by spatial mapping of the ESR signal between the atoms~\cite{Kovarik2022}.

The focus of ESR-STM experiments then shifted from single-spin spectroscopy toward architectures that enable remote sensing and multi-spin control. Electric control of single-atom spin transitions was demonstrated for Ti on MgO by exploiting junction electric fields to tune both $g$-factor and effective tip field~\cite{Kot2023ElectricControl}. 
Universal quantum control of a single surface spin qubit was achieved using phase-controlled RF pulses to realize arbitrary rotations~\cite{Wang2023Universal}. In addition, exchange coupling \corr{with} a nearby single-atom magnet can generate a local magnetic field that enables electric-field-driven ESR even when the STM tip is retracted far from the spin~\cite{Phark2023electric}. 
Double electron-electron spin resonance experiments demonstrated coherent driving of \textit{remote} Ti spin (located outside of the tunnel junction) and readout its state via \textit{sensor} Ti spin (located inside of the tunnel junction)~\cite{Phark2023double}. These capabilities paved the way for the development of multi-spin architectures in ESR-STM. Building on these advances, ESR-STM was extended to multi-qubit platforms in which a Ti sensor qubit in the junction interacts with several remote Ti spins outside the junction. Using pulsed double electron spin resonance, coherent single-, two-, and three-qubit gate operations were realized at the atomic scale~\cite{Phark2023multi-qubit}. More broadly, ESR-STM has also been extended to complex molecular systems, including spatially resolved measurements of $\pi$-radicals in single-molecule magnets~\cite{Kawaguchi2023}.

Recent experiments have pushed ESR-STM toward increasingly complex spin systems and sensing modalities. ESR-STM was extended to coherent control of highly localized $4f$ spins, where ESR transitions of an individual Er atom were driven and detected through its magnetic coupling to a nearby Ti sensor spin~\cite{Stefano2024driving4f}.  
Time-domain also resolved coupled electron-nuclear spin dynamics in single hydrogenated Ti atoms, where tuning to hyperfine avoided crossings enabled coherent electron-nuclear flip-flop oscillations and multi-frequency beating observed on nanosecond timescales~\cite{Veldman2024Coherent}. 
Multi-frequency ESR schemes further enabled the creation and readout of dressed spin states, where continuous RF driving produced Autler-Townes splittings and Mollow triplets associated with the AC-Stark effect in coupled Ti systems~\cite{Hong2024stark}. 

In parallel, ESR-STM was applied to molecular radicals such as pentacene anions, where spatially resolved measurements mapped the distribution of the molecular spin density~\cite{Stepan2024pentacene}. 
In addition, a mobile quantum sensor based on a PTCDA molecule attached to the STM tip enabled combined electric and magnetic field sensing with sub-angstrom spatial resolution~\cite{Taner2024}. \corr{At a larger scale, ESR-STM has also been used to resolve many-body excitations in atomically engineered 1D and 2D Ti spin lattices on MgO, revealing strongly localized, spin-1/2-like resonances at edges and defects whose spatial profiles directly image topological boundary and corner modes~\cite{HaoWang2024topo}.}

Recent work has also demonstrated increasing electrical control over the local spin environment, as well as high-fidelity spin readout. Bias-dependent exchange fields generated by spin-polarized STM tips were shown to \corr{induce strongly nonlinear shifts} in ESR frequencies, enabling electrical control over spin–tip coupling in both atomic and molecular systems~\cite{zhang2025controllingexchangefieldsurface}.
Spin-electric coupling was quantified for single FePc molecules and Fe-FePc complexes, where large nonlinear frequency shifts enabled all-electrical detuning and Rabi control of single and coupled molecular spins~\cite{Greule2025spincontrol}. \corr{In addition, bias-modulated ESR-STM enabled electrically tunable Landau–Zener–Stückelberg–Majorana (LZSM) interferometry in Ti spins and dimers, revealing bias-dependent transitions between transport regimes and providing a route to extract exchange interactions~\cite{HaoWang2025LZSM}.}
ESR-STM was further applied to fluorene-derived organic radical anions, where coherent driving of delocalized $\pi$-spins was achieved, and frequency-resolved magnetic resonance imaging mapped ESR responses into sub-molecular spatial regions of the radical wavefunction~\cite{Greg2025magneticresonance}. Molecular ferrimagnetic structures were engineered on surfaces by combining FePc with Fe($\mathrm{C}_6\mathrm{H}_6$) units, forming mixed-spin dimers with a well-isolated correlated ground-state doublet. These systems exhibited extended spin lifetimes and tunable ferro- or antiferromagnetic coupling~\cite{Huang2025Ferrimagnets}. 

Single-shot ESR-STM readout of the nuclear spin of an individual ${}^{49}$Ti atom achieved real-time detection with high fidelity and nuclear spin lifetimes on the order of seconds, substantially exceeding the associated electron spin lifetime~\cite{Evert2025nuclearspin}. This demonstrates the potential of ESR-STM for nuclear-spin-based quantum sensing. \saba{Electron–nuclear double resonance 
(ENDOR)}-style protocols further extend these capabilities by enabling nuclear-spin spectroscopy through RF-driven nuclear transitions detected via ESR~\cite{Manassen2018ENDOR,Otte2025ENDOR}.
ESR-STM studies of single Ti atoms on ultrathin MgO films demonstrated site- and thickness-dependent spin states, with reversible switching between $S=1/2$ and $S=1$ configurations achieved through atom manipulation~\cite{phark2025Ti3ML}. \corr{Finally, it was demonstrated that ESR-STM can also spectroscopically probe charged point defects such as sulfur vacancies ($\mathrm{V_S^-}$) and carbon substitution ($\mathrm{C_S^-}$) in MoS$_2$~\cite{Willke2026}, extending ESR-STM to a novel class of systems that is distinct from adsorbed atoms and molecules}.

Taken together, these experiments chart the evolution of ESR-STM from single-spin spectroscopy to coherent control, high-resolution imaging, and scalable multi-spin platforms on surfaces. This progress highlights the power of ESR-STM as a tool for quantum spin science at the atomic scale. These experimental advances have, in turn, motivated extensive theoretical efforts to understand the microscopic mechanisms underlying spin driving and control in the STM junction.

\section{ESR-STM junction models}
\label{sec:sensor_control}

We now discuss the theoretical models of ESR-STM. The STM junction in Fig.~\ref{fig:fig1}(a) can be treated as an open quantum system, the general Hamiltonian can be written as:
\begin{equation}
H=H_{\mathrm{QI}}+H_{\mathrm{B}}+H_{\mathrm{C}},
    \label{eq:H_Tot}
\end{equation}
where $H_{\mathrm{QI}}$ represents the QI with electron and nuclear spin degrees of freedom, $H_{\mathrm{B}}$ is the bath Hamiltonian corresponding to the electronic reservoirs of the STM tip and substrate, and $H_{\mathrm{C}}$ represents the coupling between the bath and QI. 

To describe ESR in STM, different models have been employed with increasing \jose{levels of complexity}. At the most fundamental level, where both the QI and the tip are treated as fully localized spins, the Heisenberg model provides an effective description of their spin–spin interaction (Sec.~\nameref{subsec:Heisenberg}). Incorporating itinerant electrons in the junction leads to \jose{the Anderson impurity model (AIM), which can be transformed into a Kondo model via the Schrieffer--Wolff transformation~\cite{Schrieffer1966}. The former} captures both spin and charge degrees of freedom (Sec.~\nameref{subsec:AIM}), while \jose{the latter} captures only spin fluctuations with the charge degree of freedom effectively frozen (Sec.~\nameref{subsec:Kondo}). These models are schematically illustrated in Fig.~\ref{fig:fig1}(c--e), and each captures different aspects of the physics in the STM junction and has been used to explain a range of experimental observations in ESR-STM. 


We first describe the quantum impurity in the STM junction as a localized magnetic moment. For the purpose of ESR experiments the two lowest eigenstates of the system are \jose{the most} relevant, and the resulting QI can be modeled as an effective spin-1/2.
To split the ground state doublet, a magnetic field is applied which splits the QI states. In absence of external driving, the spin undergoes Larmor precession described by
\begin{equation}
\label{eq:H-QI}
    H_{\mathrm{QI}}
    = g \mu_{\mathrm{B}} B S_z = \hbar\omega_0 S_z.
\end{equation}
Here, $g$ is the effective $g$-factor of the impurity, $\mu_{\mathrm{B}}=e\hbar/(2m_e)$ is the Bohr magneton \jose{with $m_e$ the bare electron mass}, and $B$ denotes the effective magnetic field acting on the spin, which may include contributions from both the static external magnetic field and static exchange fields induced by the STM tip. For simplicity, we assume that the direction of the effective magnetic field is along $z$. 
In reality, to create a quantum two-level system, an external magnetic field is applied $B_{\mathrm{ext}}$ along a direction that sets the quantization axis $\hat{z}$ of the QI in absence of any substantial on-site anisotropies\footnote{This is generally true for $S=1/2$ impurities where only $g$-factor anisotropies play a role; for $S>1/2$ crystal field-induced anisotropies may play a role}. This will induce dynamics on the spin which we will discuss later.

We would like to note that in addition to the electron spin, a nuclear spin (see Fig.~\ref{fig:fig1}(a)) may also be present. This introduces an additional hyperfine interaction term in the impurity Hamiltonian
$\mathbf{S}\cdot \mathbf{A}_{\rm dip}\cdot \mathbf{I}$,
where $\mathbf{S}$ is the electronic spin operator and $\mathbf{A}_{\rm dip}$ is the hyperfine coupling tensor. However, in the following sections, we neglect the hyperfine interaction and focus exclusively on the electron spin dynamics; its role in enabling coupled electron–nuclear spin control will be discussed in detail in Sec.~\ref{sec:electron-nuclear}.

\begin{figure*}[htbp!]
\centering
\includegraphics[width=0.8\columnwidth]{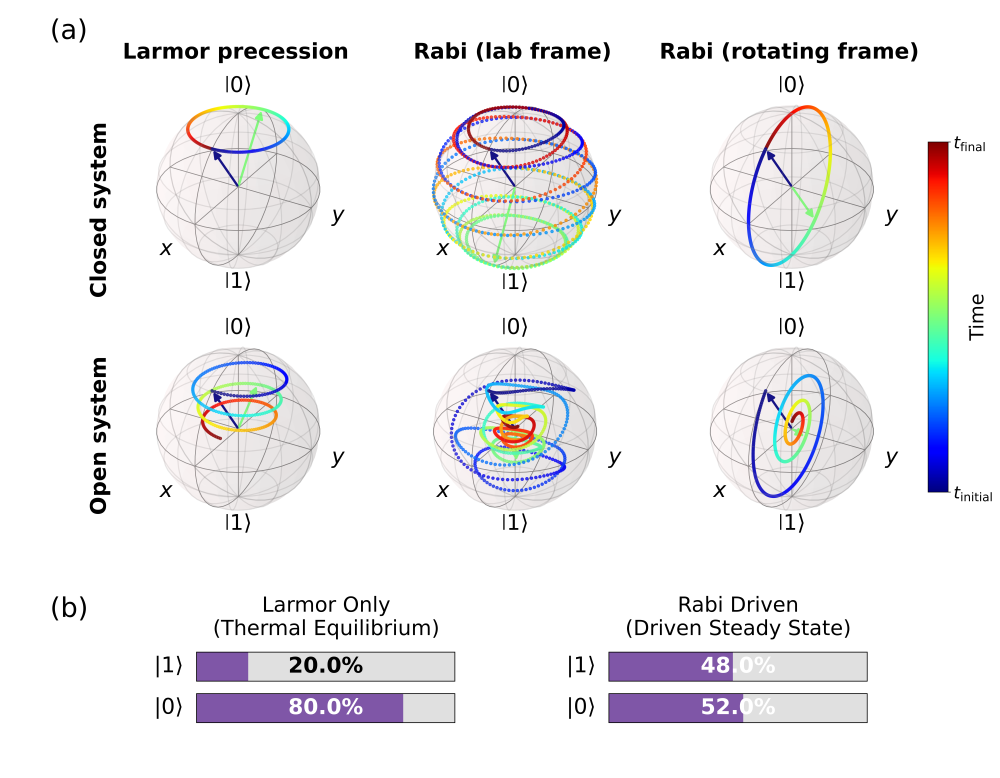}
\caption{\textbf{Spin dynamics and ESR-STM readout} (a) The time evolution of a spin in a magnetic field. The columns show free time evolution (Larmor precession), and driven transitions Rabi Oscillations in the lab frame and in the rotating frame. For the driving, a magnetic field $B(t)$ along $y$ direction was applied on-resonance. The upper row shows a closed quantum system whilst the lower row shows an open quantum system where the spin decays towards a steady state. (b) The population difference between the thermalized system (left) and the driven system (right) enables magnetoresistive readout. On the left, $p_0-p_1=60\%$ on the right $p_0-p_1\approx 0\%$.}
\label{fig:fig2}
\end{figure*}

The state of impurity spin $\ket{\psi}$ can be represented by the eigenbasis of $H_{\rm QI}$, which are $\ket{0}$ and $\ket{1}$ with eigenenergies of $\mp \hbar \omega_0 /2 $. It can be written as:
\begin{equation}
\label{eq:psi}
    \ket{\psi} = \cos \frac{\theta}{2} \ket{0} + e^{i \phi} \sin \frac{\theta}{2} \ket{1},
\end{equation}
where $(\theta, \phi)$ is the polar coordinates in the Bloch sphere. In the absence of its coupling to the bath (closed system), the time evolution of $|\psi(t)\rangle$ is governed by $H_{\rm QI}$, \corr{leading to a time-dependent relative phase between the two eigenstates.  The state can then be written as
\begin{equation}
\ket{\psi(t)} = \cos \frac{\theta}{2} \ket{0} + e^{i \phi(t)} \sin \frac{\theta}{2} \ket{1},
\end{equation}
where the phase evolves as $\phi(t) = \phi - \omega_0 t$, which describes Larmor precession of the Bloch vector around the $z$-axis, as shown schematically in the first row of Fig.~\ref{fig:fig2}(a).

}
In the presence of a driving field, the Hamiltonian of QI now reads
\begin{equation}
    H_{\rm QI} (t) = \hbar\omega_0 S_z + 2 \hbar\Omega S_x \cos (\omega_\mathrm{RF} t).
\end{equation}
\corr{When on-resonance ($\omega_{\mathrm{RF}} = \omega_0$), the} system additionally undergoes Rabi rotations with frequency $\Omega$, i.e. it continuously changes between $|0\rangle \leftrightarrow |1\rangle$.
\corr{In the rotating frame for a closed system this time evolution of the spin is described by
\begin{equation}
\ket{\psi(t)}_{\rm rot} = 
\sin\left(\frac{\Omega t}{2}\right)\ket{1}
+ i \cos\left(\frac{\Omega t}{2}\right)\ket{0}.
\end{equation}
Transforming into the lab frame introduces the fast Larmor precession,
leading to
\begin{equation}
\ket{\psi(t)} = e^{-i \omega_0 t/2} \sin\left(\frac{\Omega t}{2}\right)\ket{1} + i\, e^{+i \omega_0 t/2} \cos\left(\frac{\Omega t}{2}\right)\ket{0}.
\end{equation}
The expectation values of the spin components in the lab frame are then
\begin{equation}
\begin{aligned}
\label{eq:expectation}
\langle S_x \rangle &= -\frac{1}{2} \sin(\Omega t)\, \sin(\omega_0 t), \\
\langle S_y \rangle &= -\frac{1}{2} \sin(\Omega t)\, \cos(\omega_0 t), \\
\langle S_z \rangle &= \frac{1}{2} \cos(\Omega t).
\end{aligned}
\end{equation}
These expressions show that the spin dynamics consists of slow Rabi oscillations at frequency $\Omega$, modulated by a fast Larmor precession 
at frequency $\omega_0$.} Due to the fast Larmor precession (assuming $\Omega \ll \omega_0)$, the trajectory in the lab frame is complicated (center panel of Fig.~\ref{fig:fig2}(a)). However, it is straightforwardly understood in the rotating frame of the spin (right side of Fig.~\ref{fig:fig2}(a)), where the spin performs a full circle around the equator of the Bloch sphere.

In the presence of coupling to the bath (open system), it is often more convenient to study the reduced density matrix of the QI , i.e. the part of interest after tracing out the bath degrees of freedom. For a pure state, the density matrix of QI is defined as $\rho(t) \equiv |\psi(t)\rangle \langle \psi(t)|$. The diagonal elements of $\rho$ represent populations and off-diagonal elements are the quantum coherence. 
Note that the density matrix can be also represented in the Bloch sphere via Bloch vector $\vec{r}$ as:
\begin{equation}
\label{eq:rho}
    \rho = \frac{1}{2} \left( \mathds{1} + \vec{r} \cdot \vec{\sigma} \right),
\end{equation}
where $\mathds{1}$ and $\vec{\sigma}$ are identity operator and Pauli matrices, respectively. Time evolution of an open quantum system is shown schematically in the second row of Fig.~\ref{fig:fig2}(a). Each panel describes the same dynamics as in the closed-system case, but with additional decay towards a steady state. The steady state is a mixed state, i.e. $\text{Tr} (\rho^2) < 1$ or equivalently $|\vec{r}| < 1$, so it lies inside the Bloch sphere.
The rate with which this steady state is reached depends on the decoherence time $T_2$, which can be written as:
\begin{equation}
\label{eq:1/T2}
\frac{1}{T_2} = \frac{1}{2T_1} + \frac{1}{T_2^{*}},
\end{equation}
where $T_1$ is the characteristic life time (also longitudinal relaxation time) and $T_2^*$ is the pure dephasing characteristic time which leads to a loss of phase coherence without changing the spin populations~\cite{FernandoNico2021}. For single adatoms in ESR-STM experiments one typically finds $T_2 =2T_1$, indicating that decoherence is dominated by tunneling electrons in the junction~\cite{Phark2023double,Phark2023multi-qubit}. Consequently, the decay of Rabi oscillations and the ESR linewidth observed experimentally are primarily governed by $T_2$, as illustrated by the relaxation toward a driven steady state in the second row of Fig.~\ref{fig:fig2}(a).

By applying a radio frequency modulated bias voltage $V_\mathrm{RF}$ to the direct current component either via a bias or an antenna, ESR was evidenced by magnetic readout in the current. This can be intuitively understood as follows: in the absence of resonant driving, the QI will remain in a ground state population $p_0$ with a certain spin polarization $\Delta P=p_1-p_0$ given by the ratio of thermal energy and Zeeman level split of the first excited state $p_1$ or by the steady state solution governed by the DC bias depending on the tunneling regime, cotunneling or sequential respectively. This ratio will result in a certain off-resonance current $I^\mathrm{off}$ at finite polarization $P_\mathrm{T}$ of the tip. When driven on-resonance and in the long-term limit, both levels are populated equally for strong enough driving, resulting in zero net spin in the QI which leads to a current $I^\mathrm{res}\neq I^\mathrm{off}$. Taking the difference provides the ESR current signal $\Delta I^\mathrm{ESR}=I^\mathrm{res}-I^\mathrm{off}$ which can be experimentally verified.

Fig.~\ref{fig:fig2}(b) illustrates \textit{how} magnetoresistive readout can evidence Rabi driving. In the experiment, the current is sensitive to the polarization of the QI, i.e. $\Delta P$.
When the system is not driven, the polarization will have a certain value, for example $|\Delta P|=0.8-0.2=0.6$, ultimately determined by the Zeeman energy, temperature and voltage. When driven on-resonance and in the steady state, the polarization is close to zero, $|\Delta P|= 0.52-0.48 = 0.04$. The (time-averaged) current difference between these two measurements is the experimental evidence for ESR in the STM. 

\begin{figure*}[htbp!]
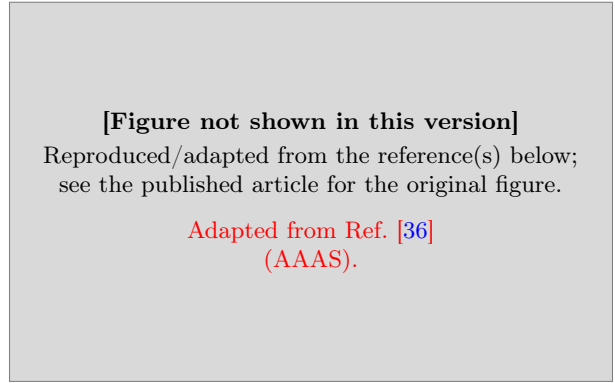

\centering
\reprofig{0.5\columnwidth}{5cm}{\rev{Adapted from Ref.~\cite{KaiYang_science2019}\\(AAAS).}}
\caption{\textbf{Experimental ESR Spectrum.} ESR Spectrum of Ti on Ag/MgO. The measured spectrum is separated into asymmetric (green) and symmetric (blue) components highlighting the complex shape in experiments. The total fit (red) is described by Eq.~\eqref{eq:fano-lineshape}. Measurement condition: $V_\mathrm{DC}=50$ mV, $I_\mathrm{DC}=5$ pA, $V_\mathrm{RF}=50$ mV, $B_\mathrm{ext}=0.82$ T. The resonant frequency is \saba{$f_{\rm res}=20.69$ }GHz. The extracted fit parameters are $\Delta I_s = 17.34$ fA, $\Delta I_a = 47.99$ fA, and $W = 0.031$ GHz. Adapted from Ref.~\cite{KaiYang_science2019} with permission from the American Association for the Advancement of Science.}
\label{fig:fig-spectra}
\end{figure*}

A magnetic STM tip allows magneto-resistive \textit{readout} if the spin state of the QI has a non-zero projection onto the magnetic axis of the tip. Since the magnetic tip is prepared stochastically and is magnetized in an arbitrary direction~\cite{Yu2023}, this is generally true. The angle between the spin of the QI and the magnetization axis of the tip plays a significant role in the shape of the ESR signal (see also Sec.~\ref{sec:multi_control} for a detailed discussion). \corr{Whilst the population difference is the origin of the ESR signal, the experimental reality is more complex. In particular, the non-collinear alignment between tip magnetization vector and surface spin leads to a strong dependence of the signal \textit{shape}, which can vary from nearly Lorentzian (peak or dip like) to strongly asymmetric (see Fig.~\ref{fig:fig-spectra})\cite{Bae2018ScienceAdvance,KaiYang_science2019, JK2021vectormagneticfield}. This can be conceptually understood by considering the magnetic contrast between an effective tip spin placed in a fixed plane, e.g. the $xz$ plane, described by $\langle S_x^\mathrm{T} \rangle, \langle S_z^\mathrm{T} \rangle $, and the impurity spin, which has time-dependent expectation values described by Eq.~\eqref{eq:expectation}. Since  experimentally only the time-average current can be obtained, one needs to consider how the adsorbate spin evolves relative to the tip spin. In general, this yields a DC component and an RF component (for a detailed derivation see, for example, Refs.~\cite{KaiYang_science2019, Stepan2024pentacene}), and the result is a strongly asymmetric line shape in the ESR signal due to the \textit{homodyne} contribution (see also Eq.~\eqref{eq:RabiFLT}), thus, the ESR signal is given by~\cite{Kovarik2022}
\begin{align}
\label{eq:fano-lineshape}
    \Delta I^\mathrm{ESR}=I^\mathrm{res}-I^\mathrm{off}=\Delta I_s \frac{1}{1+\delta^2}-\Delta I_a\frac{2\delta}{1+\delta^2}, 
\end{align}
where 
$\Delta I_s$ ($\Delta I_a$) is the symmetric (asymmetric) contribution to the ESR signal, and \saba{$\delta=2(f-f_{\rm res})/W$ }is the normalized frequency with $W$ the full width at half maximum of the ESR signal and \saba{$f_{\rm res}=2\pi\omega_{\rm res}$} the resonance of the QI, \saba{already including Lamb-shift effects or dressing of the energy levels induced by the electrodes, i.e. $f_{\rm res}=f_0+\Delta f$, with $f_0$ the bare resonance frequency and $\Delta f$ the reservoir-induced shift~\cite{reinagalvez2025contrasting}.} 

An additional complication stems from the fact that the current signals are small, generally in the pico-Ampere regime which requires lock-in detection to reliably extract the signal. In this context, a two-cycle approach is used where in the A-cycle the AC voltage source is turned on which gives the ESR current $I^\mathrm{ESR}$, whilst in the B-cycle the AC voltage is turned off, yielding the background current ($I^\mathrm{off}$). Typically, the largest energy scale is set by the Larmor frequency ($\sim10$ GHz or 41 $\mu$eV). The coupling amplitudes are of the order of $<0.01\ \mu$eV. We note that the dynamics of the QI set by the Larmor frequency appear on a time-scale of $\sim$ ns, however the STM is intrinsically slow due to the integration time of the amplifiers which usually sets the time-resolution to the kHz-MHz regime~\cite{Bastiaans2018}. All current values measured in the STM are therefore time-averages or time-ensemble measurements~\cite{Chen2023}.

To achieve a current through the QI, a DC bias ($V_\mathrm{DC}$) is applied between tip and substrate. When the adsorbate has sufficient density of states within the Fermi energy and the applied bias electrons can hop on the impurity and off, creating a current flow through sequential tunneling. In most experiments measurements are performed at low bias, where transport can still occur through higher-order processes (co-tunneling \jose{or Kondo}) through virtual states~\cite{Delgado_2011}.

}

In ESR-STM experiments, spin resonance is driven \textit{electrically} rather than magnetically which has lead to the terminology \textit{all-electrical} ESR~\cite{FernandoNico2021, reinagalvez2025contrasting}. A central question is \textit{how} the effective ESR drive arises from the modulation by an oscillating electric field applied across the junction. In the following, we review several microscopic mechanisms that explain how such the applied radio-frequency voltage results in spin resonance.

\subsection*{Heisenberg Model}
\label{subsec:Heisenberg}

The Heisenberg model uses a fully localized picture in which the spin $S$ of the QI and the tip spin are treated as two localized spin centers coupled by Heisenberg exchange 
\begin{equation}
    H_{\rm C}= J \mathbf{S}_{\rm T} \cdot \mathbf{S},
    \label{eq:HeisExchange}
\end{equation}
where $\mathbf{S_{\rm T}}$ represents the effective spin vector of the STM tip. The exchange coupling strength $J$ between tip and QI depends on the orbital overlap and decays exponentially with the distance $z$ between two spin centers $ J \propto \exp{(-z/d_{\rm dec})}$, where $d_{\rm dec}$ sets the length scale over which the coupling decays~\cite{Bae2018ScienceAdvance, Yang2019Tuning}. For a Ti adatom and Fe tip, the decay length was found to be $ d_{\rm dec} = 0.42 \pm 0.02~\angstrom$~\cite{Yang2019Tuning}. 


\begin{figure}
\centering
\reprofig{0.6\columnwidth}{5.5cm}{\rev{Adapted from Ref.~\cite{Lado_Ferron_prb_2017}\\(Copyright 2017 American Physical Society).}}
\caption{
(a) Schematic of an Fe atom on MgO in the STM junction including the tip. The variation in the RF bias leads to a displacement $z(t)$ via piezoelectric coupling. This displacement in turn can (b) modulate the tip-adatom exchange coupling $J(z)$, driving the surface spin through a time-dependent exchange field and altering its magnetic moment via magnetoresistance, and/or (c) modulate the on-site anisotropy through $g$-tensor changes. Adapted from Ref.~\cite{Lado_Ferron_prb_2017} with permission. {\color{black} Copyright 2017 American Physical Society}
}
\label{fig:fig3}
\end{figure}

Based on the Heisenberg model, several driving mechanisms were proposed. It was first proposed that oscillating electric fields induces atomic height variations through \textit{piezo-electric coupling} (PEC), \corr{as illustrated in Fig.~\ref{fig:fig3}(a),} which modulates the tip-adatom exchange coupling, see Fig.~\ref{fig:fig3}(b)~\cite{Lado_Ferron_prb_2017}. This mechanism relies on the fact that the spin interaction between the tip and adatom is sensitive to their separation. Modulating the position of the adatom with an electric field leads to a change in the spin-spin interaction, which drives the surface spin and alters the occupation of its spin states, thereby modifying its average magnetic moment. 
Mathematically, it assumes that the RF bias on the STM tip is $E(t) = (V_\mathrm{RF}/d) \cos{(\omega_\mathrm{RF}t)}$, and the piezoelectric vertical displacement of the adatom can be written as:
\begin{equation}
\label{eq:gfactor1}
    z(t) = \frac{q V_{\rm RF}}{k d} \cos{(\omega_\mathrm{RF}t)},
\end{equation}
where $q $ is the effective adatom charge, $k$ the effective elastic constant, and $d$ is the tip–surface separation. The resulting Rabi rate is given by:
\begin{equation}
\label{eq:Omega-heisenberg}
    \Omega = \frac{1}{\hbar} \frac{\partial J}{\partial z} \langle\mathbf{S}_T \rangle \cdot \langle 0|\mathbf{S}|1 \rangle \frac{q V_{\rm RF}}{k d},
\end{equation}
where the tip spin is treated in a mean-field approximation.

A complementary mechanism that drives the surface spin is the electric modulation of the $g$-factor anisotropy caused by piezoelectric displacement of the adatom~\cite{Ferron2019}, as illustrated in Fig.~\ref{fig:fig3}(c). This mechanism requires an anisotropic Zeeman interaction of impurity spin and external magnetic field. Assuming different out-of-plane $z$ and in-plane $xy$ components, the Hamiltonian of the QI is expressed as:
\begin{equation}
\label{eq:H-QI-vect}
H_{\rm QI} = g_x\mu_B B_x S_x + g_z \mu_B B_z S_z.
\end{equation}
In the presence of external driving field with frequency $\omega_\mathrm{RF}$, the modulations of $g$-factors read as:
\begin{equation}
\label{eq:g_x}
    g_x \rightarrow g_x + \delta g_x \cos(\omega_\mathrm{RF} t), \; g_z \rightarrow g_z + \delta g_z \cos(\omega_\mathrm{RF} t),
\end{equation}
These modulations induce ESR transitions if their relative strengths are different (Eq.~\eqref{eq:g-mod})
\begin{equation}
\label{eq:g-mod}
\frac{\delta g_z}{g_z}\neq \frac{\delta g_x}{g_x}.
\end{equation}
The Rabi rate is directly proportional to the difference in modulation strength and is given by:
\begin{equation}
\label{eq:g-mod-rabi}
\Omega=\frac{\mu_B  \mathbf{g B}}{4 \hbar} \sin(2\Theta) \left(
\frac{\delta g_z}{g_z}- \frac{\delta g_x}{g_x}\right), 
\end{equation}
where the polar coordinate is $\Theta = \arctan (g_xB_x/g_z B_z)$ and $\vert \mathbf{g B}\vert = \sqrt{(g_x B_x)^2 + (g_z B_z)^2}$.

It is worth discussing in more detail the microscopic origin of the $g$-tensor modulation, where we should view the impurity as an atom sitting on insulating layers instead of a simple two-level system. Its general Hamiltonian includes both orbital spin $\mathbf{S}$ and angular momentum $\boldsymbol{\ell}$, and their coupling with the substrate beneath (i.e. crystal field). The Hamiltonian reads
\begin{equation}
\label{eq:g-mod-Fe-MgO}
H_{\rm QI} =(g\mathbf{S}+\boldsymbol{\ell})\mu_B\mathbf{B} + \lambda  \mathbf{S} \cdot \boldsymbol{\ell}-|D| \ell_z^2+F  \left[(\ell^{(+)})^4 +(\ell^{(-)})^4 \right],
\end{equation}
where $\lambda$, $D$ and $F$ are the spin-orbit coupling, axial and transversal crystal field parameters, respectively. The piezoelectric displacement results in modulations of $F$ and $D$, which all lead to modulation of $g$ (Eq.~\eqref{eq:gfactor2})
\begin{equation}
\label{eq:gfactor2}
    \delta g_a = \left( \frac{\partial g_a}{\partial D} \frac{d D}{d z} +\frac{\partial g_a}{\partial F} \frac{d F}{d z} \right) \frac{q V_{\rm RF}}{k d} \cos(\omega_{\rm RF} t) \quad (a = x, z).
\end{equation}
Ultimately, the efficiency of these modulation mechanisms sensitively depend on the \textit{magnitude} of displacement for a certain adsorbate as well as the \textit{symmetry properties} of the adsorption site on the substrate which enter through the crystal field Hamiltonian. As we will discuss later in Sec.~\ref{subsec:comparison} subsequent experiments have indicated that crystal field modulation contributes most likely negligibly to the overall Rabi rate for adsorbed adatoms~\cite{Seifert2020LongitudinalMicroscope}.

A limitation of the Heisenberg model is that it assumes localized magnetic moments and does not explicitly describe the \jose{itinerant electrons from the electrodes that interact with the QI, which are essential for a proper description of the tunneling junction, as argued in Ref.~\cite{Ternes2015Spinexcitations}. This framework naturally gives rise to key effects such as the exchange interaction and spin-transfer torque within the Anderson impurity model~\cite{reinagalvez2025contrasting}, which are supported by experimental evidence~\cite{Stepan2024pentacene,zhang2025controllingexchangefieldsurface,Greule2025spincontrol}.}
In this picture, tunneling electrons play a central role in exciting, relaxing, and detecting the impurity spin. Consequently, the localized-spin approximation of the Heisenberg model may overlook the dynamical role of conduction electrons and their influence on spin dynamics within the junction.

\subsection*{Kondo Model}
\label{subsec:Kondo}

\jose{On a fundamental level, as briefly mentioned above, the Kondo model can be derived from the Anderson impurity model. To achieve this, the STM junction is first described as an Anderson impurity coupled to a spin-polarized tip and the substrate reservoirs, and then mapped onto an effective spin-$1/2$ Kondo Hamiltonian via a Schrieffer--Wolff transformation~\cite{Schrieffer1966}. In this mapping, virtual charge fluctuations involving empty and doubly occupied impurity states are integrated out to second order in the tunneling, which yields a form analogous to a Heisenberg-type coupling
\begin{equation}
H_{\rm C} = \sum_{\alpha\alpha'} J_{\rm K,\alpha\alpha'} \mathbf{S}_{\alpha\alpha'} \cdot \mathbf{S},
\end{equation}
where $\mathbf{S}$ is the impurity spin operator. $J_{\rm K,\alpha\alpha'}$ characterizes the strength of the exchange-assisted Kondo coupling, 
as depicted in Fig.~\ref{fig:fig1}(d), and has the form
\begin{equation}
J_{\text{K},\alpha\alpha'}=-2 t_{\alpha} t_{\alpha'}^* \left(\frac{1}{\varepsilon_d+U}-\frac{1}{\varepsilon_d} \right)<0,
\label{eq:kondo_coupling}
\end{equation}
where $\alpha=\mathrm{T,S} $ labeling the tip and the substrate, with $t_\alpha$ denoting the tunneling amplitude between the impurity and reservoir $\alpha$, $\varepsilon_d$ the impurity level energy, and $U$ the on-site Coulomb repulsion. $\mathbf{S}_{\alpha\alpha'}$ denotes the spin operator of the reservoirs,
\begin{equation}
\mathbf{S}_{\alpha\alpha'} = \frac{1}{2} \sum_{k k' \sigma \sigma'} c^\dagger_{\alpha k \sigma} \boldsymbol{\sigma}_{\sigma \sigma'} c_{\alpha' k' \sigma'},
\end{equation}
where $c^\dagger_{\alpha k \sigma}$ ($c_{\alpha k \sigma}$) are creation (annihilation) operators of an electron with spin $\sigma$ in reservoir $\alpha$. $\boldsymbol{\sigma}$ is the vector of Pauli matrices. Under typical STM conditions $t_S > t_T$, implying $J_{\rm K,SS} \gg J_{\rm K,TT}$, so $J_{\rm K,SS}$ governs substrate-driven spin relaxation and Kondo physics, $J_{\rm K,TS}$ controls transport and current-induced spin dynamics, and $J_{\rm K,TT}$ is often negligible due to weak back-action from the tip. If one only considers the spin dynamics and neglects transport-induced processes, the model reduces to an effective Heisenberg-type exchange interaction, recovering Eq.~\eqref{eq:HeisExchange}.

Since Eq.~\eqref{eq:kondo_coupling} includes the dependence on the itinerant electrons, t}he Kondo model provides an extension of the localized-spin description by explicitly incorporating the interaction between a quantum impurity and itinerant conduction electrons in the STM tip and substrate. This extension allows a more in-depth analysis of the electric current through the quantum impurity. Ref.~\cite{J_Cuevas_C_Ast_ESR_theory_2024} comprehensively studied ESR-STM based on Kondo model, with the coupling Hamiltonian written in the following \jose{equivalent form}:
\begin{equation}
\label{eq:H-tunnel}
H_{\rm C}(t) =
\tau e^{i \phi(t)}\sum_{\sigma \sigma'}c_{\text{S} \sigma}^\dagger\left(\delta_{\sigma \sigma'}+\lambda\, \boldsymbol{\sigma}_{\sigma \sigma'} \cdot \mathbf{S}\right)c_{\text{T} \sigma'}+\mathrm{h.c.},
\end{equation}
The first term with the Kronecker delta $\delta_{\sigma \sigma'}$ describes spin-conserving elastic tunneling; while the term $\lambda\,\boldsymbol{\sigma}\cdot\mathbf{S}$ describes inelastic spin-flip processes analogous to perturbative Kondo scattering. Assuming the bias voltage in the junction consists of DC and AC components
\begin{equation}
    V(t) = V_{\rm DC} + V_{\rm RF} \cos (\omega_{\rm RF} t),
    \label{eq:voltage_junction}
\end{equation}
the hopping amplitude $\tau$ is accompanied by a time-dependent phase 
\begin{equation*}
    \phi(t)= \frac{e V_{\rm DC}}{\hbar} t+ \frac{e V_\mathrm{RF}}{\hbar \omega_{\rm RF}} \sin(\omega_{\rm RF} t).
\end{equation*}
By using a transport formalism that combining nonequilibrium Green's functions for electron tunneling with a Lindblad master equation for the driven impurity spin (which is represented by a density matrix $\rho$), Ref.~\cite{J_Cuevas_C_Ast_ESR_theory_2024} showed that the DC current decomposes into elastic $I_{\rm el}$, interference $I_{\rm int}$, and inelastic $I_{\rm inel}$ contributions, with the interference term dominating the ESR signal. The interference term can be explicitly written as:
\begin{equation}
    I_{\rm int} (V) = \lambda \left[ I_{\rm el}^{\sigma = \uparrow} (V) - I_{\rm el}^{\sigma = \downarrow} (V) \right] 
    \left( p_1 - p_0 \right).
\end{equation}
The above equation implies two important points. First, the interference term is proportional to the spin polarization of the impurity, $\langle S_z \rangle  = \hbar \left( p_1 - p_0 \right)/2$, thus \textit{ESR spectra is the measure of the change in magnetoresistance}. Second, the elastic tunneling current must be spin polarized to generate an ESR signal, which \textit{requires the STM tip to be spin polarized to observe ESR}. \jose{In this framework, the driving mechanism is associated with the relation between the RF electric field and the effective AC magnetic field it induces, leading to a Rabi frequency proportional to the magnetic field generated in the junction between the tip and substrate.}

\jose{Another model based on a Kondo Hamiltonian is detailed in~\cite{Shakirov_spin_torque,Role_coh_shakirov_2016}. It employs a similar form for the coupling Hamiltonian as in Eq.~\eqref{eq:H-tunnel}, but uses a quantum master equation with Redfield-type superoperators to describe the spin dynamics, which provides a more microscopic treatment of system–bath interactions and is well suited for capturing frequency-dependent responses.}
Recalling that $\hbar\omega_0$ denotes the Zeeman splitting of the impurity (see Eq.~\eqref{eq:H-QI}), in both the adiabatic limit $(\omega_{\rm RF} \ll \omega_0)$ and the fast-driving limit $(\omega_{\rm RF} \gg \omega_0)$, the differential response to the AC voltage is purely Ohmic (Eq.~\eqref{eq:Ohmic_STT}):
\begin{equation}
\label{eq:Ohmic_STT}
i(t) = \frac{dI}{dV} \times V_{\rm RF} \cos (\omega_{\rm RF} t) \propto V_{\rm RF}.
\end{equation}
At resonance, $\omega_{\rm RF} = \omega_0$, they found a nontrivial non-Ohmic response with the induced current scaling as $V_{\rm RF}^2$, giving rise to the ESR signal. In addition, the trajectory of spin operator $\mathbf{S}$ forms a circular motion in the plane parallel to the $xy$ plane, with a radius proportional to $V_{\rm RF}$. This behavior was interpreted as a spin-transfer torque exerted by the perpendicular spin-polarized current, which drives ESR and motivates the name of the driving mechanism.


Within a Kondo-coupling framework, and in the presence of a bias voltage as described in Eq.~\eqref{eq:voltage_junction}, Ref.~\cite{Ye2024theory} argued that the mean-field effect of a spin-polarized tip generates an effective alternating magnetic field,
\begin{equation}
\label{eq:B-shak}
    B^{\rm eff}_{\rm tip}(t) \simeq B_0 + B_1 \sin{(\omega_\mathrm{RF}t)},
\end{equation}
which drives ESR signal as long as the frequency matches the Zeeman splitting of the impurity, i.e. $\omega_{\rm RF} = \omega_0$. \jose{As we will discuss below, this effective field corresponds to a first-order coherent coupling term rather than to a second-order contribution, such as that arising from Eq.~\eqref{eq:kondo_coupling}, since the transition rates will depend quadratically on the Kondo coupling. In other words, although the Schrieffer--Wolff transformation removes first-order contributions from the Hamiltonian $H_{\rm C}$, a consistent treatment still allows first-order terms to survive, such as the effective field in Eq.~\eqref{eq:B-shak}. This aspect was not included in Refs.~\cite{J_Cuevas_C_Ast_ESR_theory_2024,Shakirov_spin_torque}, with important consequences for the resulting physical predictions.}

Experimentally, the influence of local anisotropy and inter-spin coupling on the Kondo effect has been probed using STM. Magnetic anisotropy was shown to modify the Kondo resonance of individual transition-metal atoms~\cite{Otte2008MagneticAnisotropy}, while exchange coupling between two magnetic atoms was demonstrated to perturb the Kondo singlet and generate distinct spin-excitation features~\cite{Otte2009Kondo}.

Despite their success, both the Heisenberg and Kondo models remain effective low-energy descriptions in which the exchange interaction is treated as an effective parameter. Consequently, variations in the exchange strength primarily influence the tunneling current and the Kondo temperature but do not modify the resonance frequency or lift the spin degeneracy of the impurity. Since these spin-only descriptions preserve time-reversal symmetry, they cannot account for bias-induced spin splitting~\cite{zhang2025controllingexchangefieldsurface}, motivating the introduction of the Anderson impurity model.

\subsection*{Anderson Impurity Model}
\label{subsec:AIM}

The Anderson impurity model 
explicitly includes both spin and quantum charge fluctuations at the impurity, along with tunneling processes between the impurity and spin-polarized leads. This model captures Coulomb charging energy, bias-dependent tunnel rates, and non-equilibrium transport effects, which are essential for explaining bias-dependent spin splitting and resonance frequency shifts observed in ESR-STM experiments.

In the Anderson impurity model the total Hamiltonian of Eq.~\eqref{eq:H_Tot} can be written as follows:
\begin{equation}
\label{eq:H-QI-AIM}
    H_\text{QI} = \sum_{\sigma = \uparrow, \downarrow} \varepsilon_d d^\dagger_\sigma d_\sigma + U \hat{n}_{d \uparrow} \hat{n}_{d \downarrow} + g \mu_B \mathbf{B} \cdot \mathbf{S},
\end{equation}
where $d^\dagger_\sigma$ $(d_\sigma)$ creates (annihilates) an electron of spin $\sigma$ on the impurity orbital with energy $\varepsilon_d$, $U$ denotes the on-site Coulomb repulsion, and $\hat{n}_{d \sigma}=d^\dagger_\sigma d_\sigma$. The impurity spin operator is given by $\mathbf{S}= \hbar/2  \sum_{\sigma, \sigma'} d^\dagger_\sigma \boldsymbol{\sigma}_{\sigma \sigma'} d_{\sigma'}$, and the Zeeman term accounts for coupling to an external magnetic field $\mathbf{B}$. 
\jose{In this model, the effective exchange interaction emerges dynamically from virtual charge fluctuations mediated by spin-polarized itinerant electrons, as will be discussed below. Therefore, an explicit Heisenberg exchange term, Eq.~\eqref{eq:HeisExchange}, need not be introduced in Eq.~\eqref{eq:H-QI-AIM}.}
The electronic reservoirs are described as fermionic baths as:
\begin{equation}
\label{eq:H-B-AIM}
    H_\mathrm{B} = \sum_{\substack{\alpha={\mathrm{T,S}} \\ k,\sigma = \uparrow, \downarrow} } \varepsilon_{\alpha k} c^\dagger_{\alpha k \sigma} c_{\alpha k \sigma}.
\end{equation}
Here, $\alpha=\mathrm{T, S}$ labels the STM tip and substrate, respectively. The operators $c^\dagger_{\alpha k \sigma}$ and $c_{\alpha k \sigma}$ create and annihilate electrons with momentum $k$, spin $\sigma$, and energy $\varepsilon_{\alpha k}$ in reservoir $\alpha$.

\jose{Each electrode in Eq.~\eqref{eq:H-B-AIM} is characterized by an electrochemical potential $\mu_{\alpha}$ and a temperature $T_\alpha$. As is commonly assumed in the literature for simplicity~\cite{reinagalvez2025contrasting,J_Cuevas_C_Ast_ESR_theory_2024,Shakirov_spin_torque,Ye2024theory,J_Reina_Galvez_2019}, the spin polarization of the tip is taken to point along the $z$ direction, although it can be generalized to an arbitrary direction~\cite{Braun_Konig_prb_2004,Piotr_Martinek_Hanle_effect_2023,Busz_PRB_2025}. Thus, the spin asymmetry $D_{\alpha \uparrow} \neq D_{\alpha \downarrow}$ in the spin-dependent density of states $D_{\alpha \sigma}$ at the Fermi level can be characterized by the spin polarization $P_{\alpha} \equiv (D_{\alpha \uparrow}-D_{\alpha \downarrow})/(D_{\alpha \uparrow}+D_{\alpha \downarrow})$, where $D_{\alpha \sigma} = D_{\alpha} \left(1/2 + \sigma P_{\alpha} \right)$ with $P_{\alpha} \in [-1,1]$, and the extreme cases represent fully spin-polarized electrodes pointing down ($-1$) or up ($+1$). This density of states enters the rates of the QME derived from the AIM, leading to spin-dependent tunneling rates that can be interpreted as Rabi-like flip–flop processes if a RF driving is introduced.}

The coupling between the impurity and the reservoir is described by:
\begin{equation}
\label{eq:H-C-AIM}
    H_\text{C} = \sum_{\substack{\alpha={\mathrm{T,S}} \\ k,\sigma = \uparrow, \downarrow} } \left[ t_\alpha c^\dagger_{\alpha k \sigma} d_\sigma + \text{h.c.} \right],
\end{equation} 
where $t_\alpha$ denotes the tunneling amplitude between the impurity and reservoir $\alpha$. \jose{Notably, the coupling Hamiltonian is now linear in the tunneling amplitude, in contrast to the quadratic dependence in Eq.~\eqref{eq:kondo_coupling}. As reminder, t}he asymmetry condition of STM junction is captured by $\left|t_\mathrm{T}\right| < \left|t_\mathrm{S}\right|$.

\jose{For the Anderson model, various approaches have been considered in the literature to construct the rates in the QME, depending on whether sequential tunneling rates (first-order perturbation theory) or cotunneling rates (second-order processes) are used.  Cotunneling transport relies on second-order perturbation in the couplings ($ \propto |t_{\alpha}|^2 $) between the electrodes and the quantum impurity, leading to rates proportional to $ |t_{\alpha}|^4$. The resulting cotunneling description coincides with the effective low-energy theory (lowest-order perturbation) of the Kondo model when applied to a single-orbital impurity~\cite{Delgado_2011}.  In cotunneling, transport proceeds via virtual processes in which an electron tunnels virtually onto the impurity from one electrode and immediately exits into either the same or the opposite lead. This mechanism allows spin fluctuations to be incorporated in the form of first-order coherent Rabi rates that give rise to ESR~\cite{J_Reina_Galvez_2019}. However, a comprehensive ESR-STM theory that consistently treats sequential (first-order) and cotunneling (second-order) processes together with coherent first-order contributions such as the exchange field is currently lacking.


A meaningful approach was developed in~\cite{Weymann_2006_seq_cotu}, where a smooth crossover between sequential and cotunneling regimes is introduced, avoiding the divergences that typically arise when treating both regimes. This method, along with the approach of~\cite{Galperin_2016}, employs diagrammatic techniques to ensure consistency. Other works~\cite{Kosov_prb_2018,Kosov_prb_2019} incorporate both sequential and cotunneling rates using regularization procedures~\cite{Averin_physics_B_1994,Koch_Oppen_2004,Koch_Oppen_2006,Turek_2002}, again to remove divergences in the $T$-matrix. 
Finally, in~\cite{Wacker_cotu_2010}, a second-order von Neumann approach is applied, employing hierarchical equations of motion (HEOM\footnote{HEOM systematically expands the exact bath influence functional into a hierarchy of auxiliary density operators, thereby capturing memory effects, non-Markovian dynamics, and strong system–bath coupling beyond perturbative treatments. Upon truncation at sufficient depth, the hierarchy converges to the full quantum dynamics.}). This method captures both transport regimes, albeit at a high computational cost, and provides a nonperturbative description valid even in the presence of strong coupling and bath–impurity correlations. None of these approaches explicitly includes a driving mechanism, which complicates the development of a complete theory. Only recently, time-domain HEOM simulations of STM currents have been implemented~\cite{Ye2024theory,Cao2025}. These studies report a linear increase of both the ESR linewidth (FWHM) and the Rabi frequency $\Omega$ with increasing $V_\mathrm{RF}$, and explicitly resolve Larmor precession and Rabi nutation driven by the oscillating exchange field generated by the spin-polarized tip, albeit with some remaining open questions regarding homodyne detection~\cite{reinagalvez2025contrasting}. Alternatively, using sequential tunneling rates has been shown to lead to a consistent QME framework in which coherent and dissipative processes emerge naturally once a driving mechanism is introduced~\cite{reinagalvez2025contrasting}. The drawback of this approach is that it is unable to capture higher-order (cotunneling, Kondo) contributions, which have been shown to coexist with ESR~\cite{zhang2025controllingexchangefieldsurface, J_Cuevas_C_Ast_ESR_theory_2024, Chen2025FePc}. 

Focusing on a first-order theory, a linear QME is obtained under the assumptions of weak coupling and fast-decaying baths (Born-Markov approximation). Unlike in cotunneling, the rates here are proportional to $  |t_{\alpha}|^2 $.}
To achieve ESR within this model, it has been shown that the RF voltage $V_\mathrm{RF}$ can modulate the tunneling amplitudes. This effect has been physically interpreted as a modulation of the tunneling barrier \corr{height or width} (the so-called “barrier modulation mechanism”) induced by the RF drive~\cite{reinagalvez2025contrasting,J_Reina_Galvez_2019,J_Reina_Galvez_2021,J_Reina_Galvez_2023}. When $V_\mathrm{RF}$ is applied, the tunneling amplitudes acquire a periodic time dependence of the form $t_\alpha(t)=t_\alpha^0[1+A_\alpha \cos(\omega t)]$, which effectively converts the electrical excitation into a spin torque capable of coherently manipulating the impurity spin. 
\jose{An equivalent description can be formulated in terms of time-dependent chemical potentials generated by the RF voltage, which, via a gauge transformation, produce the same effective dependence of the tunneling amplitudes \corr{through the \textit{phase} of the tunneling operators, which leads, in general, to Bessel function in the tunneling Hamiltonian}~\cite{arrachea-moskalets-2006,caso_model_2014}. This means that the simple harmonic modulation of the hopping amplitudes can be regarded as an approximation to a richer time-dependent structure involving Bessel functions, as discussed in driven transport theories~\cite{jauho_wingreen_prb_1994}. }

The driven dynamics of the Anderson impurity model (AIM) have therefore often been analyzed using a Bloch–Redfield quantum master equation~\cite{reinagalvez2025contrasting,J_Reina_Galvez_2021,J_Reina_Galvez_2023}. To access the long-time limit, the QME can be solved using a Floquet formalism~\cite{J_Reina_Galvez_2021}. Alternatively, the time evolution can be investigated within the same framework via direct time propagation~\cite{Eric2025entanglement,TimeESR_github} \corr{(see Fig.~\ref{fig:fig4}(d))}. As an important result of this model, it was found that virtual charge fluctuations at the impurity give rise to a bias-tunable exchange field $B_\mathrm{exch}$ \corr{(Fig.~\ref{fig:fig4}(c)), }\jose{as well as a dissipative spin-transfer torque, similar to that discussed in Ref.~\cite{Shakirov_spin_torque}}. The exchange field, a many-body effect analogous to a Lamb shift \jose{that emerges from the imaginary part of the rates}, renormalizes the spin energy levels and leads to ESR frequency shifts beyond a single-particle tunneling description~\cite{reinagalvez2025contrasting}.

The AIM exhibits two distinct charge transport regimes. When the bias window lies within $\varepsilon_d$ and $\varepsilon_d+U$, the QI is in the Coulomb blockade (CB) regime, and transport is exponentially suppressed, \jose{unless cotunneling processes are included}. When the bias exceeds these threshold energies, transport proceeds via sequential tunneling \corr{(Fig.~\ref{fig:fig4}(a))}. In previous approaches, a smooth transition between these regimes was captured by introducing an effective broadening of the impurity levels, corresponding to a finite lifetime~\cite{reinagalvez2025contrasting}. The analysis reveals a crossover in the spin dynamics controlled by the applied bias.

\begin{figure}[ht!]
\centering
\reprofig{\columnwidth}{6.5cm}{\rev{%
(a),(d) Adapted from Ref.~\cite{reinagalvez2025contrasting} (Copyright 2025 American Physical Society).\\
(b) Reproduced from Ref.~\cite{Corina2025} (Copyright 2025 The Authors, CC BY 4.0).\\
(c) Adapted from Ref.~\cite{zhang2025controllingexchangefieldsurface} (Copyright 2025 The Authors, CC BY 4.0).}}
\caption{\textbf{Anderson Impurity Model}
(a) Energy diagram of a quantum impurity (QI). Filled arrows indicate occupied single-electron states at $\varepsilon_d$ defining the QI spin, while hollow arrows show unoccupied states at $\epsilon + U$, corresponding to double occupancy. Only the tip (left) electrode is polarized in this example $(P_\mathrm{T} > 0)$ and modulated $(t_\mathrm{T} = t_\mathrm{T}(t))$ with $(t_\mathrm{T} < t_\mathrm{S})$. Depending on the bias, the Rabi process is dominated by either spin-transfer torque (STT) or field-like torque (FLT), with electron tunneling from the right lead producing incoherent resonance signatures. (b) Simulated figure of merit $\Omega T_2$ for quantum coherent operations as a function of DC bias voltage $V_{\mathrm{DC}}$. The blue region corresponds to the STT regime, where the spin polarizes toward the tip and external magnetic field. The pink region corresponds to the FLT regime, where the spin undergoes coherent Rabi oscillations. Both illustrated by the corresponding Bloch-sphere representations. Reproduced from Ref.~\cite{Corina2025}. Copyright 2025 The Authors. Licensed under CC BY 4.0. (c) Resonance frequency shift of a Ti atom measured with a bistable tip. Red dashed line shows the bare resonance; blue curves show calculations for opposite polarizations. \saba{In this review, $f_0$ is actually addressed as $f_{\rm res}$.} Adapted from Ref.~\cite{zhang2025controllingexchangefieldsurface} with permission. Copyright 2025 The Authors. Licensed under CC BY 4.0. (d) Time evolution of the current in the FLT regime, showing six Rabi oscillations before relaxing to the steady state. (a) and (d) Adapted from Ref.~\cite{reinagalvez2025contrasting} with permission. Copyright 2025 American Physical Society.}
\label{fig:fig4}
\end{figure}

In the CB regime, $(\varepsilon_d < eV_\mathrm{DC} < \varepsilon_d + U)$, the response is dominated by coherent spin dynamics driven by virtual charge fluctuations, leading to well-defined Rabi oscillations \corr{(Fig.~\ref{fig:fig4}(d))}. \jose{In this regime, the QI is effectively shielded from strong incoherent spin-polarized electrons from the tip.} This behavior is often referred to as field-like torque (FLT). Within the CB regime, virtual tunneling processes give rise to an effective exchange field $\vec{B}_\mathrm{exch}$. Its component parallel to the quantization axis results in a spin-dependent energy shift with a characteristic nonlinear dependence, \jose{while the perpendicular component drives coherent spin dynamics if RF driving is introduced.} 

\jose{The expression for the exchange field for a $V_\mathrm{DC}$ applied to the sample\footnote{We note that in Ref.~\cite{reinagalvez2025contrasting} the DC voltage was applied to the tip, which slightly modifies Eq.~\eqref{eq:bexch} although the main dependencies remain unaltered.}, under the assumption that $\varepsilon_d+U, |\varepsilon_d| \gg$ than the Zeeman splitting, reads
\begin{equation}
    \vec{B}_\mathrm{exch} \propto \Gamma_\mathrm{ QI \leftrightarrow T} P_\mathrm{T} 
    \ln{\left| \frac{eV_\mathrm{DC}-\varepsilon_d}{eV_\mathrm{DC}-(\varepsilon_d+U)} \right|} \hat{n}_{\rm T},
    \label{eq:bexch}
\end{equation}
where $\hat{n}_{\rm T}$ denotes the tip direction.} We note that the sharp logarithmic divergence in Eq.~\eqref{eq:bexch} at $eV_\mathrm{DC}=\varepsilon_d$ and $\varepsilon_d+U$ is a consequence of \jose{computing the exchange field at zero temperature and without effective broadening. Finite temperature or broadening (effective one or from higher order processes) prevent this divergence~\cite{Braun_Konig_prb_2004, Busz_PRB_2025}.} 
According to Eq.~\eqref{eq:bexch} the exchange field is linear with the coupling $\Gamma_\mathrm{ QI \leftrightarrow T}$ between tip and impurity and the polarization $P_\mathrm{T}$ of the tip. We would like to note that the local exchange field can be determined by measuring the Hanle effect, as proposed by Ref.~\cite{Piotr_Martinek_Hanle_effect_2023}.

\jose{Since both the frequency shift and the Rabi rate originate from projections of the same exchange field, they are geometrically related: \saba{
\begin{equation}
    h\Delta f\sin \Theta = h(f_{\rm res}-f_0)\sin \Theta= 2\hbar \Omega_\mathrm{FLT}\cos\Theta = g\mu_B B_{\rm exch} \sin\Theta \cos\Theta,
    \label{eq:RabiFLT}
\end{equation} }
Here, $\Theta$ denotes the angle between the quantization axis and the tip polarization, referred to as the homodyne angle, and coincides with the polar angle defined previously. This angle gives rise to a distinct asymmetric line shape in certain field–spin geometries. When a hopping modulation is introduced, Eq.~\eqref{eq:RabiFLT} yields a time-dependent exchange field as a consequence of the modulation of the coupling $\Gamma_\mathrm{ QI \leftrightarrow T}(t)$. Different harmonics emerge since $\Gamma_\mathrm{ QI \leftrightarrow T}(t)\propto |t_\alpha(t)|^2$, with the component oscillating as $\cos(\omega_{\rm RF} t)$ giving rise to conventional ESR. Alternatively, a time-dependent chemical potential can be considered which leads to Bessel-function contributions to the tunneling amplitudes~\cite{arrachea-moskalets-2006,jauho_wingreen_prb_1994}, and general higher-harmonic responses akin to photon-assisted sideband tunneling~\cite{Tien1963, Foden1998, Grifoni1998}.}

\jose{Reference~\cite{Ye2024theory} derived a result similar to Eq.~\eqref{eq:bexch} for their effective magnetic driving field. However, as discussed in~\cite{reinagalvez2025contrasting}, their approach appears to fail to reproduce the experimentally observed homodyne detection: a slightly tilted $\Theta$ angle with respect to the polarization direction yields the largest ESR signal~\cite{JK2021vectormagneticfield,Phark2023electric}, whereas in-plane angles in that configuration produce the smallest signal, contrary to what is shown in Fig.~2b of Ref.~\cite{Ye2024theory}. This homodyne response arises from the coexistence of exchange-field and dissipative torque driving within a current-based readout scheme, as these distinct mechanisms impose different phase relations on the spin relative to the $\cos(\omega_{\rm RF} t)$ drive. The exchange-field contribution is phase shifted by $90^\circ$, whereas the dissipative torque contribution remains in phase with the drive~\cite{Stepan2024pentacene}, giving rise to asymmetric ESR signals and the characteristic homodyne dependence.

Such spin-transfer torque effects emerge in both sequential~\cite{reinagalvez2025contrasting} and cotunneling descriptions~\cite{Shakirov_spin_torque}, but appear to be absent in Ref.~\cite{Ye2024theory}. A likely reason for the absence of the homodyne response in that work is the implementation of the readout current. In a spin-polarized STM setup, the current should be proportional to the tip spin polarization (taken along the $z$ axis) and therefore probe the $S_z$ component of the impurity spin, i.e., it is proportional to the scalar product between the tip spin polarization and the adatom spin. As a consequence, one expects a near-zero ESR signal at $\Theta=90^\circ$, when the magnetic field and the spin-$1/2$ are oriented in-plane. However, Ref.~\cite{Ye2024theory} instead assumes that the ESR current is proportional to the projection of the spin onto a direction perpendicular to the external magnetic field, as indicated in Eq.~(6) of their work, which we rewrite here for convenience using our notation:
\begin{equation*}
\Delta I^{\mathrm{ESR}}(\omega_{\rm RF}) \propto \mathbf{S}\cdot \mathbf{e}_\perp \sin\Theta .
\end{equation*}
In contrast, approaches based on the AIM, or even Kondo models such as in Ref.~\cite{J_Cuevas_C_Ast_ESR_theory_2024}, yield
\begin{equation}
\Delta I^{\mathrm{ESR}}(\omega_{\rm RF}) \propto P_T\left[(p_1-p_0) \cos\Theta + 4 A_T\mathrm{Re}(\rho_{\mathrm{off}}) \sin\Theta\right],
\label{eq:homo_current_final}
\end{equation}
where $p_1-p_0$ denotes the population difference of the QI and $\rho_{\mathrm{off}}$ the coherence. This expression naturally reproduces the observed homodyne detection as shown in~\cite{reinagalvez2025contrasting}.}

Because the exchange field survives even when the current is strongly suppressed \jose{in the CB regime}, $\Omega_{\mathrm{FLT}}$ does not vanish at low bias, while the coherence time $T_2$ is maximized in this regime, leading to $\Omega_{\mathrm{FLT}} T_2 \gg 1$ and well-resolved Rabi oscillations suitable for qubit operations. Above the charging thresholds, \jose{sequential processes govern and FLT driving acts too slowly to influence the spin dynamics. However, another driving mechanism emerges: spin-transfer torque (STT). In a similar spirit to Ref.~\cite{Shakirov_spin_torque}, but in the sequential regime, this mechanism is governed by a dissipative spin-accumulation torque $\langle\dot{\mathbf{S}}\rangle_{\mathrm{acc}}$, whose associated Rabi rate $\Omega_{\mathrm{STT}}$ originates from the real part of the transition rates (i.e., it is not a Lamb-shift contribution) and scales with the polarized current, and thus with the tunnel coupling:
\begin{equation*}
\hbar\Omega_{\mathrm{STT}}(t)\propto P_T\Gamma_\mathrm{ QI \leftrightarrow T}(t)\sin\Theta 
\left[\frac{f_T^-(\varepsilon_d)-f_T^+(\varepsilon_d + U) }{2}\right] ,
\end{equation*}
with $f_T^{\pm}(\epsilon)$ defined as
$f_T^{+}(\epsilon)=f_T(\epsilon)$ and $f_T^{-}(\epsilon)=1-f_T(\epsilon)$, where $f_T(\epsilon)=1/(1+e^{(\epsilon-\mu_T)/k_B T})$ is the Fermi distribution function.}

The same current also strongly shortens $T_2$, so that $\Omega_{\mathrm{STT}} T_2 \lesssim 1$ and the spin dynamics become overdamped and incoherent. Consequently, FLT provides a coherent, exchange-field-driven control knob, whereas STT yields fast initialization and polarization at the cost of a small quality factor $\Omega T_2$. The calculated bias dependence of $\Omega T_2$ for FePc directly illustrates how these two regimes are realized~\cite{Corina2025}. In Fig.~\ref{fig:fig4}(b), the quantity $\Omega T_2$ plays the role of a dimensionless quality factor that counts how many coherent Rabi cycles fit into a single decoherence time. In the blue STT-dominated window around $V_{\mathrm{DC}} \approx -425\,\mathrm{mV}$, $\lvert \Omega T_2 \rvert$ is well below unity, showing that the spin barely completes a fraction of a Rabi period before decaying, which is reflected in the Bloch-sphere 
trajectory as a short, overdamped excursion toward the steady state~\cite{reinagalvez2025contrasting}. By contrast, in the red FLT window near  $V_{\mathrm{DC}} \approx -200\,\mathrm{mV}$, $\lvert \Omega T_2 \rvert$ reaches values of several units, indicating multiple coherent Rabi oscillations within $T_2$, correspondingly, the Bloch-sphere path shows Rabi oscillations driven by the exchange field.

\subsection{Phenomenological models}
\label{subsec:Phenom}
A computationally inexpensive method to describe ESR-STM experiments within an open quantum system framework are Markovian master equations such as the Lindblad and Bloch-Redfield quantum master equation. This is valid as long as the coupling to the baths is weak and the bath states decay fast enough, i.e. the Born-Markov approximation. \corr{Extensions to stronger coupling regimes have also been proposed in the context of ESR-STM~\cite{Shavit2019}.} Here, the spin dynamics under RF excitation are described by the time evolution of the density matrix, and parameterized decay channels between the quantum impurity $H_{\rm QI}$ coupled and baths. To describe the evolution of the QI, the bath degrees of freedom are traced out and the reduced density matrix of the system  describes the evolution of the QI under the influence of the baths. The Lindblad formalism uses the Gorini–Kossakowski–Sudarshan–Lindblad (GKSL) equation~\cite{BreuerPetruccione2002}. This can be derived intuitively from the time-dependent Schrodinger equation
\begin{equation}
i\hbar \frac{d}{dt} |\psi(t)\rangle = H |\psi(t)\rangle,
\label{eq:tdseq}
\end{equation}
by replacing the quantum state with its density matrix representation $\rho (t) = |\psi(t) \rangle\langle\psi(t)|$. Taking the time-derivative results in the Liouville-von Neumann equation (Eq. \eqref{eq:LvNeq}).
\begin{equation}
\frac{d\rho}{dt} = -\frac{i}{\hbar} [H, \rho].
\label{eq:LvNeq}
\end{equation}
To account for the environment an additional term, often called dissipator, $\mathcal{D}(\rho)$ is added, which leads to the master equation in Lindblad form
\begin{equation}
    \frac{d\rho}{dt}= -\frac{i}{\hbar} \left[ H , \rho \right]+ \sum_k \left(L_k \rho L_k^\dagger- \frac{1}{2} L_k^\dagger L_k \rho -\frac{1}{2} \rho L_k^\dagger L_k\right).
\label{eq:lind_master}
\end{equation}
The Lindblad form has the advantage of being trace-preserving and positive, i.e. it always results in physical populations which sum up to 1. In the presence of an RF excitation, the Hamiltonian becomes explicitly time dependent and can be written as:
\begin{equation}
\label{eq:H-t}
    H(t)=H_{\mathrm{QI}}+H_{\mathrm{control}}(t),
\end{equation}
where $H_{\mathrm{QI}}$ is the static quantum impurity Hamiltonian and $H_{\mathrm{control}}(t)$ is a phenomenological time-dependent driving term that represents a $B_x(t)$ field. For a single driving frequency, 
\begin{equation}
\label{eq:H-control}
    H_{\mathrm{control}}(t)=A \cos(\omega_{\mathrm{RF}} t) \sigma_x
\end{equation}
describes the coupling of the spin to the oscillating RF field, with amplitude $A$ and frequency $\omega_{\mathrm{RF}}$. This can be easily generalized for multi-spin control and multi-frequency driving~\cite{Hong2024stark, Broekhoven2024}.

The last term in Eq.~\eqref{eq:lind_master} contains the collapse operators. $L_k$ is the collapse operator of the $k$th relaxation channel, which is defined by all relaxation mechanisms. There is some freedom in designing the collapse operators. Since in most experiments the eigenstates were designed to be Zeeman-like a natural choice for the collapse operators is to act on the individual spins~\cite{Phark2023double,Hong2024stark}. 
We first consider energy relaxation for a single spin-$1/2$ qubit. Energy relaxation between the two Zeeman states $|0\rangle$ and $|1\rangle$ can be described by the collapse operators $L_{+}=\sqrt{\Gamma_{+}}\sigma_{+}$ and $L_{-}=\sqrt{\Gamma_{-}}\sigma_{-}$ where $\sigma_{+}$ and $\sigma_{-}$ are the spin raising and lowering operators, and $\Gamma_{+}$ and $\Gamma_{-}$ denote the excitation and relaxation rates, respectively. These rates define the energy relaxation time $T_1=\frac{1}{\Gamma_{+}+\Gamma_{-}}$.
This description can be straightforwardly generalized to a two-spin system. For example, if only the second spin undergoes energy relaxation, transitions between the states $|00\rangle$ and $|01\rangle$ are implemented by the collapse operators
\begin{equation}
L_{2+}=\mathds{1}\otimes \sqrt{\Gamma_{2+}}\sigma_{2+} \text{ and } L_{2-}= \mathds{1} \otimes  \sqrt{\Gamma_{2-}}\sigma_{2-},
\end{equation} 
where $\mathds{1}$ acts on the first spin and $\sigma_{2\pm}$ act on the second spin. The corresponding relaxation time of the second spin is given by $ T_1^{2}=\frac{1}{\Gamma_{2+}+\Gamma_{2-}}.
$

As for pure dephasing, the collapse operator can be represented via Pauli-$z$ matrix. For example, consider the pure dephasing noise on $k$-th spin of a system with $N$ spins, the collapse operator is be written as:
\begin{equation}
L^{\phi}_{k}=\sqrt{\frac{1}{2 T_{\phi,k}}}\,\sigma_{z,k},
\label{eq:dephase}
\end{equation}
where $\sigma_{z,k}=\mathds{1}  \otimes_1 \mathds{1} \cdots \mathds{1} \otimes_{k-1} \sigma_z \otimes_{k}\mathds{1} \cdots \otimes_{N-1} \mathds{1}$ and $T_{\phi,k}$ is the pure dephasing time associated with spin $k$.

However, the situation is different if the eigenstates of the system are far from Zeeman product states. In such a case, a more natural choice for the collapse operators is to act on the actual eigenstates~\cite{Tupkary2022}. In such a scenario, the collapse operator connecting eigenstates $|m\rangle, |n\rangle$ can be written as~\cite{Broekhoven2024}:
\begin{equation}
\label{eq:collapse}
L_{m,n}=\sqrt{\sum_{k}\left|J_k\sum_{s_i,s_f}\langle m, s_i |\mathbf{s}\!\otimes \!\mathbf{S}_k| n, s_f \rangle\right|^{2}\;\frac{\varepsilon_{mn}}{e^{\varepsilon_{mn}/k_B T}-1}}\;|m\rangle\langle n|.
\end{equation}
The sums are over the different spins $k$ and the initial and final state of the electron spin interacting with these spins $s_i$ and $s_f$. The strength of the interaction with each spin is defined by $J_k$ and the energy difference between $m$ and $n$ states of the three-spin system is given by $\varepsilon_{mn}$.


While the Lindblad formalism provides a widely used description of dissipative spin dynamics in ESR-STM experiments~\cite{Phark2023double,Hong2024stark}, other open quantum system approaches have also been employed. In particular, the Redfield master equation offers an alternative in which dissipation arises from an explicit system–reservoir coupling described by the matrix elements of the system–bath interaction~\cite{BreuerPetruccione2002}.

In the Redfield approach, the dynamics of the reduced density matrix are obtained by starting from an explicit decomposition of the total Hamiltonian. In the notation of Eq.~\eqref{eq:H_Tot}, this corresponds to separating the quantum impurity, bath, and coupling degrees of freedom.
Tracing out the reservoir degrees of freedom within the Born–Markov approximation leads to the Redfield master equation for the reduced density matrix $\rho$~\cite{BreuerPetruccione2002}.
In the eigenbasis of $H_{\rm QI}$, the Redfield equation can be written as:
\begin{equation}
\label{eq:redfield}
    \frac{d\rho_{mn}}{dt} = - i \omega_{mn} \rho_{mn} + \sum_{k,l} R_{mnkl}\, \rho_{kl},
\end{equation}
where $\omega_{mn} = (E_m - E_n)/\hbar$ are the Bohr frequencies of the quantum impurity and $R_{mnkl}$ denotes the Redfield tensor. The Redfield tensor is determined by the matrix elements of the system–reservoir coupling operators and by the reservoir correlation functions evaluated at the corresponding transition frequencies~\cite{BreuerPetruccione2002}. In contrast to the Lindblad formalism, dissipation in the Redfield equation is not introduced through phenomenological collapse operators. Instead, relaxation and dephasing processes arise from an underlying microscopic coupling between the quantum impurity and the electronic reservoirs, treated perturbatively within the Born–Markov approximation, together with the spectral properties of the bath~\cite{FernandoNico2021,BreuerPetruccione2002}. As a result, population and coherence dynamics may be coupled, reflecting interference between different transition pathways induced by the environment~\cite{BreuerPetruccione2002}. As a drawback, the Redfield equation is not guaranteed to be norm conserving and can be numerically more challenging than the Lindblad equation.

\subsection{Comparing experiments and models}
\label{subsec:comparison}
In the following we will compare experimental observations and model predictions for the dependence of physical observables on experimentally adjustable variables. Some models suggest physically different coupling mechanisms (see Tab.~\ref{tab:Driving}), leading to different predictions on the relationships between experimentally adjustable variables and the system response (see Tab.~\ref{tab:comparison}). We note that the model used to describe the QI does not necessarily predicate a specific physical coupling mechanism, for example in the Heisenberg model the physical coupling between RF field and the spin could stem from mechanical displacement of the tip or $g$-factor modulation or a combination of both. 

\begin{table*}[t]
\scriptsize
\centering
\caption{Driving mechanisms proposed for ESR-STM in different theoretical models. Representative references for each approach are listed. All formulas in the table are defined in the main text, see Secs.~\nameref{subsec:Heisenberg},~\nameref{subsec:Kondo},~\nameref{subsec:AIM}, and~\nameref{subsec:Phenom}.
}\vspace*{2mm}
\begin{tabularx}{\textwidth}{l X X c}
\toprule
\textbf{Model} & \textbf{Driving mechanism} & \textbf{Driving term} & \textbf{Ref.} \\
\midrule \midrule

\multirow{2}{*}{Heisenberg}
& Piezoelectric displacement of the adatom
& 
$ \dfrac{\partial J}{\partial z} \langle\mathbf{S}_T \rangle \cdot \langle 0|\mathbf{S}|1 \rangle \dfrac{q V_{\rm RF}}{k d} \cos(\omega_{\rm RF}t)$
& \cite{Lado_Ferron_prb_2017}  \\
\cmidrule(lr){2-4}
& Anisotropic modulation of the $g$-factor
& 
$\dfrac{\mu_B |\mathbf{g B}|}{4} \sin(2\Theta) \left( \dfrac{\delta g_z}{g_z}- \dfrac{\delta g_x}{g_x}\right) \cos(\omega_{\rm RF}t)$
& \cite{Ferron2019} \\

\midrule

\multirow{3}{*}{Kondo}
& Modulation of tip electrochemical potential via RF voltage, considering only the real part of the rates (spin-transfer torque)
& $\sum_{m=-\infty}^{\infty}i^{m}e^{i m \omega_{\rm RF} t}\exp\left({i \dfrac{\Delta E}{\hbar\omega_{\rm RF}} \cos(\omega_{\rm RF} t)}\right)$
$\times J_m\left(-\dfrac{\Delta E}{\hbar \omega_{\rm RF}}\right)$ $^{\mathrm{b}}$ 
& \cite{Shakirov_spin_torque}  \\
\cmidrule(lr){2-4}
& Magnetic component of the electromagnetic field enhanced by the tip-sample gap
& $\dfrac{g \mu_\mathrm{B}}{\hbar}B_\mathrm{gap}(t)$
& \cite{J_Cuevas_C_Ast_ESR_theory_2024} \\
\cmidrule(lr){2-4}
& Modulation of tip electrochemical potential via RF voltage taking only the imaginary part (exchange field)
& $\dfrac{P_T \Gamma_{\mathrm{ QI \leftrightarrow T}}}{\pi g\mu_B}\bigintssss \dfrac{D_T(\epsilon)}{\epsilon - \varepsilon_d} f_T'\left(\epsilon + eV_{\mathrm{\rm DC}}+eV_{\rm RF}(t)\right) d\epsilon $
& \cite{Ye2024theory} \\

\midrule

Cotunneling
& Barrier modulation leading to virtual particle exchange
& $\dfrac{A_{\rm T}\Gamma_{\mathrm{ QI \leftrightarrow T}}P_{\rm T}}{h}\times \mathcal{F} \cos(\omega_{\rm RF}t)$ $^{\mathrm{a}}$
& \cite{J_Reina_Galvez_2019} \\

\midrule

\multirow{2}{*}{Anderson}
& $B_{\rm exch}$ from virtual particle exchange (field-like torque)
& $\dfrac{\Gamma_\mathrm{ QI \leftrightarrow T}(t) P_\mathrm{T}}{2\pi\hbar} \sin\Theta  \ln{\left| \dfrac{eV_\mathrm{DC}-\varepsilon_d}{eV_\mathrm{DC}-(\varepsilon_d+U)} \right|}$
& \cite{reinagalvez2025contrasting}  \\
\cmidrule(lr){2-4}
& Spin-transfer torque from the spin accumulation
& $\dfrac{\Gamma_\mathrm{ QI \leftrightarrow T}(t) P_\mathrm{T}}{2\pi\hbar} \sin\Theta \left(f_T^-(\varepsilon_d)-f_T^+(\varepsilon_d + U) \right) $
& \cite{reinagalvez2025contrasting,J_Reina_Galvez_2021, J_Reina_Galvez_2023} \\

\midrule

Phenomenological
& None
& 
$A\cos(\omega_{\mathrm{RF}} t) \sigma_x$
& \cite{Broekhoven2024} \\

\bottomrule
\end{tabularx}
\newline
\vspace*{2mm}
\begin{minipage}{\textwidth}
\scriptsize
$^{\mathrm{a}}$ $\mathcal{F}$ is a function of the rates similar to the one found in $B_{\mathrm{exch}}$, i.e. the logarithmic  dependence, but does not include $\sin\Theta$.\newline
$^{\mathrm{b}}$ $\Delta E$ is the energy difference between two pairs of states.
\end{minipage}
\label{tab:Driving}
\end{table*}

\begin{table*}[ht]
\renewcommand{\arraystretch}{1}
\centering
\caption{Reported scaling relations between ESR-STM observables and experimental control parameters. The table lists representative experimental systems and theoretical models in which the same functional dependence is observed. The observables are the Rabi rate $\Omega$, the ESR resonance frequency \saba{$f_{\rm res}$}, and the Rabi coherence time $T_2$, while the control parameters include the RF voltage amplitude $V_\mathrm{RF}$, the DC tunneling current $I_\mathrm{DC}$, and the DC bias voltage $V_\mathrm{DC}$. Theoretical models are denoted by the prefixes H (Heisenberg), K (Kondo), and AIM (Anderson impurity model).} 
\vspace*{2mm}
\begin{tabularx}{\textwidth}{l l X X}
\toprule
\textbf{Property} & \textbf{Relation} & \textbf{Experiments} & \textbf{Theory} \\
\midrule \midrule
\multirow{1}{*}{$\Omega \propto V_\mathrm{RF}$} 
 & Linear & Ti\cite{KaiYang_science2019, Wang2023Universal, Hong2024stark,Seifert2020LongitudinalMicroscope}, FePc\cite{Huang2025Ferrimagnets,Willke2021ControlMolecules}, Fe\cite{Seifert2020LongitudinalMicroscope}, $\mathrm{V_S^{-}}$ (sulfur vacancy)\cite{Willke2026}  & H\cite{Lado_Ferron_prb_2017,Ferron2019}, K\cite{Ye2024theory,J_Cuevas_C_Ast_ESR_theory_2024}, AIM\cite{reinagalvez2025contrasting, Cao2025} \\
\midrule

\multirow{2}{*}{$\Omega \propto I_\mathrm{DC}$} 
 & Linear & Ti\cite{KaiYang_science2019} & K\cite{Ye2024theory}, AIM\cite{reinagalvez2025contrasting, Cao2025} \\
 & None &  & H\cite{Lado_Ferron_prb_2017,Ferron2019}, K\cite{J_Cuevas_C_Ast_ESR_theory_2024} \\
\midrule

\multirow{2}{*}{$ 1/T_2 \propto I_\mathrm{DC}$} 
 & Linear & Ti\cite{KaiYang_science2019}, FePc\cite{Willke2021ControlMolecules}  &  K\cite{Ye2024theory, J_Cuevas_C_Ast_ESR_theory_2024}\\
 & Other &   & AIM\cite{reinagalvez2025contrasting} \\
 & None &  & H\cite{Lado_Ferron_prb_2017,Ferron2019} \\
\midrule

\multirow{3}{*}{\saba{$f_{\rm res} \propto V_\mathrm{DC}$}} 
 & Linear & PTCDA\cite{Taner2024}  & H\cite{Ferron2019} \\
 & Other & Ti\cite{zhang2025controllingexchangefieldsurface,Kot2023ElectricControl}, FePc\cite{zhang2025controllingexchangefieldsurface,Greule2025spincontrol}, $\mathrm{V_S^{-}}$\cite{Willke2026} &H\cite{Lado_Ferron_prb_2017}, K\cite{Ye2024theory}, AIM\cite{reinagalvez2025contrasting}  \\
 & None &  &K\cite{J_Cuevas_C_Ast_ESR_theory_2024}  \\
\midrule
\multirow{3}{*}{$\Omega \propto V_\mathrm{DC}$} 
 & Linear &  & K\cite{Ye2024theory}  \\
 & Other & Pentacene\cite{Stepan2024pentacene}, FePc\cite{Greule2025spincontrol} & AIM\cite{reinagalvez2025contrasting}  \\
  & None &  & H\cite{Lado_Ferron_prb_2017,Ferron2019}, K~\cite{J_Cuevas_C_Ast_ESR_theory_2024}  \\
\midrule
\multirow{2}{*}{$T_2 \propto V_\mathrm{DC}$} 
 & Other & Pentacene\cite{Stepan2024pentacene} &K\cite{J_Cuevas_C_Ast_ESR_theory_2024}, AIM\cite{reinagalvez2025contrasting}  \\
  & None &  & H\cite{Lado_Ferron_prb_2017,Ferron2019}, K\cite{Ye2024theory} \\
\bottomrule
\end{tabularx}
\label{tab:comparison}
\end{table*}

\subsubsection{Continuous-wave ESR spectra}
It is natural to start by comparing continuous-wave (CW) ESR spectra which are widely available and a standard characterization method in ESR-STM. Simulated ESR spectra obtained from diagonalization of a Heisenberg Hamiltonian, Kondo, and Anderson impurity models all compare well with the measured ESR signal \corr{measured on} a Ti atom \corr{in a Ti-Ti dimer} on an MgO/Ag substrate~\cite{Yang2017Engineering}, see Fig.~\ref{fig:fig5}. All approaches reproduce the experimental signal in terms of its line shape, peak intensity and peak splittings. We note that in particular the line shape depends sensitively on the geometric relation between the QI spin and the tip magnetization axis and is therefore not a good metric to compare different experiments~\cite{JK2021vectormagneticfield,reinagalvez2025contrasting}. Despite their different physical assumptions, the three models yield nearly identical ESR spectra in Fig.~\ref{fig:fig5}. This indicates that CW spectroscopy of the ESR signal alone is insufficient to determine which microscopic model best describes the ESR driving mechanism in ESR-STM. It is therefore necessary to compare scaling relations between observables \corr{of ESR-STM experiments} and external control parameters with the corresponding predictions of theoretical models. 

\begin{figure}[ht!]
\centering
\begin{tabular}{cc}
\reprobox{0.23\columnwidth}{2.8cm}{\rev{\textbf{(a)} Heisenberg model\\ simulation\\ (this work)}} &
\reprobox{0.23\columnwidth}{2.8cm}{\rev{\textbf{(b)} Kondo model\\ Adapted from\\ Ref.~\cite{Ye2024theory}\\ (APS, 2024)}} \\[4pt]
\reprobox{0.23\columnwidth}{2.8cm}{\rev{\textbf{(c)} Anderson impurity\\ model simulation\\ (this work)}} &
\reprobox{0.23\columnwidth}{2.8cm}{\rev{\textbf{(d)} Experimental Ti\\ Adapted from\\ Ref.~\cite{Yang2017Engineering}}} \\
\end{tabular}
\caption{\textbf{Comparison of Simulated and Experimental ESR Signals.}
Simulated ESR spectra based on
(a) the Heisenberg model;
\saba{(b)} the Kondo model and (c) the Anderson impurity model. (d) Experimentally measured ESR signal for a Ti atom~\cite{Yang2017Engineering}.
Panel (b) is dapted from Ref.~\cite{Ye2024theory} with permission (Copyright 2024 American Physical Society).
}
\label{fig:fig5}
\end{figure}

\subsubsection{Dependence of \texorpdfstring{$\Omega$}{W} on \texorpdfstring{$ V_\mathrm{RF}$}{VRF}.}

In ESR-STM, detection of coherent spin dynamics relies on monitoring changes in the tunneling current when the frequency of the RF-modulated bias matches the Larmor frequency of the spin in the junction~\cite{Yang2017Engineering}. Under resonant excitation, coherent Rabi oscillations between two spin states are induced, and the Rabi frequency $\Omega$ is extracted from time-resolved or pulsed measurements while systematically varying the amplitude of the RF-modulated bias voltage~\cite{KaiYang_science2019, Willke2021ControlMolecules}.

\begin{figure}[ht!]
\centering
\includegraphics[width=0.7\columnwidth]{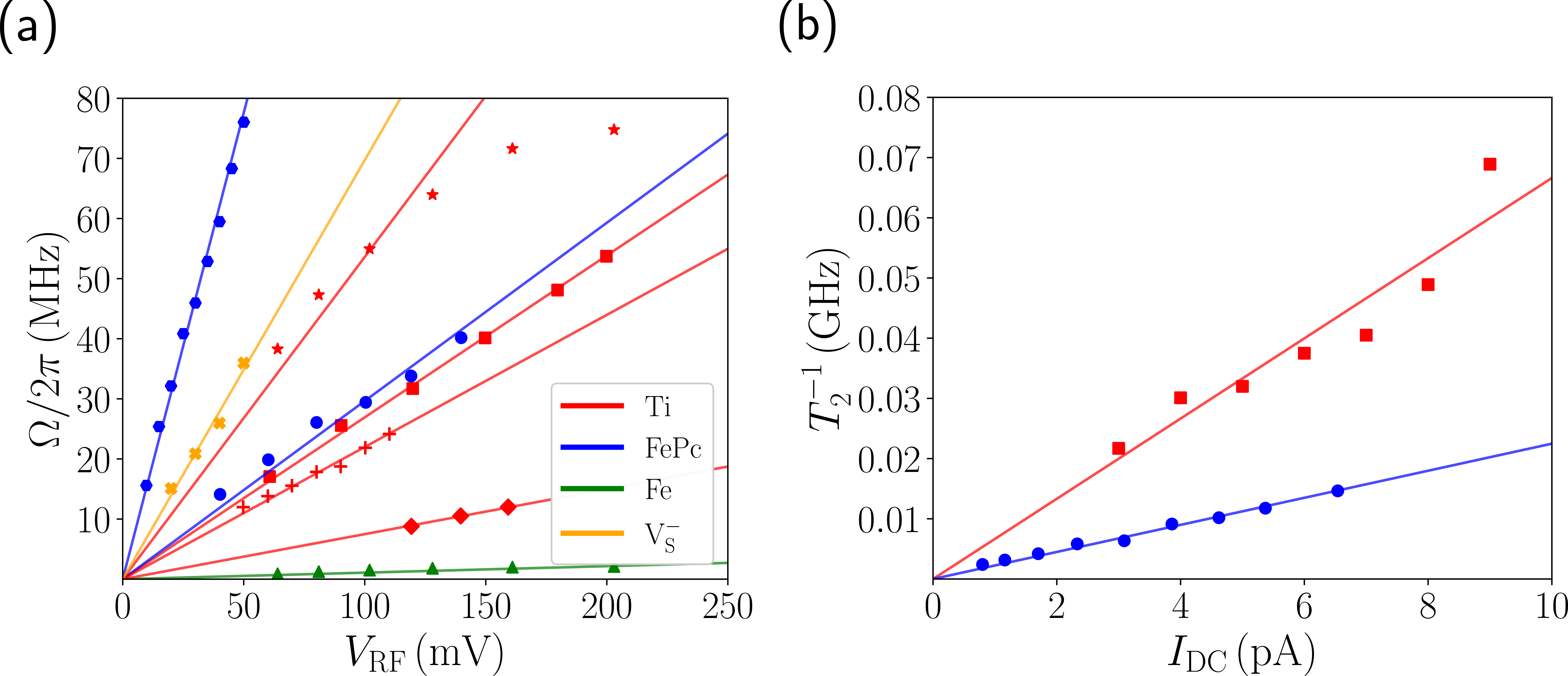}
\caption{
(a) Rabi frequency $\Omega/2\pi$ as a function of the applied radio-frequency voltage $V_{\mathrm{RF}}$ for individual Ti and Fe atoms, and FePc molecules. Solid lines are linear fits constrained to pass through the origin, illustrating the proportionality $\Omega \propto V_{\mathrm{RF}}$. The extracted slopes are
$0.075$ (Ti~\cite{KaiYang_science2019}),
$0.27$ (Ti~\cite{Wang2023Universal}),
$0.22$ (Ti~\cite{Hong2024stark}),
$0.54$ (Ti~\cite{Seifert2020LongitudinalMicroscope}),
$0.30$ (FePc~\cite{Willke2021ControlMolecules}),
$1.55$ (FePc~\cite{Huang2025Ferrimagnets}), and
$0.0108$ (Fe~\cite{Seifert2020LongitudinalMicroscope}),
$0.70$ ($\mathrm{V_S^-}$ (sulfur vacancy)~\cite{Willke2026}),
all in units of $\mathrm{MHz/mV}$.
(b) Decoherence rate $T_2^{-1}$ as a function of the DC tunneling current $I_{\mathrm{DC}}$ for Ti atoms and FePc molecules. Linear fits constrained to pass through the origin yield slopes of
$6.7\times10^{-3}~\mathrm{GHz/pA}$ (Ti~\cite{Wang2023Universal}) and
$2.25\times10^{-3}~\mathrm{GHz/pA}$ (FePc~\cite{Willke2021ControlMolecules}).
}
\label{fig:fig6}
\end{figure}

\corr{The Rabi frequency was found to scale linearly with the RF voltage amplitude in all experiments, virtually independent of the adsorbate and substrates. This includes measurements on $S=1/2$, such as Ti atoms~\cite{KaiYang_science2019, Wang2023Universal, Hong2024stark},  FePc molecules~\cite{Willke2021ControlMolecules, Huang2025Ferrimagnets}, and charged point defects in MoS$_2$, indicative of a general mechanism that does not depend on a specific system. Interestingly, Fe, a high-spin ($S=2$) system with significant anisotropy~\cite{Baumann_Paul_science_2015,Baumann2015PRL, wolfdelgado2020} also showed strict linear dependence even in a DC bias regime exceeding the magnetic anisotropy energy ($V_\mathrm{DC}>20 $ mV)~\cite{Willke2018Probing,Seifert2020LongitudinalMicroscope}, where it has been suggested that the excited states of the atom influence the line-shape of the ESR signal~\cite{Shavit2019}.  These results are} summarized in Fig.~\ref{fig:fig6}(a). 

As for Heisenberg-based models, both piezoelectric model~\cite{Lado_Ferron_prb_2017} and $g$-factor modulation mechanism~\cite{Ferron2019} predict that the Rabi rate scales linearly with $V_{\rm RF}$. Specifically, Eq.~\eqref{eq:Omega-heisenberg} derived from the piezoelectric model and Eqs.~\eqref{eq:g-mod-rabi} and \eqref{eq:gfactor2} from the $g$-factor modulation mechanism lead to the scaling relation $\Omega \propto V_{\rm RF}.$

In Kondo-based models, the same linear scaling arises either from a direct coupling of the microwave field to a transverse magnetic field in the STM junction or from the mean-field exchange interaction with a spin-polarized tip. In antenna-driven formulations, electromagnetic simulations show the following relation between Rabi frequency and magnitude of magnetic field in the junction (see section VI of Ref.~\cite{J_Cuevas_C_Ast_ESR_theory_2024})
\begin{equation}
\label{eq:omega-kondo1}
    \Omega= \frac{g \mu_\mathrm{B}}{\hbar}B_\mathrm{gap} \propto \frac{g \mu_\mathrm{B}}{\hbar} \frac{V_{\rm RF}}{c d} \propto V_\mathrm{RF},
\end{equation}
where $c$ denotes speed of light and $d$ is the gap size. In another work based on the Kondo model, Eq.~\eqref{eq:B-shak} shows that the Rabi rate is determined by the amplitude of the oscillating field $B_1$. The field strength can be expressed as (see Eq.~(S15) of Ref.~\cite{Ye2024theory})
\begin{equation*}
B_1 \sim \tau P_\mathrm{T}  D_{\rm T} V_\mathrm{RF},
\label{eq:B1-kondo}
\end{equation*}
where $P_\mathrm{T}$ is the spin-polarization of the tip and $D_T$ is DOS of the tip at the Fermi level. It is straightforward to derive $\Omega \propto V_{\rm RF}$.

An equivalent linear dependence emerges within the Anderson impurity model in all driving regimes~\cite{reinagalvez2025contrasting}. The RF voltage modulates the spin-dependent tunneling amplitude via the time-dependent hopping $t_\mathrm{T}(t)=t^{0}_{\rm T}\left[1+A_T\cos(\omega t)\right]$, producing first-harmonic Rabi rates as:
\begin{equation}
\label{eq:omega-first}
\hbar \Omega = g \mu_{\mathrm{B}} B_{\mathrm{exch},0} A_T \sin{\Theta},
\end{equation}
where $B_{\mathrm{exch},0} \equiv B_{\mathrm{exch}}(A_T=0)$ \jose{with $A_T$ being the tip modulation amplitude.}
While its exact microscopic dependence on the externally applied RF bias $V_{\mathrm{RF}}$ is not yet established, 
the observed $\Omega\propto A_T$ implies $A_T$ scales at most linearly with $V_{\mathrm{RF}}$. 
Ref.~\cite{J_Reina_Galvez_2019} was the first to introduce a closely related form of the Rabi frequency as the primary driving mechanism on top of a cotunneling QME \jose{and also led to $\Omega \propto A_{\rm T}$}.


Even though Heisenberg, Kondo, and Anderson impurity models suggest physically distinct microscopic coupling mechanisms between the RF bias and the spin (see also Tab.~\ref{tab:Driving}), they all predict an effective transverse driving field whose amplitude scales linearly with the RF voltage. Consequently, the measured Rabi frequency follows the universal linear relation \corr{$\Omega= k V_\mathrm{RF}$ (where the proportionality constant $k$ depends on the measurement conditions, see Fig.~\ref{fig:fig6}(a))}. The observation of this linear dependence alone therefore does not uniquely identify the dominant driving mechanism (see the first row of Tab.~\ref{tab:comparison}).

\subsubsection{Dependence of \texorpdfstring{$\Omega$}{W} on \texorpdfstring{$I_\mathrm{DC}$}{IDC}.}

The Rabi frequency is also found to increase linearly with the DC tunneling current when $V_\mathrm{DC}$ and $V_\mathrm{RF}$ are fixed, as observed for a Ti atom~\cite{KaiYang_science2019}. This observation indicates that increasing the tunneling current enhances the effective driving strength acting on the localized spin.
Heisenberg-based models cannot derive this relation between $\Omega$ and $I_{\rm DC}$ \jose{since they do not contain any explicit current dependence.}

In the spin-polarized Kondo model, the relation arises because both the DC current and the RF-induced driving field are controlled by the same exchange-mediated tunneling processes. In the low-bias, low-temperature regime, the DC current originated from elastic cotunneling scales as:
\begin{equation}
    I_\mathrm{DC} \propto J_{\rm ST}^2 D_{\rm S} D_{\rm T} V_\mathrm{DC}
    \label{eq:I_DC_phenomenological}
\end{equation}
where $J_{\rm ST}$ is the Kondo exchange between the substrate and tip, $D_{\rm S}$ and $D_{\rm T}$ are density of states of the substrate and tip reservoirs, respectively. Equation~\eqref{eq:I_DC_phenomenological} can be phenomenologically derived from the Ohm's law, $I = G V$, where $G$ is the conductance which is proportional to DOSs of the tip and the adatom. Also, the conductance is proportional to the square coupling strength due to higher order tunneling in Kondo model.
This scaling reflects the opening of a bias-dependent energy window for cotunneling processes~\cite{Ye2024theory}. Variations in tip height at fixed bias simultaneously modify $I_\mathrm{DC}$ and the effective transverse field, producing the following linear relation
\begin{equation}
    \Omega\propto I_\mathrm{DC}.
\end{equation}

A similar linear relation can be derived from the effective mean field effect of a spin-polarized tip~\cite{Ye2024theory}. There, the RF drive generates an effective oscillating magnetic field proportional to the spin-polarized tip hybridization, 
\begin{equation}
\label{eq:B1-cao}
    B_1^\mathrm{T} \propto \left(P_{\rm T},\Gamma_{\rm T},D_{\rm T}\right) \, V_\mathrm{RF} \, f'(eV_\mathrm{DC}),
\end{equation}
where $P_{\rm T}$ is the tip spin polarization, $\Gamma_{\rm T} $ the tip–impurity hybridization strength, $D_{\rm T}$ the tip density of states in dimensionless units and $f'$ is the derivative of the Fermi function encoding the temperature and energy dependence of the transport integral~\cite{Ye2024theory}. This yields the following Rabi rate:
\begin{equation}
\label{eq:Omega-AIMCao}
    \Omega \sim \frac{g \mu_\mathrm{B}}{\hbar}B_1^\mathrm{T} \propto D_T V_\mathrm{RF} f'(eV_\mathrm{DC}) 
\end{equation}
At the same time, the DC tunneling current scales as:
\begin{equation}
\label{eq:I-prop}
    I_\mathrm{DC} \propto \Gamma_{\rm T} D_{\rm T} F(V_\mathrm{DC}, T),
\end{equation}
with $\Gamma_{\rm T}$ the tip–impurity hybridization strength and $F(V_\mathrm{DC}, T)$ a transport integral over the tip and substrate densities of states weighted by Fermi functions. At fixed $V_\mathrm{DC}$, variation in junction transparency (e.g. tip height) primarily modifies $\Gamma_T$, and thus $P_{\rm T} \Gamma_{\rm T} D_{\rm T}$, simultaneously tuning $I_\mathrm{DC}$ and the effective driving field and producing an linear relation $\Omega \propto I_\mathrm{DC}$~\cite{Cao2025}.

Unlike the  RF voltage amplitude dependence, a linear scaling of the Rabi frequency with the DC tunneling current is not a generic feature of all driving mechanisms. It arises specifically when transport and spin driving are governed by the same spin-dependent tunneling processes, as in spin-polarized Kondo and Anderson impurity models, as shown in the second row of Tab.~\ref{tab:comparison}.  The observation of $\Omega \propto I_\mathrm{DC}$ therefore points to an exchange-mediated origin of the driving field, in contrast to mechanisms where the RF excitation is decoupled from DC transport.
In this sense, the current dependence provides a more selective experimental signature of the microscopic driving mechanism than RF-voltage scaling alone.

\subsubsection{Dependence of \texorpdfstring{$1/T_2$}{1/T2} on \texorpdfstring{$I_\mathrm{DC}$}{IDC}.}

The main source of decoherence for the QI are the tunneling electrons in the STM junction. It was observed that the Rabi coherence time $T_2$ decreases linearly with increasing DC tunneling current $I_\mathrm{DC}$ at fixed $V_\mathrm{RF}$ and $V_\mathrm{DC}$. Consequently, the decoherence rate $\Gamma=1/T_2$ scales linearly with $I_\mathrm{DC}$, as shown in Fig.~\ref{fig:fig6}(b) for a Ti atom~\cite{KaiYang_science2019} and FePc molecule~\cite{Willke2021ControlMolecules}. Due to back-scattering from the metallic substrate the coherence time is generally limited, even in the absence of tunneling current, and increasing the thickness of the insulator can significantly increase the lifetime of the QI~\cite{phark2025Ti3ML,Paul2017}.

In the Heisenberg model the spins are fully localized and no tunneling electrons are taken into account. Therefore, there is no clear way how to calculate the current or the influence it has on the coherence time in this model. 

Within the Kondo model, the transverse decoherence rate arises from the same exchange-mediated tunneling processes that generate the DC current. If pure dephasing is absent $T_2^*=0$, and thus $T_2=2T_1$ (See Eq.~\eqref{eq:1/T2}), or the loss of coherence in the system is solely caused by electrons tunneling through with a fixed polarization/orientation that destroys your quantum behavior. Since the current is a measurement of the flow of electrons passing through the junction, it is expected to be proportional to the rate $1/T_2$ at which coherence is destroyed. We obtain the following relation:
\begin{equation}
\label{eq:1/T2-Ye}
    T_2^{-1} = \alpha I_\mathrm{DC} 
\end{equation}
where $\alpha$ is a weakly geometry- and temperature-dependent constant of order unity~\cite{Ye2024theory}.

Alternatively, the total decoherence rate can be decomposed as 
\begin{equation}
\label{eq:T2-Ast1}
    T_2^{-1} = \Gamma_{ab}^{ad}+\frac{1}{2}(\Gamma_{ab}+\Gamma_{ba})+ \Gamma_0^\phi,
\end{equation}
where $\Gamma_{ab}^{ad}$ describes pure dephasing arising from elastic cotunneling processes, $\Gamma_{ab}$ and $\Gamma_{ba}$ are inelastic spin-flip relaxation rates, and $\Gamma_0^\phi$ accounts for intrinsic dephasing mechanisms such as substrate phonons or spontaneous emission~\cite{J_Cuevas_C_Ast_ESR_theory_2024}. As evident from Eq.~\eqref{eq:T2-Ast1}, the dominant current-dependent terms originate from elastic cotunneling–induced pure dephasing and inelastic spin-flip tunneling, while $\gamma_0$ provides a small, current-independent background. All tunneling-induced terms scale with the junction transmission and therefore with $I_\mathrm{DC}$, explaining the observed linear current dependence of ESR linewidths~\cite{J_Cuevas_C_Ast_ESR_theory_2024}.

At fixed bias and for weak energy dependence of the density of states, this yields
\begin{equation}
\label{eq:T2-Ast2}
    T_2^{-1} \propto \Gamma_\mathrm{tunnel} \propto I_\mathrm{DC}.
\end{equation}
In the low-current STM regime (pA–nA), tunneling-induced dephasing dominates over intrinsic contributions, resulting in a linear increase of the decoherence rate with tunneling current, as expressed in Eq.~\eqref{eq:T2-Ast2}, in agreement with experimentally observed current-dependent ESR coherence times~\cite{J_Cuevas_C_Ast_ESR_theory_2024}. 

For the Anderson impurity model and in the sequential transport regime, the decoherence rate takes the form
\begin{eqnarray}
\label{eq:T2-jose1}
     T_2^{-1} &=&\frac{\Gamma_{\mathrm{ QI \leftrightarrow T}}}{h}
     \left(1+\frac{A_T^2}{2} \right)
     \left[\pi+\arctan\left(\frac{eV_\text{DC}+\varepsilon_d}{\gamma_c}\right)-\arctan\left(\frac{eV_\text{DC}+\varepsilon_d+U}{\gamma_c} \right)\right]\nonumber \\
     &+&\frac{\Gamma_\mathrm{ QI\leftrightarrow S}}{h}\left[\pi+\arctan\left(\frac{\varepsilon_d}{\gamma_c}\right) -\arctan\left(\frac{\varepsilon_d+U}{\gamma_c} \right)\right],
\end{eqnarray}
where $\Gamma_{\mathrm{ QI \leftrightarrow T/S}}$ denote the undriven tip and surface couplings, $A_T$ is the modulation amplitude, $U$ the Coulomb repulsion energy, $\varepsilon_d$ the ionization energy, and $\gamma_c$ a finite lifetime broadening, see Eq.~(26) of Ref.~\cite{reinagalvez2025contrasting}. As captured by Eq.~\eqref{eq:T2-jose1}, the DC current follows the same arctangent bias dependence through sequential tunneling rates, yielding a functional, though generally nonlinear, relation between $1/T_2$ and $I_\mathrm{DC}$ consistent with dependencies of the form \saba{ $I_\mathrm{DC}\propto \Gamma_{\mathrm{ QI \leftrightarrow T}}\Gamma_{\mathrm{ QI \leftrightarrow S}}/(\Gamma_{\mathrm{ QI \leftrightarrow T}}+\Gamma_{\rm QI\leftrightarrow S})$}~\cite{jauho_wingreen_prb_1994}, where the current is linear in the tip coupling when $\Gamma_{\mathrm{ QI \leftrightarrow T}}<\Gamma_{\rm QI\leftrightarrow S}$ implying 
\begin{equation}
T_2^{-1}= \frac{1}{2T_1}  \propto I_\text{DC}.
\end{equation}

The linear scaling of the decoherence rate with tunneling current reflects a more general consequence of electron-induced noise in the STM junction. In exchange-based descriptions, elastic cotunneling and inelastic spin-flip processes contribute directly to dephasing and relaxation, making the tunneling current the dominant control parameter for $T_2^{-1}$. As a result, increasing $I_\mathrm{DC}$ uniformly enhances both pure dephasing and relaxation rates, yielding the experimentally observed linear behavior in the low-current regime. The Anderson impurity model reproduces this trend at low bias, while allowing for deviations from strict linearity at higher bias or stronger driving. As shown in the third row of Tab.~\ref{tab:comparison}, the Kondo and Anderson impurity models (in the low bias regime) are consistent with this observation.

Phenomenological models based on the Lindblad formalism did not account for current-induced decoherence. However, good agreement has been found in several simulations for Ti spins on Ag/MgO by assuming that the coherence time is lifetime-limited and $T_2=2 T_1$~\cite{Phark2023double, Hong2024stark}.

\subsubsection{Dependence of \texorpdfstring{\saba{$f_{\rm res}$}}{f0} on \texorpdfstring{$V_\mathrm{DC}$}{VDC}.}

\begin{figure}
\centering
\reprofig{0.45\columnwidth}{6cm}{\rev{%
(a) Adapted from Ref.~\cite{zhang2025controllingexchangefieldsurface}\\(CC BY 4.0).\\
(b),(c) Adapted from Ref.~\cite{Greule2025spincontrol}\\(CC BY 4.0).}}
\caption{\textbf{Resonance frequency dependency to $V_\mathrm{DC}$} (a) Universal, rescaled Ti resonance shift described by the exchange-field model. Adapted from Ref.~\cite{zhang2025controllingexchangefieldsurface} {\color{black} with permission. Copyright 2025 The Authors. Licensed under CC BY 4.0.}. 
(b) Non-linear spin-electric coupling in the ESR spectra on FePc molecule. Colormap of the ESR signal $\Delta I$ vs. $V_\mathrm{DC}$ and $f$ shows experimental data (left) and simulation (right) using the exchange-bias model; inset depicts the spin-polarized STM tip. Adapted from Ref.~\cite{Greule2025spincontrol} (c) Voltage dependence of $\Delta f = f_{\mathrm{res}}-g\mu_B B/h$, labeled as $\delta f$ in panel (a),with experimental data (blue) and model fits (black), right axis shows relative frequency change  $\Delta f/f_0$ with $hf_0=g\mu_B B$. Adapted from Ref.~\cite{Greule2025spincontrol} {\color{black} with permission. Copyright 2025 The Authors. Licensed under CC BY 4.0}.
}
\label{fig:fig7}
\end{figure}

In several ESR-STM experiments, including individual Ti atoms \cite{Kot2023ElectricControl}, FePc molecules~\cite{zhang2025controllingexchangefieldsurface,Greule2025spincontrol}, and PTCDA molecule~\cite{Taner2024}, the resonance frequency \saba{$f_{\rm res}$} exhibits a pronounced dependence on the DC bias voltage $V_\mathrm{DC}$. 
For Ti atoms and FePc molecules, experiments reveal nonlinear bias-induced shifts of \saba{$f_{\rm res}$}, as illustrated in Fig.~\ref{fig:fig7}.
For Ti atoms, Fig.~\ref{fig:fig7}(a) demonstrates a universal bias dependence of the resonance frequency across different spin-polarized tips. After rescaling to account for variations in the external magnetic field, $g$-factor, and tip polarization amplitude, all datasets collapse onto a single curve, indicating a common underlying mechanism governing the voltage-controlled resonance shift~\cite{zhang2025controllingexchangefieldsurface}. For FePc molecules, differential conductance spectra recorded at varying DC bias reveal highly nonlinear resonance shifts, as shown in Fig.~\ref{fig:fig7}(b). The extracted frequency shifts, summarized in Fig.~\ref{fig:fig7}(c), further highlight the bias dependence of \saba{$f_{\rm res}$}, consistent with an exchange-field-driven origin mediated by the magnetic STM tip~\cite{Greule2025spincontrol}.
By contrast, ESR measurements on PTCDA molecules show an approximately linear dependence of the resonance frequency on $V_\mathrm{DC}$, which could be attributed to the narrow DC voltage range~\cite{Taner2024}.


Within a Heisenberg exchange description of STM-ESR, the resonance frequency \saba{$f_{\rm res}$} exhibits a nonlinear and non-monotonic dependence on the tip and QI distance. In constant-current, variation of $V_\mathrm{DC}$ are compensated by adjustments of the distance, so increasing $\left| V_\mathrm{DC} \right|$ corresponds to an increased separation.
The resonance frequency is determined by an effective longitudinal field 
\begin{equation}
\label{eq:B-Ast1}
B_z^\mathrm{eff}(z)=B_z+B_z^\mathrm{exch}(z)+B_z^\mathrm{dip}(z),
\end{equation}
where $z$ is the distance between QI and tip (see also Eq.~(A1) of Ref.~\cite{Lado_Ferron_prb_2017}). The relation contains two competing distance-dependent contributions. We have
\begin{equation}
\label{eq:B-Ast2}
    B_z^\mathrm{exch}\propto -\exp{(-z/d_{\rm dec})}, B_z^\mathrm{dip}\propto z^{-3}
\end{equation}
\jose{with $d_{\rm dec}$ setting the length scale over which the exchange coupling decays.} $B_z^\mathrm{exch}$ dominates at short distances and reduces \saba{$f_{\rm res}$}, while a ferromagnetic dipolar field $B_z^\mathrm{dip}$ becomes increasingly relevant at larger separations and increases \saba{$f_{\rm res}$}. As implied by Eqs.~\eqref{eq:B-Ast1} and~\eqref{eq:B-Ast2}, the competition between these distance-dependent fields leads to a non-monotonic dependence of the ESR frequency on the tip and QI separation, and thus, in constant-current ESR-STM, on the applied bias..

\jose{The above considerations apply broadly to all models discussed so far, as they incorporate distance dependence either through the coupling $\Gamma_{\mathrm{ QI \leftrightarrow T}}$ or via the exchange interaction $J$. On the other hand, when measurements are performed at constant height, the analysis of the resonance position can help discriminate between different models. We now discuss the predictions of the present models in this specific scenario.}

According to the piezoelectric $g$-factor modulation mechanism~\cite{Ferron2019}, the $V_\mathrm{DC}$ alone produces a static piezoelectric displacement of the adatom, which shifts the anisotropic $g$-tensor in a similar way described in Eq.~\eqref{eq:gfactor2}
\begin{equation}
\label{eq:delta-g}
    \delta g_a =\left( \frac{\partial g_a}{\partial D} \frac{d D}{d z} + \frac{\partial g_a}{\partial F} \frac{d F}{d z}\right) \frac{q V_{\rm DC}}{k d} \propto V_{\rm DC} \quad (a = x, z),
\end{equation}
where $a=x,z$. This static modulation changes the Zeeman energy splitting and produces a linear shift in the resonance frequency as follows
\begin{equation}
\label{eq:deltaf}
   \saba{ \Delta f }\propto \frac{\mu_B}{h \vert \mathbf{gB}\vert} \left(g_x \delta g_x B_x^2 + g_z \delta g_z B_z^2\right) \propto V_\mathrm{DC}.
\end{equation}


\jose{In the model of Ref.~\cite{J_Cuevas_C_Ast_ESR_theory_2024}, the resonance position is treated as an input parameter, and the theory does not account for any shift induced by a DC electric field. This can be traced back to the fact that their Kondo description does not include the first-order terms that survive the Schrieffer--Wolff transformation when one of the electrodes is spin polarized.
In contrast, as discussed throughout this review, Kondo, cotunneling, and Anderson impurity models yield an effective exchange field generated by spin-polarized nonequilibrium electrons when treated consistently. This, in turn, produces a DC-bias-dependent shift of the resonance frequency \saba{$f_{\rm res}$} through the static component of the exchange field~\cite{reinagalvez2025contrasting,Ye2024theory,J_Reina_Galvez_2019,Braun_Konig_prb_2004,Weymann_2006_seq_cotu}. For the case of Ref.~\cite{Ye2024theory}, this contribution can be written as Eq.~\eqref{eq:B0}}
\begin{equation}
\label{eq:B0}
    B_0 = P_{\rm T} \frac{\Gamma_{\rm QI\leftrightarrow T}}{\pi g \mu_\mathrm{B}} D_T \int_{-\infty}^{\infty} \frac{1}{\epsilon - \varepsilon_d} f_T(\epsilon + eV_\mathrm{DC}) d\varepsilon,
\end{equation}
where $P_{\rm T}$ is the tip spin polarization, $D_T$ is the dimensionless density of states of the tip, considered constant in the wide-band limit, \saba{and $f_T$ is the Fermi distribution function \jose{defined in Ref.~\cite{Ye2024theory} as $f_T(\epsilon)=1/(1+e^{\epsilon/k_BT})$}. The resonance frequency is then given by $hf_{\rm res} \propto g \mu_\mathrm{B} (\mathbf{B}_\mathrm{ext} + \mathbf{B}_0 \cos\Theta )$}.
At zero or low temperatures, i.e. $k_B T\ll \varepsilon_d+eV_{\rm DC}$, the Fermi distribution function becomes a step function:
\begin{equation*}
    f_T(\epsilon + eV_{\rm DC}) \approx 1-\Theta(\epsilon+eV_{\rm DC}),
\end{equation*}
implying that
\begin{equation}
\label{eq:B0_T0}
    B_0 = P_{\rm T} \frac{\Gamma_{\rm QI\leftrightarrow T}}{\pi g \mu_\mathrm{B}} D_T \ln\left|\frac{eV_{\rm DC}+\varepsilon_d}{E_c+\varepsilon_d}\right|,
\end{equation}
where a cutoff energy $E_c\gg|\varepsilon_d|$ must be introduced to avoid divergences in the integration. In general, Eq.~\eqref{eq:B0_T0} shows a nonlinear dependence on $V_\mathrm{DC}$ unless $|eV_{\rm DC}|\ll |\varepsilon_d|$, a result more compatible with recent experiments~\cite{zhang2025controllingexchangefieldsurface,Greule2025spincontrol}. However, the exchange field in Eq.~\eqref{eq:B0} is computed in the limit $U\rightarrow+\infty$, which requires the introduction of the cutoff energy and neglects the exchange contribution for the opposite sign of the DC bias.

As mentioned above, the Anderson impurity model also provides an effective magnetic field that produces a frequency shift. However, in Ref.~\cite{reinagalvez2025contrasting}, the resonance frequency exhibits a logarithmic dependence that includes the Coulomb repulsion energy $U$ and, by extension, it covers both signs of the DC bias.
The frequency shift arising from virtual spin-flip charge fluctuations for finite $U$ takes the form
\begin{equation}
\label{eq:f0-AIM}
  \saba{h\Delta f}=g\mu_\text{B} {B}_\text{exch}\cos\Theta = \frac{\cos\Theta}{2\pi} \left[P_\mathrm{T}  \Gamma_{\rm QI\leftrightarrow T} \ln \left|\frac{eV_\text{DC}-\varepsilon_d-U}{eV_\text{DC}-\varepsilon_d}\right| \right] .
\end{equation}
Here, there is no longer any need to introduce a cutoff energy, since for $E_c\gg |\varepsilon_d|,\ \varepsilon_d+U$, the cutoff-dependent contribution cancels out in Eq.~\eqref{eq:f0-AIM}. This is one of the advantages of considering finite $U$: the model does not depend on an artificial cutoff energy. The bias dependence is antisymmetric around the symmetry point $eV_\mathrm{DC}=\varepsilon_d+U/2$. At low bias ($|eV_\mathrm{DC}|\ll|\varepsilon_d|,\ \varepsilon_d+U$), the shift appears approximately linear (see Fig.~\ref{fig:fig4}), while showing a logarithmic increase near the charge-degeneracy thresholds, where $eV_\mathrm{DC}\approx\varepsilon_d,\ \varepsilon_d+U$. In the STT regime, these virtual processes are suppressed, leading to a reduced bias dependence of the resonance frequency~\cite{reinagalvez2025contrasting}. Equation~\eqref{eq:f0-AIM}, as well as frequency shift given by $B_0$, depends on the homodyne angle $\Theta$, which allows one to infer the relative orientation of $P_{\rm T}$ with respect to the external magnetic field by measuring the frequency shift for two magnetic-field orientations differing by $90^\circ$, as explicitly shown in Ref.~\cite{zhang2025controllingexchangefieldsurface}. The same angular dependence also predicts that no frequency shift occurs when the tip polarization is perpendicular to the external magnetic field. 

The DC-bias dependence at constant height of the ESR resonance frequency reflects the transport regime that governs the coupling between the localized spin and tunneling electrons. 
When the many-body effects from a polarized tunneling current within the Kondo or Anderson impurity model are included, the resonance frequency follows the linear dependency to $V_\mathrm{DC}$ near the electron–hole symmetry point, while it is logarithmic near the charging thresholds. Mechanisms based on static electric-field effects, such as piezoelectric $g$-factor modulation, instead yield linear bias-induced shifts (see the fourth row of Tab.~\ref{tab:comparison}). The strong non-linear features of some experiments for FePc~\cite{zhang2025controllingexchangefieldsurface,Greule2025spincontrol} are a strong indicator that the exchange-bias model captures the essential physical processes for quantum impurities in the CB (FLT-driven) regime.

\subsubsection{Dependence of \texorpdfstring{$\Omega$}{W} and \texorpdfstring{$T_2$}{T2} on \texorpdfstring{$V_\mathrm{DC}$}{VDC}.}


In ESR-STM experiments on a single pentacene molecule \cite{Stepan2024pentacene}, the coherence time $T_2$ was found to decrease rapidly as the DC bias voltage approaches the singly occupied molecular orbital (SOMO). This behavior was attributed to the sharp increase in tunneling current through the molecule and the associated enhancement of current-induced decoherence processes~\cite{Stepan2024pentacene}. Related behavior has been observed in ESR-STM measurements on FePc molecule, where a bias-dependent exchange field was shown to induce a nonlinear dependence of the Rabi frequency $\Omega$ on $V_\mathrm{DC}$ once virtual tunneling into higher-lying molecular orbitals becomes efficient~\cite{Greule2025spincontrol}. Concurrently, the bias dependence signals a crossover from field-like torque (FLT) to spin-transfer torque (STT) dominated driving once tunneling into molecular orbitals becomes energetically allowed, leading to a distinct bias dependence of both the Rabi frequency $\Omega$ and the coherence time $T_2$.



There is no explicit dependence of the Rabi frequency $\Omega$ or the coherence time $T_2$ on the DC bias voltage based on the Heisenberg \jose{model, and therefore it cannot reproduce the experimental evidence.} 

 
\jose{There is no explicit dependence on $V_{\rm DC}$ for a Rabi frequency generated by an AC magnetic field in the junction region produced by the radiation field~\cite{J_Cuevas_C_Ast_ESR_theory_2024}. However, as introduced previously, a consistent treatment of the Kondo model leads to a Rabi frequency $\Omega$ that originates from the alternating component $B_1$ of the effective magnetic field exerted by the spin-polarized STM tip, with $\hbar\Omega = g \mu_\mathrm{B} B_1$.} Here, $B_1$ arises from the mean-field exchange interaction between the localized spin and the RF-modulated spin polarization of the tip~\cite{Ye2024theory}. The effective magnetic field can be written explicitly as (Eq.~\eqref{eq:B-eff})
\begin{equation}
\label{eq:B-eff}
B_\mathrm{T}^\mathrm{eff}(t) = \frac{P_{\rm T} \Gamma_{\mathrm{ QI \leftrightarrow T}}D_T}{\pi g \mu_\mathrm{B}} \int d\epsilon \,\frac{1}{\epsilon - \varepsilon_d} \,
f_T \bigl(\epsilon + eV_\mathrm{DC} + eV_\mathrm{RF}\sin(\omega_\mathrm{RF} t)\bigr),
\end{equation}
Expanding to first order in the RF voltage yields $B_\mathrm{T}^\mathrm{eff}(t) \simeq B_0 + B_1 \sin(\omega_\mathrm{RF} t)$ and, here $B_1$ is written as in Eq.~\eqref{eq:B1-prop}
\begin{equation}
\label{eq:B1-prop}
B_1 = \frac{P_{\rm T} \Gamma_{\mathrm{ QI \leftrightarrow T}} D_T\, eV_\mathrm{RF}}{\pi g\mu_B} \int d\epsilon \,\frac{1}{\epsilon - \varepsilon_d}\, \frac{\partial f_T(\epsilon + eV_\mathrm{DC})}{\partial \epsilon}.
\end{equation}
The Rabi frequency therefore scales linearly with $V_\mathrm{RF}$, while its dependence on the DC bias is encoded in the energy window sampled by the integral. \jose{At zero temperature, $\partial f_T(\epsilon + eV_\mathrm{DC})/\partial \epsilon$ is sharply peaked at $\epsilon=- eV_{\rm DC}$, leading to
\begin{equation}
\label{eq:B1-prop_T0}
B_1 = -\frac{P_{\rm T} \Gamma_{\mathrm{ QI \leftrightarrow T}} D_{\rm T}\, eV_\mathrm{RF}}{\pi g\mu_B} \frac{1}{eV_{\rm DC} + \varepsilon_d}
\underbrace{\implies}_{eV_{\rm DC}\ll \varepsilon_d}
B_1 \approx -\frac{P_{\rm T} \Gamma_{\mathrm{ QI \leftrightarrow T}} D_{\rm T}\, eV_\mathrm{RF}}{\pi g\mu_B}
\left(\frac{1}{\varepsilon_d} - \frac{eV_{\rm DC}}{\varepsilon_d^2}\right).
\end{equation}
\saba{Therefore, in the low-bias regime, the $B_1$ field consists of a constant term plus a linear correction in the DC bias, whereas at larger bias the integral tends to zero for $ eV_{\rm DC}\gg |\varepsilon_d|$, effectively suppressing the Rabi frequency.}}

A complementary Kondo-based treatment focuses on the dissipative dynamics of the localized spin under finite bias, rather than on the coherent exchange field responsible for ESR driving~\cite{J_Cuevas_C_Ast_ESR_theory_2024}.
Recall from Eq.~\eqref{eq:T2-Ast1}, the total decoherence rate can be expressed as:
\begin{equation}
\label{eq:1T2-Ast}
    T_2^{-1} = \Gamma_{ab}^{ad}(V_\mathrm{DC}) + \frac{1}{2}\bigl(\Gamma_{ab} + \Gamma_{ba}\bigr) + \Gamma_0^\phi,
\end{equation}
where $\Gamma_{ab}^{ad}(V_\mathrm{DC})$ describes dephasing induced by elastic cotunneling, $\Gamma_{ab}$ and $\Gamma_{ba}$ are inelastic spin-flip excitation and relaxation rates, and $\Gamma_0^\phi$ accounts for bias-independent intrinsic dephasing. Both elastic and inelastic tunneling rates are governed by spectral functions involving products of lead Green’s functions weighted by shifted Fermi distributions~\cite{J_Cuevas_C_Ast_ESR_theory_2024}. In the wide-band, high-bias limit with energy-independent densities of states, the tunneling-induced contributions vanish at $V_\mathrm{DC}=0$ and increase approximately linearly and symmetrically with $\lvert V_\mathrm{DC}\rvert$, reflecting the opening of a nonequilibrium energy window for cotunneling and spin-flip processes~\cite{J_Cuevas_C_Ast_ESR_theory_2024}. In this regime, one finds an approximate scaling (Eq.~\eqref{eq:1/T2-Ast2})
\begin{equation}
\label{eq:1/T2-Ast2}
    T_2^{-1} \simeq \Gamma_0^\phi + c\,\lvert eV_\mathrm{DC}\rvert,
\end{equation}
with $c$ set by the microscopic tunneling parameters.

\jose{In the Anderson impurity model, the bias dependence of both the Rabi frequency and the coherence time is governed by the specific transport regime~\cite{reinagalvez2025contrasting} and follows a similar principle to that in~\cite{Ye2024theory}, while simplifying the modeling by considering only a tunneling modulation. The introduction of Bessel functions in such a model would be a natural step toward a more complete description of the driving mechanism and would likely lead to a result closer to that in Eq.~\eqref{eq:B1-prop_T0}, with the additional inclusion of a finite Coulomb repulsion energy $U$. In the current theoretical description, within the FLT regime, when the DC bias lies inside the Coulomb-blockade window $(\varepsilon_d < eV_\mathrm{DC} < \varepsilon_d + U)$, the corresponding Rabi rate $\Omega_\mathrm{FLT}$ originates from virtual spin-flip processes mediated by the spin-polarized tip and is encoded in the imaginary part of a single complex spin-flip rate. We write it here again for convenience:
\begin{equation}
\label{eq:omega-flt1}
    2\hbar\Omega_\mathrm{FLT} = g\mu_\text{B} B_\text{exch}\sin\Theta = \frac{\sin\Theta}{2\pi} \left[P_\mathrm{T} \Gamma_{\rm QI\leftrightarrow T}(t) \ln \left|\frac{eV_\text{DC}-\varepsilon_d-U}{eV_\text{DC}-\varepsilon_d}\right| \right].
\end{equation}
T}his leads to logarithmic enhancements as the bias approaches the charge thresholds at $\varepsilon_d$ and $\varepsilon_d + U$. \jose{On the other hand, the decoherence rate follows Eq.~\eqref{eq:T2-jose1}}, which is minimized at electron-hole symmetry, $eV_{\rm DC}=\varepsilon_d + U/2$, where contributions from virtual charge fluctuations at the ionization and charging thresholds cancel out, and $T_2$ is maximal.

Beyond the charge thresholds, $(eV_\mathrm{DC} \ll \varepsilon_d$ or $eV_\mathrm{DC} \gg \varepsilon_d + U)$, the system enters the STT-driven regime. In this regime, the relevant Rabi rate is set by the real part of the same complex spin-flip rate and becomes proportional to the spin-polarized current through the impurity:
\jose{\begin{equation}
\label{eq:omega-stt1}
    \Omega_\mathrm{STT} = \dfrac{\Gamma_\mathrm{QI \leftrightarrow T}(t) P_\mathrm{T}}{2\pi \hbar} \sin\Theta \left(f_T^-(\varepsilon_d)-f_T^+(\varepsilon_d + U) \right),
\end{equation}}
where $I_\mathrm{pol}(V_\mathrm{DC})$ denotes the spin-polarized component of the tunneling current, proportional to the spin polarization of the tip $P_\mathrm{T}$ and the bias-dependent sequential tunneling rate. As the bias crosses the charge thresholds, $\Omega_\mathrm{STT}$ rises sharply and then saturates together with the current at large $\left|V_\mathrm{DC}\right|$. In this high-bias regime, the coherence time collapses to a short, nearly bias-independent value set by the saturated sequential tunneling rate:
\jose{\begin{equation}
\label{eq:1T2-AIM}
    T_2^{-1} \simeq \frac{\Gamma_\mathrm{QI \leftrightarrow T}}{2\hbar} \left(1+\frac{A_T^2}{2}\right) \simeq \mathrm{const},
\end{equation}
where $\Gamma_\mathrm{QI \leftrightarrow T}$} is the 
coupling to the tip in the absence of AC modulation~\cite{reinagalvez2025contrasting}. This regime corresponds to overdamped spin dynamics and the loss of coherent ESR control.

The DC bias voltage primarily controls ESR-STM dynamics by selecting the transport regime of the junction. In the FLT regime, where tunneling proceeds via virtual charge fluctuations, the spin experiences an effective field-like exchange interaction with the spin-polarized tip, enabling coherent Rabi driving while preserving relatively long coherence times. As the bias crosses the charge thresholds, the system enters the STT regime, in which spin-polarized currents transfer angular momentum irreversibly, leading to a rapid increase and eventual saturation of the Rabi frequency together with a collapse of $T_2$, signaling overdamped spin dynamics and loss of coherent ESR control. 

Although the Heisenberg model captures the existence of exchange-driven ESR, only Kondo and Anderson impurity models reproduce the experimentally observed nonlinear bias dependence of $\Omega$ and the suppression of $T_2$ at finite bias (see Tab.~\ref{tab:comparison}). The bias dependence of $\Omega$ and $T_2$ therefore provides a direct experimental signature of the transition from exchange-driven, field-like control to current-driven spin-transfer torque.

\subsection{Key insights from comparisons between experiments and models}

CW spectroscopy merely establishes that a localized spin can be coherently driven, but it does not constrain how the driving field is generated, how it couples to the spin, or how non-equilibrium tunneling processes modify the spin Hamiltonian. Discriminating between \corr{the suggested} mechanisms therefore requires examining the systematic dependencies of ESR observables on experimental control parameters such as RF amplitude, DC bias, tunneling current, and tip and QI separation. 

\corr{A key insight from comparing different theoretical models is that not all observables are equally useful for distinguishing between different models. In particular, the linear dependence of the Rabi rate on the RF voltage amplitude which is observed in experiments (see the first row of Tab.~\ref{tab:comparison}), is predicted by all three theory models. This universality implies that $\Omega-V_{\mathrm{RF}}$ scaling does not provide information about the underlying driving mechanism. 
In contrast, the DC bias dependence of ESR observables reveals qualitative differences between models. 

While all models predict nonlinear shifts of the resonance frequency with DC bias, only exchange field models capture the logarithmic and threshold-like behavior emerging near charge degeneracy points, including divergences associated with ionization and charging energies. Furthermore, the Anderson impurity model uniquely predicts a crossover between distinct driving regimes as a function of bias voltage, reflecting a transition from field-like torque to spin-transfer torque. This behavior has been observed experimentally for both Rabi rate and decoherence time by bringing the DC bias close to the SOMO charge resonance in pentacene~\cite{Stepan2024pentacene}. This behavior cannot be described by Heisenberg or Kondo models which are restricted to localized spin-spin interactions or spin fluctuations.}

\subsection{Figure of merit for coherent control in STM junction}

The comparison discussed above clarifies the driving mechanisms in ESR-STM and their experimental signatures. To assess how effectively these mechanisms enable coherent spin control in the STM junction, it is useful to condense the combined effects of drive strength and decoherence into a single figure of merit. 
In the context of quantum information the $S=1/2$, QI represents a prototypical quantum two-level system and can be therefore considered a qubit. For driven spin qubits, this figure of merit is given by the product $\Omega T_2$, which quantifies how many coherent Rabi oscillations can be executed within the decoherence time $T_2$~\cite{reinagalvez2025contrasting,Xuan2025}. Achieving large $\Omega T_2$ requires the combination of fast Rabi driving and long-lived coherence, and thus directly benchmarks the quality of quantum control in a given platform.

\begin{figure}
\centering
\reprofig{0.7\columnwidth}{5cm}{\rev{%
(a) Adapted from Ref.~\cite{KaiYang_science2019} (AAAS, 2019).\\
(b) Adapted from Ref.~\cite{Willke2021ControlMolecules}\\(Copyright 2021 The Authors, CC BY 4.0).}}
\caption{\textbf{Quantum Coherent Manipulation.}
Rabi oscillations measured via the tunneling current as a function of pulse width for varying $V_{\mathrm{RF}}$:
(a) \saba{Ti atom,} Adapted from~\cite{KaiYang_science2019}, {\color{black} with permission. Copyright 2019 AAAS.}
(b) FePc molecule. Adapted from Ref.~\cite{Willke2021ControlMolecules}. {\color{black} Copyright 2021 The Authors. Licensed under CC BY 4.0.}
}

\label{fig:fig8}
\end{figure}

Qubit platforms such as NV centers, superconducting circuits, trapped ions, and Ge hole quantum-dot qubits, which are capable of high-fidelity quantum operations, routinely achieve large Rabi rates and/or long coherence times, resulting in $\Omega T_2 > 10^2$ (Tab.~\ref{tab:OmegaT2_comparison}).
For a single FePc molecule on MgO/Ag substrate, the maximum $\Omega T_2 \approx 5$~\cite{Willke2021ControlMolecules}. Combining DFT calculations with Anderson impurity model simulations yields predictions that closely align with these experimental values~\cite{Corina2025}. Therefore, alternative approaches that enable longer coherence times with constant value of $\Omega$ are required to achieve larger values of $\Omega T_2$.

It has been demonstrated that for spins inside the STM junction Rabi oscillations can be achieved, but they are limited by decoherence arising from the tunneling current. The Ti atom (Fig.~\ref{fig:fig8}(a)) beneath the tip exhibits damped oscillations. The Rabi rate increases with the applied RF voltage $V_\mathrm{RF}$, while the envelope decays with a roughly constant timescale of only a few hundred nanoseconds. Fig.~\ref{fig:fig8}(b) presents analogous data for a single FePc molecule which shows a comparable  modest figure of merit $\Omega T_2=5$. These measurements quantitatively illustrate how tip‑proximal driving in the junction can reach a Rabi rate of tens of MHz but current‑induced relaxation places a significant constraint on coherence~\cite{Willke2021ControlMolecules}.

These limitations motivate alternative strategies in which the coherently driven spin is placed outside the STM junction. In such \textit{remote spin} approaches, the qubit is spatially separated from the tunneling current, thereby reducing measurement-induced decoherence, tip-induced environmental noise, and life time of the spin is only limited by the remaining scattering by the metal substrate which can further reduced by the use of thicker decoupling layers~\cite{phark2025Ti3ML,Paul2017}.


\begin{table*}[t]
\centering
\caption{Typical achievable values of the figure of merit $\Omega T_2$ for leading qubit platforms. Here $\Omega$ is the single-qubit Rabi rate and $T_2$ decoherence time.}
\label{tab:OmegaT2_comparison}
\begin{tabularx}{\textwidth}{l X X X c}
\toprule

\textbf{Platform} 
& $\boldsymbol{\Omega/2\pi}$ 
& $\boldsymbol{T_2}$ 
& $\boldsymbol{\Omega T_2}$
& \textbf{References}\\
\hline
\hline
NV center 
& $1$--$10$ MHz 
& $0.5$--$5$ ms 
& $10^{3}$--$10^{4}$ 
&\cite{BarGill2013NV}\\

Superconducting qubits (transmon) 
& $20$--$100$ MHz 
& $20$--$100$ $\mu$s 
& $10^{3}$--$10^{4}$
&\cite{Krantz2019superconducting}\\

Trapped-ion qubits 
& $10$--$500$ kHz 
& $1$--$100$ s 
& $10^{6}$--$10^{9}$ 
&\cite{Bruzewicz2019Trappedion}\\

Ge hole quantum-dot qubits 
& $10$--$200$ MHz 
& $1$--$20$ $\mu$s 
& $10^{2}$--$10^{3}$ 
&\cite{Jirovec2021Ge}\\
\bottomrule
\end{tabularx}
\end{table*}

\section{Multi-spin control}
\label{sec:multi_control}

\begin{figure*}[htbp!]
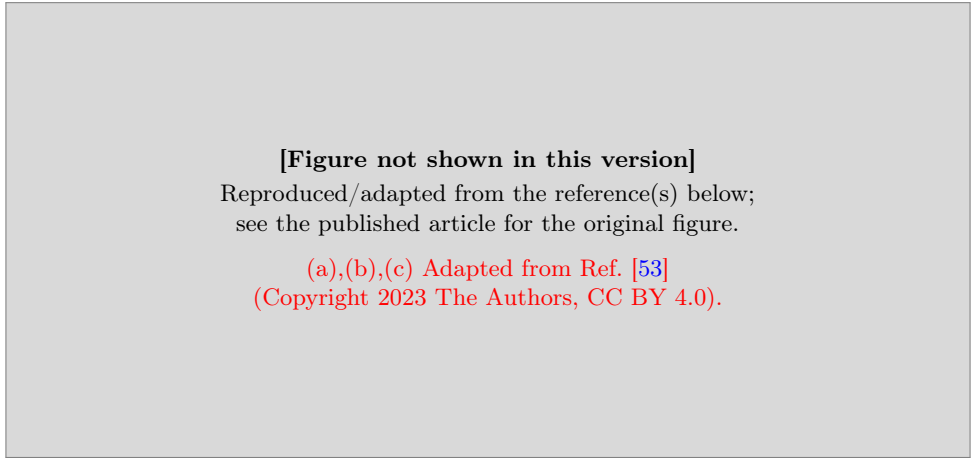

\centering
\reprofig{0.8\columnwidth}{6cm}{\rev{(a),(b),(c) Adapted from Ref.~\cite{Phark2023electric}\\(Copyright 2023 The Authors, CC BY 4.0).}}
\caption{\textbf{Multi-spin control.} 
(a) Experimental setup for electron spin resonance of a Ti spin exchange-coupled to a nearby Fe single-atom magnet on an MgO substrate. 
(b) Angular dependence of the peak splitting in continuous-wave ESR (CW-ESR) spectra of Ti–Fe pairs, defined as the frequency separation between the two resonance peaks arising from the exchange interaction between the Ti spin and the Fe atom.
(c) Measured Rabi rates extracted from Rabi oscillation measurements for an isolated Ti atom (gray) and two different Ti–Fe pairs (red and blue). 
(a), (b), and (c) adapted from Ref.~\cite{Phark2023electric}. {\color{black} Copyright 2023 The Authors. Licensed under CC BY 4.0.}
(d) Simulated angular dependence of the ESR frequency splitting for different exchange coupling strengths $J_0$, shown for comparison with the experimental data in panel (b).}
\label{fig:fig9}
\end{figure*}

After demonstrating coherent control of individual spins in ESR-STM, an important next step is the manipulation of multiple spins using a single STM junction. While spins located directly beneath the tip (sensor spin) can be efficiently driven through their interaction with the STM tip, spins positioned outside the tunnel junction (remote spin) couple only weakly to the tip because tunneling interactions decay exponentially with distance.
Remote spin control can nevertheless be achieved by placing a single-atom magnet (SAM) in close proximity to the target spin~\cite{Phark2023electric}, as illustrated in Fig.~\ref{fig:fig9}(a).
In this setup, the SAM provides a local magnetic field for the target spin, creating an additional driving channel that can be comparable to the tip contribution. The dependence of the ESR peak splitting on external magnetic field's angle $\theta_\text{ext}$ reveals an isotropic exchange interaction between the target spin and the SAM, as shown in Fig.~\ref{fig:fig9}(b).
The key signature of this mechanism is when the tip is retracted, ESR Rabi rates drop to zero for isolated spins but remain finite for spins coupled to the nearby SAM, demonstrating that the SAM provides a robust additional channel for coherent spin control beyond the junction (Fig.~\ref{fig:fig9}(c)). Fig.~\ref{fig:fig9}(d) shows a simulated angular dependence of the ESR frequency splitting for different values of the exchange coupling strength $J_0$ between SAM and the spin. The simulation reproduces the overall symmetry and angular modulation observed in the experimental data shown above, confirming that the exchange interaction with the nearby SAM is the dominant mechanism underlying the observed splitting. In the experiment, data points in the vicinity of $\theta_\mathrm{ext}=0$ are not accessible, as a result, the experimental curves do not resolve the vanishing behavior predicted by the simulation.

When the tip is retracted and sufficiently far from the target spin so that the magnetic exchange field between tip and spin becomes negligible, coherent control remains possible because the oscillating electric field from the tip modulates the exchange interaction between the spin and the single-atom magnet~\cite{jose_anh2025}. This modulated interaction becomes the dominant driving mechanism, which results in Rabi rates on the order of tens of MHz, consistent with experimental observations.
The microscopic origin of such modulated exchange interaction is the electron hopping between the adsorbate and substrate orbitals. Let $\Delta$ denote the energy difference between these orbitals in the absence of electric field. When an electric field $\mathbf{E}$ is applied, the energy offset is modified by the work done on an electron during the hopping process, as written as follows 
\begin{equation}
\label{eq:deltaE}
    \Delta (\mathbf{E}) = \Delta - e \ell |\mathbf{E}|
\end{equation}
with $\ell$ being the distance between the orbitals. One can show that the amplitude of modulated exchange coupling takes the following form:
\begin{equation}
\label{eq:J1}
    J_1 = \frac{J_0}{2} \left( \frac{3}{U + \Delta} + \frac{1}{2U + \Delta} \right)e \ell |\mathbf{E}|,
\end{equation}
where $U$ is the on-site repulsion when the orbital is doubly occupied, and 
\begin{equation}
\label{eq:J0}
    J_0 = \frac{4 |t_1 t_2|^2}{(U + \Delta)^2} \left( \frac{1}{U} + \frac{1}{U + \Delta}\right)
\end{equation}
is the coupling strength in the absence of electric field. The Rabi rate is proportional to the modulation amplitude $J_1$. Interested readers can visit Supplemental Material of Ref.~\cite{jose_anh2025} for detailed derivations.

The approach of using spins outside the junction called \textit{remote spin} mitigates strong measurement-induced decoherence, tip-induced environmental noise, and rapid loss of phase coherence intrinsic to spin control in the STM junction. This enables longer coherence times and high-fidelity operations while preserving substantial Rabi rates, thereby enhancing the figure of merit $\Omega T_2$ for quantum coherent control.

\subsection{Multi spin continuous-wave spectroscopy}

\begin{figure*}[htbp!]
\centering
\reprofig{0.55\columnwidth}{6cm}{\rev{%
(a) Adapted from Ref.~\cite{Phark2023double}\\(Copyright 2023 American Chemical Society).\\
(b),(c) Adapted from Ref.~\cite{Hong2024stark}\\(Copyright 2024 American Chemical Society).}}
\caption{\textbf{Multi-spin spectroscopy} 
(a) \saba{Experimental} setup for double-resonance spectroscopy of two coupled Ti spins on a surface. Ti-1 is the sensor and the Ti-2 is the remote spin, which are coupled by $J_{1,2}$. The Fe atom nearby the Ti-2 provides a strong magnetic field gradient for the remote spin (left). Energy level diagram corresponding to the dressing of the Ti-2 and probing using Ti-1 (right). Adapted from Ref.~\cite{Phark2023double}, {\color{black} with permission from the American Chemical Society. Copyright 2023 American Chemical Society.} 
(b) Double-resonance spectra measured with RF2 frequency fixed at a resonance frequency of Ti-2 and RF1 frequency swept over the frequency range of the Ti-1 transitions. The simulated spectrum using the peak splitting as Rabi rate, is shown as a solid gray curve. (c) Dependence of the splitting $\Delta f$ on the driving amplitude $V_{\mathrm{RF2}}$ obtained from the fitting of the spectra in (b). The Rabi rate of the transition $(f_3)$ is given by the linear fit (solid line). (b) and (c) are adapted from Ref.~\cite{Hong2024stark}. {\color{black}  Copyright 2024 American Chemical Society.}
}
\label{fig:fig10}
\end{figure*}

The double-resonance spectroscopy can be performed on a system of two exchange-coupled Ti atoms on MgO/Ag substrate (sensor and remote spins), as illustrated in Fig.~\ref{fig:fig10}(a). In this setup, the sensor spin is driven by the magnetic field from the coupling with the tip, while the remote spin is driven by the magnetic field from the nearby Fe atom. In the weak coupling regime (exchange coupling strength is much smaller than Zeeman frequencies), the eigenstates of system is approximated by Zeeman product states of the sensor and remote spin, which results in four possible ESR transitions. Two RF voltages with different frequencies $f_{\rm RF1(2)}$ are employed in the double-resonance spectroscopy. By fixing the $f_\mathrm{RF2}$ at resonance $f_3$ of the Ti-2 and the sweeping $f_\mathrm{RF1}$ over the frequency range of the Ti-1 transitions, the achieved spectra in presented in Fig.~\ref{fig:fig10}(b). In addition, there exists an energy splitting $\Delta f$ between a pair of dressed states, and its strength is determined by the coupling strength between the two-level system and the driving field. This splitting is a manifestation of the AC Stark effect, as the strong driving field dresses the spin states. The energy splitting increases linearly with $V_\mathrm{RF2}$, resulting in a Rabi rate for the Ti-2 spin, as shown in Fig.~\ref{fig:fig10}(c).

\subsection{Quantum Coherent Control}

Multi-spin spectroscopy provides detailed access to the energy level structure, transition frequencies, and interaction landscape of coupled spins, which are essential for implementing quantum coherent operations~\cite{DJ2025review}. The observation of dressed states and their controllable splitting under strong driving fields indicates that the spin states can be coherently manipulated~\cite{Hong2024stark}. Building on this spectroscopic framework~\cite{Phark2023double}, the ability to drive remote spins~\cite{Phark2023electric} enables multi-spin coherent control in on-surface spin systems. 

Atomic and molecular spins on insulating surfaces can serve as electron-spin qubits. Universal control of a surface spin requires the ability to perform arbitrary rotations of the qubit state. This capability was demonstrated for an individual Ti atom on MgO/Ag using phase-controlled RF pulses that enable coherent rotations about arbitrary axes~\cite{Wang2023Universal}. Also, the implementation of two-axis control allows both single-axis and composite rotations, enabling arbitrary quantum state preparation.

Engineering the structure of three Ti atoms on the MgO/Ag surface, with two remote spins exchange-coupled to each other and to a sensor spin, enabled the implementation of coherent multi-spin control~\cite{Phark2023multi-qubit}. Fig.~\ref{fig:fig11}(a) illustrates the control scheme for the three-qubit structure, composed of two remote qubits and a sensor qubit. The remote qubits are coupled with corresponding SAMs, which allows them to be driven by RF electric fields without tunnel current passing through. The structure should be designed so that transition frequencies are unique, thereby the remote qubits can be selectively excited by tuning resonant RF frequency, allowing for multi-qubit gate operations. 

\begin{figure*}[ht!]
\centering
\reprofig{0.8\columnwidth}{6cm}{\rev{%
(a),(b) Adapted from Ref.~\cite{Phark2023multi-qubit} (AAAS, Copyright 2023 The Authors).\\
(c),(d) Adapted from Ref.~\cite{le2025overcoming}\\(Copyright 2025 The Authors, CC BY 4.0).}}
\caption{\textbf{Multi spin operations.} 
(a) A multi-qubit structure including two remote qubits and a sensor qubit for a control scheme. (b) Driving the transition $\lvert 0 \rangle \lvert 00 \rangle \leftrightarrow \lvert 0 \rangle \lvert 10 \rangle$ showing the CCNOT operation of remote qubit 1. (a) and (b) adapted from Ref.~\cite{Phark2023multi-qubit} , {\color{black} with permission from the American Association for the Advancement of Science. Copyright 2023 The Authors.} (c) Coherent and diagonal fidelity of NOT operation implemented on two remote qubits, while the DC bias is used during both the gate operation and measurement. The average gate fidelity is $F_{\mathrm{avg}} = 0.446$. (d) Coherent and diagonal fidelity with using DC bias just for the measurement. The average gate fidelity is $F_{\mathrm{avg}} = 0.887$. Adapted from Ref.~\cite{le2025overcoming} {\color{black} with permission. Copyright 2025 The authors. Licensed under CC BY 4.0}. 
}
\label{fig:fig11}
\end{figure*}

A demonstration of multi-qubit gate is presented in Fig.~\ref{fig:fig11}(b). CCNOT operation on remote qubit 1 is performed by driving the transitions from $\lvert 0 \rangle \lvert 00 \rangle \leftrightarrow \lvert 0 \rangle \lvert 10 \rangle$ state. The sensor qubit’s ESR transitions are used to sense the states of the remote qubits, with four different frequencies corresponding to the four possible states of the two remote qubits. When the pulse duration is varied, oscillations are observed in the populations of the $\lvert00 \rangle$ and $\lvert10 \rangle$ states of the remote qubits. The CCNOT gate time is approximately $20$ ns~\cite{Phark2023multi-qubit}.

The energy relaxation time $T_1$ of the sensor qubit and remote qubits was measured using an inversion recovery measurement. For the remote qubit, a $\pi$ pulse was first applied to invert its population, followed by a variable delay time, after which a sensing pulse was applied to the sensor qubit to measure the remote qubit’s state. The resulting signal decay over time was fitted to an exponential, yielding a $T_1$ of $166 \pm 14$ ns for the remote qubit. The sensor qubit’s $T_1$ was similarly measured, resulting in a value of about 10 ns. These measurements showed the improved stability of remote qubits due to the absence of tunnel current~\cite{Phark2023multi-qubit}.

While this experiment demonstrate fast and selective multi-qubit control, they also reveal that gate performance is constrained by decoherence, particularly the life time of the sensor qubit. This motivates the use of control methods that fight against dissipation and noise. 
Quantum Optimal Control Theory (QOCT) helps to design the pulse shapes to the specific noise sources present in the system, such as energy relaxation and pure dephasing, and exploits all degrees of freedom in the control Hamiltonian to achieve high-fidelity operations~\cite{Goerz2019krotov}. 

For the same coupled Ti-atom system discussed above, optimized pulse was designed by QOCT methods to implement CCNOT gate~\cite{le2025overcoming}. Fig.~\ref{fig:fig11}(c) shows the scheme where the DC bias is applied throughout both the gate operation and measurement, same as the experiment in~\cite{Phark2023multi-qubit}. This causes the sensor qubit to have a much shorter lifetime, while the remote qubits retain their long lifetimes. The average gate fidelity even with using an optimized pulse is $F_\mathrm{avg}=0.446$, with both coherent and diagonal fidelities showing significant variation and lower values, especially for higher-energy states. The fidelity distribution is skewed toward low-energy states due to rapid energy relaxation, highlighting the detrimental effect of the DC always-on scheme on gate fidelity. 
In contrast, when the DC bias is applied only during the measurement stage, such that all qubits including both sensor and remote qubits, experience long lifetimes during gate operation, the coherent and diagonal fidelities remain high across all basis states. This leads to a substantially improved average gate fidelity of $F_\mathrm{avg}=0.887$, shown in Fig.~\ref{fig:fig11}(d).

Designing gate sequences can result in entangled states preparation, which is a fundamental requirement for quantum information processing in atomic-scale systems. Another method for creating entanglement is to employ the tunneling current~\cite{TimeESR_github, Horovitz2021}. By performing quantum operations such as Hadamard and CNOT gates, it is possible to create entanglement between individual spins on a surface. These gate sequences exploit the weak interactions between spins to generate states that are not eigenstates of the Zeeman product basis, leading to distinct phase accumulation during free evolution. This phase can be mapped onto the population of one spin and subsequently read out using a third, weakly coupled sensor spin, enabling direct and unambiguous certification of entanglement~\cite{Broekhoven2024}. To certify entanglement in surface spin systems using ESR-STM, it is proposed to exploit the free time evolution of entangled spin states, which accumulate a phase at a rate proportional to the energy splitting between specific eigenstates, distinct from the evolution of non-entangled states. By mapping this phase onto the population of one spin and subsequently reading it out using the  sensor spin, the protocol enables direct and unambiguous detection of entanglement. The experimental scheme involves creating entanglement via quantum gates (Hadamard and CNOT), then measuring the accumulated phase through a disentangling sequence, which projects the phase onto one spin for read-out~\cite{Broekhoven2024}.

Beyond gate implementation for entangled state preparation, using tailored current pulses from the STM tip is possible to initialize entangled states through electron scattering~\cite{Veldman2021cohevolution}, which creates a non-equilibrium superposition of the coupled spin states, leading to coherent flip-flop oscillations and the emergence of quantum correlations between the spins. The coherent evolution of these multi-spin systems is typically monitored using a pump–probe technique, in which a short pump pulse initializes the spin state and a delayed probe pulse measures the resulting dynamics. This approach leverages electron spin resonance principles to trigger and read out spin transitions, but instead of microwave excitation, the initialization is achieved through direct electron scattering from the STM tip. The pump–probe scheme enables time-resolved observation of free coherent evolution, revealing the formation and propagation of quantum correlations between coupled spins~\cite{Veldman2021cohevolution}.

\section{Coupled electron-nuclear spin control}
\label{sec:electron-nuclear}

\begin{figure}[ht!]
\centering
\includegraphics[width=\columnwidth]{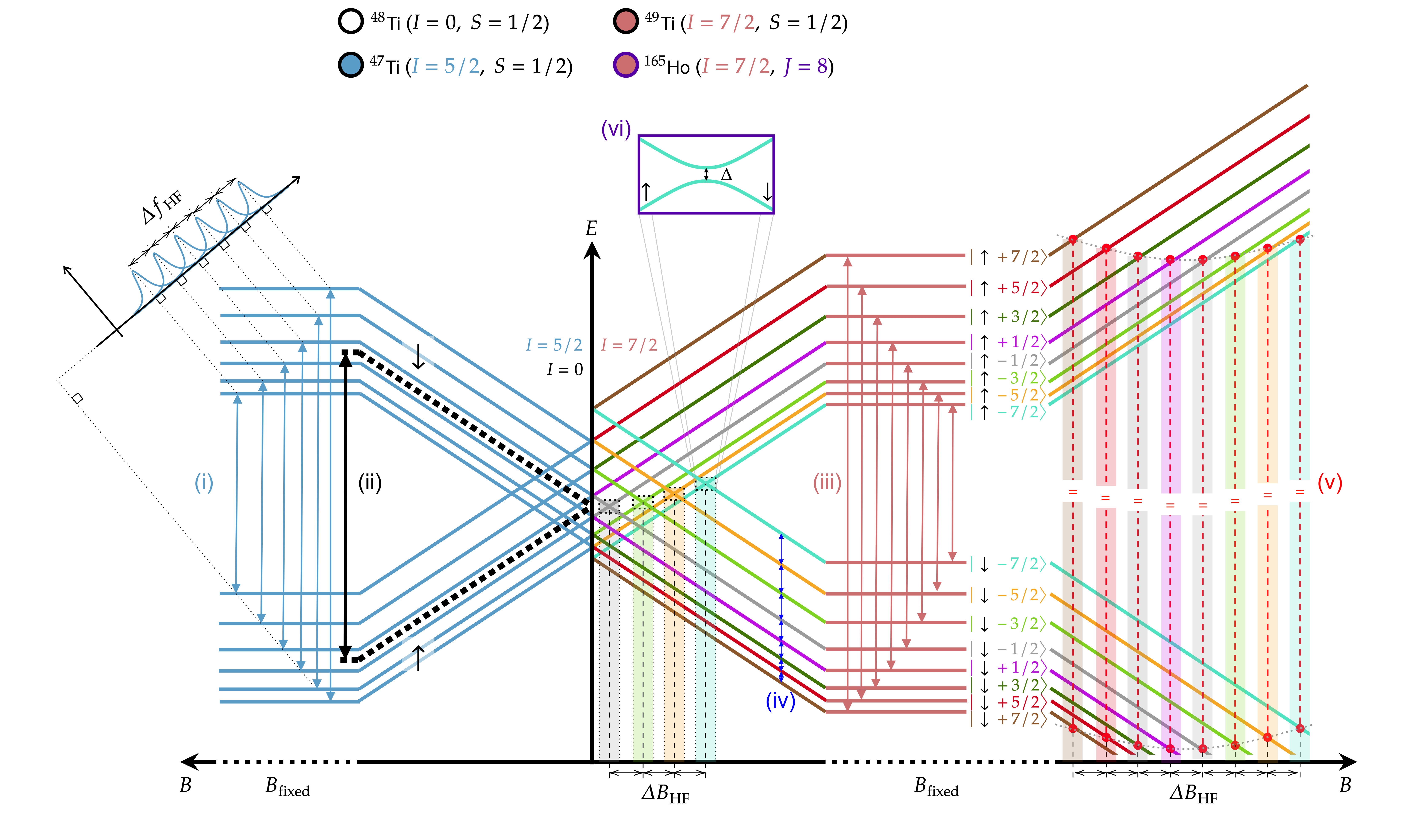}
\caption{\textbf{Hyperfine energy spectra and protocols for single-atom spin readout/control on MgO.} Energies $E$ ($y$-axis) are plotted versus applied magnetic field $B$ ($x$-axis) for Ti isotopes and Ho adsorbed on MgO. Blue/black/dark red correspond to $^{47}$Ti, $^{48}$Ti, and $^{49}$Ti; non-ESR-driven $I = 7/2$ levels can also correspond to $^{165}$Ho. The spectra are governed by the hyperfine interaction $A\,\mathbf{I}\!\cdot\!\mathbf{J}$ with an effective quadrupolar term $P_{\mathrm{quad}} I_z (I_z + 1)$.
(i) ESR-STM at a fixed operating field $B_{\mathrm{fixed}}$ resolves hyperfine-split electron spin transitions, projecting the $2I + 1 = 6$ nuclear states $I_z$ of $^{47}$Ti. The even spacings $\Delta f$ of the frequencies in the expected ESR spectrum is also shown.
(ii) $^{48}$Ti shows a single unsplit ESR line (no nuclear spin).
(iii) The same ESR-STM scheme for $^{49}$Ti yields $2I + 1 = 8$ nuclear-state-resolved ESR lines.
(iv) RF-driven $\Delta I_z = \pm 1$ nuclear transitions (NMR) address the nuclear manifold; combined with ESR readout this constitutes ENDOR (ESR-detected NMR).
(v) A constant probe ESR frequency is applied while sweeping $B$ (e.g., via junction- or tip-induced effective-field tuning); the field at which the atom flips provides nuclear-state-dependent readout, with values separated evenly by $\Delta B_{\mathrm{HF}}$.
(vi) Since the high spin of Ho renders ESR driving impossible, sweeps through avoided crossings drive Landau--Zener transitions (gap $\Delta$), enabling nuclear-spin spectroscopy via state-dependent switching probabilities without ESR driving.
}
\label{fig:fig12}
\end{figure}

\subsection{Single-atom hyperfine ESR spectroscopy}

The coupled electron--nuclear (hyperfine, $\text{HF}$) spectrum in an external magnetic field $\mathbf{B}$ is captured by the effective Hamiltonian
\begin{equation}
\label{eq:H-hy}
    H_\text{HF} = \mu_B \mathbf{B} \cdot \mathbf{g} \cdot \mathbf{J}
    + \mathbf{I} \cdot \mathbf{A}_\text{dip} \cdot \mathbf{J}
    + P_\text{quad}\, I_z(I_z+1)
    + \mu_N \mathbf{B} \cdot \mathbf{g}_I \cdot \mathbf{I},
\end{equation}
where $\mathbf{J}=\mathbf{S}+\mathbf{L}$ is the total angular momentum of the electron (often effectively $J=S=1/2$ in STM-ESR studies), $\mathbf{I}$ is the nuclear spin, and $\mathbf{g}$ and $\mathbf{g}_I$ are the electronic and nuclear $g$-tensors. The tensor $\mathbf{A}_\text{dip}$ describes the magnetic dipolar hyperfine interaction between electron and nucleus, while $P_\text{quad}$ parametrizes the electric quadrupole interaction arising from the nuclear quadrupole moment and the local electric-field gradient (EFG). The nuclear Zeeman term is frequently neglected to first approximation because it is proportional to the small nuclear magneton $\mu_N$  $(\mu_B\approx2000\mu_N)$. Diagonalizing $H_\text{HF}$ as a function of $\mathbf{B}$ for $I=0$ yields the known spectrum with two electron transitions (i.e. $\uparrow \leftrightarrow \downarrow$). However, the spectra become significantly more complex when $I=5/2,$ or $7/2$, as shown in the energy diagram shown in Fig.~\ref{fig:fig12}.

\begin{figure}[ht!]
\centering
\reprofig{0.8\columnwidth}{6cm}{\rev{%
(a),(c) Adapted from Ref.~\cite{JK2022Hyperfine} (Copyright 2022 The Authors).\\
(b),(d),(e) Adapted from Ref.~\cite{Willke2018Hyperfine} (Copyright 2018 The Authors).}}
\caption{\textbf{Hyperfine splitting modulation in Ti isotopes.}
(a) Schematic of ESR-STM measurements on the three Ti isotopes ($^{47}$Ti, $^{48}$Ti, $^{49}$Ti) on MgO in a rotating magnetic field. The MgO lattice directions are $\hat{x}$ and $\hat{y}$, with $\hat{z}$ out of plane. An external magnetic field $B_{\mathrm{ext}}$ is applied in a plane rotated by $15.5^\circ$ from the $yz$-plane about $\hat{z}$; $\theta$ denotes its angle relative to the out-of-plane direction.
(b) ESR spectra of $^{48}$Ti, $^{47}$Ti, and $^{49}$Ti on oxygen binding sites. Resonances correspond to transitions (ii), (i), and (iii) in Fig.~\ref{fig:fig12}, respectively.
(c) ESR spectra of $^{49}$Ti on an oxygen site as a function of $\theta$, corresponding to transition (iii) in Fig.~\ref{fig:fig12}.
(d) ESR spectra of $^{49}$Ti at different binding sites.
(e) Aforementioned binding-site geometries (left) and corresponding DFT-calculated spin densities (right).
(a) and (c) are adapted from Ref.~\cite{JK2022Hyperfine} {\color{black} with permission. Copyright 2022 The authors.} (b), (d) and (e) are adapted from Ref.~\cite{Willke2018Hyperfine} {\color{black} with permission. Copyright 2018 The authors.}}
\label{fig:fig13}
\end{figure}

Continuous-wave ESR-STM can directly resolve the hyperfine interaction of a single surface spin, i.e., the coupling between an electronic spin the nuclear spin on the same atom. In early demonstrations, hyperfine splittings were observed as multiple, nuclear-state-resolved ESR lines for individual Fe- and Ti atoms adsorbed on MgO~\cite{Willke2018Hyperfine}. Titanium provides a particularly clear example: $^{48}$Ti has $S=1/2, I=0$ and therefore shows a single unsplit ESR line, whereas $^{47}$Ti ($S=1/2, I=5/2$) and $^{49}$Ti ($S=1/2, I=7/2$) exhibit six and eight lines, respectively (Fig.~\ref{fig:fig13}(a, b)). Each line corresponds to a fixed nuclear projection $I_z=-I, -I+1, \dots I$, which shifts the electronic transition frequency by a discrete hyperfine offset. The ESR lines are uniformly spaced, with the separation determined by the hyperfine tensor along the field direction which sets the quantization axis 
\begin{equation}
    \Delta f_\text{HF} = 2\left(A_\text{dip}\right)_{\parallel}\, S =\left(A_\text{dip}\right)_{\parallel},
    \label{eq:hyperfine_splitting_f}
\end{equation} 
as shown in Fig.~\ref{fig:fig13}(b), where $\theta$ denotes the components along the quantization axis set by the $\mathbf{B}$-field. The peak intensities encode the nuclear-state populations, which in equilibrium follow Boltzmann statistics.

These measurements also showed that the hyperfine coupling can be strongly anisotropic. By combining ESR-STM with vector magnetic fields, varying $\theta$, and atom manipulation, the full hyperfine tensor of single Ti atoms on MgO was mapped~\cite{JK2022Hyperfine,Farinacci2022}. The out-of-plane and in-plane components can differ by order unity (often quoted as $>67\%$ anisotropy), consistent with a highly directional electron spin density in Ti $3d$ orbitals, see Fig.~\ref{fig:fig13}(c). Moreover, it has been observed that $^{49}$Ti atoms on different bonding sites, exhibit different hyperfine splittings~\cite{Willke2018Hyperfine}, showcasing the dependence of the hyperfine tensor on the environment, see Fig.~\ref{fig:fig13}(d,e). In practice, this turns the nuclear spin into a local probe of the electronic orbital character and bonding geometry at the single-atom level~\cite{JK2022Hyperfine,Farinacci2022}.

Beyond spectroscopy, ESR-STM allows the single-shot readout of the nuclear projection in ESR-addressable $S=1/2$ systems (e.g., $^{49}$Ti). One approach is to apply a fixed probe tone while sweeping the magnetic field---or, equivalently, tuning the local effective field via the STM junction---so that the ESR condition is satisfied only for selected nuclear projections (Fig.~\ref{fig:fig12}v). In this mode, real-time monitoring of the tunneling current provides a nuclear-state-dependent binary signal. With a continuous probe tone set to an ESR frequency that is resonant only for (e.g.) $I_z=-7/2$, the tunneling current becomes a binary indicator of the nuclear state: high signal when the nucleus occupies that state, and baseline otherwise as shown in Fig.~\ref{fig:fig13}(c)~\cite{Stolte2025SingleShot}. The measured time traces show random telegraph switching between the two levels, corresponding to nuclear-state changes. From the dwell-time statistics, a nuclear lifetime $T_1 \approx 5.3 \text{ s}$ was extracted, while the electron $T_1$ in the same atom is $\sim 10^{-7} \text{ s}$, i.e. a gap of about seven orders of magnitude. The single-shot assignment fidelity reached $\approx 98\%$. The same work also quantified how the nuclear switching rate increases under stronger perturbation (continuous ESR drive and/or higher tunneling current), consistent with nuclear relaxation being dominated by coupling to the electron spin through driven or stochastic hyperfine-mediated processes. 

In this field-sweep readout, the nuclear projection is inferred from the magnetic-field value at which the resonance (or a switching event) occurs. Adjacent branches are separated by
\begin{equation}
    \Delta B_\text{HF} = \left(A_\text{dip}\right)_{\parallel}/(g \mu_B),
    \label{eq:hyperfine_splitting_B}
\end{equation}
as indicated schematically in Fig.~\ref{fig:fig12}.

Importantly, the ESR-based hyperfine splittings in Eqs.~\eqref{eq:hyperfine_splitting_f} and~\eqref{eq:hyperfine_splitting_B} primarily probe the projected dipolar hyperfine coupling $\left(A_\text{dip}\right)_{\parallel}$. In the common regime where ESR transitions are predominantly $\Delta m_S=\pm 1$ with $\Delta I_z=0$, the quadrupole term $P_\text{quad}$ shifts energies within a given nuclear manifold but cancels to first order in the \emph{electronic} transition frequency, so ESR line spacings alone do not provide a robust determination of $P_\text{quad}$. Instead, quadrupolar structure appears to first-order only in nuclear resonance spectra, as discussed below. Note that since $P_\text{quad}$ reflects the local electric-field gradient at the nucleus, changes in electronic configuration in the low and high spin states could lead to $\delta P_\text{quad}\neq 0$. If sufficiently large, this could appear as a small departure from perfect equidistance of hyperfine-resolved ESR branches; isolating such effects generally requires combined ESR and nuclear-sensitive measurements and a global fit to $H_\text{HF}$.

\subsection{Nuclear polarization and driving}

ESR-STM has also been used to polarize nuclear spin states of individual atoms. In several systems the lowest-energy nuclear state is overpopulated compared to thermal equilibrium, indicating dynamic nuclear polarization driven by tunneling electrons. For single Cu, electrically controlled nuclear polarization and single-atom NMR were demonstrated by pumping the nucleus via repeated electron spin flips and then driving nuclear transitions at the nuclear Larmor frequency ~\cite{Yang2018}. For Ti on MgO, bias-dependent changes in the hyperfine-line amplitudes can be interpreted as effective nuclear spin temperatures far below the lattice temperature, reaching the $\mathcal{O}(10\,\mathrm{mK})$ range in some regimes~\cite{JK2022Hyperfine}. Rate- and master-equation models account for these observations by combining (i) electron spin pumping by a spin-polarized tunneling current, (ii) hyperfine-mediated electron--nuclear flip-flop channels, and (iii) bias-dependent tunneling rates that set the steady-state nuclear polarization and relaxation.

Microscopically, inelastic electron spin-flip scattering can be accompanied by nuclear flips (hyperfine-assisted ``flip-flops''), thereby transferring angular momentum from the polarized current to the nucleus. These processes become particularly efficient when the electron Zeeman splitting is tuned into resonance with a nuclear transition. In STM, this matching condition can be reached using the local tip field as a control knob: changing the tip height (and/or tip magnetization) shifts the local electronic Zeeman energy until it approaches the nuclear splitting, creating avoided crossings and characteristic ESR signatures~\cite{Veldman2024Coherent}. Near such resonances, the eigenstates hybridize and the coupled electron-nuclear system can be swept through the avoided crossing by STM-based tuning of the local field.

Moreover, time-domain ESR-STM has resolved coherent electron–nuclear dynamics directly. In pump–probe measurements on $^{47}$Ti, the coupled two-spin system is first initialized (electron and nucleus polarized via spin-dependent tunneling), then tuned near a flip-flop avoided crossing with the tip field, and finally perturbed by a fast pulse. The subsequent free evolution shows a beat pattern from coherent exchange of polarization between electron and nucleus~\cite{Veldman2024Coherent}. Multiple oscillation frequencies appear because several hyperfine-split flip-flop channels contribute in an $I=5/2$ nucleus. The observed frequencies and amplitudes match the hyperfine coupling strengths and detunings extracted from spectroscopy, providing a stringent check of the two-spin Hamiltonian in the STM environment.

Beyond spin pumping via inelastic tunneling, a complementary question is what microscopic mechanism actually converts the RF tip voltage into an effective transverse drive on the local spin degrees of freedom. It was proposed that an "electric $g$-tensor modulation" mechanism in which the RF tip field induces a small vertical (piezoelectric) displacement $\delta z(t)$ of the adatom, thereby modulating the local crystal field and hence the \emph{anisotropy} of the electronic $g$-tensor through spin--orbit coupling; this produces a time-dependent Zeeman Hamiltonian with non-collinear components that can directly drive ESR transitions in $S=1/2$ adatoms~\cite{Ferron2019}. For electron-nuclear control, the same framework naturally yields nuclear-spin-flip transitions in the hyperfine-coupled manifold: hyperfine mixing admixes small opposite-electron-spin components into nominally nuclear-only states, so that an electronically driven perturbation can induce $\Delta I_z=\pm 1$ transitions with matrix elements proportional to the transverse hyperfine coupling $A_\perp$~\cite{Ferron2019}. This provides a microscopic route by which purely electrical RF driving, mediated by the electron, can access nuclear dynamics in STM-ESR/STM-NMR settings and helps rationalize why nuclear control can emerge most clearly in systems where both spin-orbit-induced anisotropy and hyperfine mixing are appreciable, like the flip-flop study previously discussed.

To actively control the nucleus, ESR-STM can implement electron–nuclear double resonance (ENDOR). In STM-ENDOR, one monitors the ESR signal while sweeping a second RF tone through the nuclear transition frequency.  In STM-ENDOR, an ESR signal is monitored while a second RF tone is swept through a nuclear transition: nuclear flips shift the ESR condition and appear as a dip (or change) in the ESR response at the nuclear Larmor frequency~\cite{Manassen2018ENDOR}. This approach has been used to distinguish isotopes (e.g. $^{63,65}$Cu with $I=3/2$) and to resolve additional nuclear structure such as quadrupolar splittings (e.g. $^{14}$N in TEMPO with $I=1$). Moreover, ENDOR in ESR-STM has been recently used on $^{47}$Ti of spin $I=5/2$ to hint at a possible driving of the nuclear spin using NMR addressing~\cite{Otte2025ENDOR}. A sweeping of the ESR frequency $f_\text{ESR}$ was performed while applying a constant RF frequency $f_\text{NMR}$ corresponding to the nuclear transition from the first to second nuclear states, $I_z=-5/2 \leftrightarrow -3/2$. An equalization of the peak intensities of the two ESR lines corresponding to these nuclear states was observed when $f_\text{NMR}$ matches the nuclear transition frequency, indicating a nuclear pumping effect. However, a conclusive demonstration of coherent nuclear driving in ESR-STM remains an open goal, as this previous result is under a continuous NMR drive and lacks time-domain verification. The latter could be achieved by combining pulsed ESR-STM with synchronized NMR bursts, similar to established ENDOR protocols in bulk ESR/NMR systems~\cite{Schweiger2001Principles}.

\subsection{Higher electronic spins and forbidden ESR transitions}

So far we have focused on $S=1/2$ systems, where ESR transitions are dipole allowed with $\Delta m_s=\pm1$. Many surface spins, however, have $S>1/2$ (or more generally $J>1/2$), which introduces both additional structure and new constraints. For instance, Fe on MgO is a $J=2$ system, whilst Ho on MgO has a $J=8$ electronic ground state. Mn adatoms on Cu$_2$N can realize $J=2$ and $J=5/2$, respectively~\cite{Loth2010}. In high-$J$ systems with strong axial (Ising-like) anisotropy, the standard ESR-STM paradigm of driving $\Delta m=\pm 1$ electronic transitions can break down: within the ground doublet, the relevant transverse matrix elements can become extremely small, and microwave-driven ESR lines may be absent or exceedingly weak. For integer-$J$ systems, the absence of Kramers degeneracy further allows crystal-field terms to generate zero-field splittings and avoided crossings, which can be exploited even when conventional ESR driving is ineffective.

An on-surface example is Ho on MgO, where a ground-state doublet with $J_z=\pm 8$ is split into $2I+1=8$ sublevels, and just as many anticrossings, by the $^{165}$Ho nuclear spin ($I=7/2$). Sweeping the local magnetic field through the series of avoided crossings induces Landau--Zener transitions with a probability set by the tiny tunnel splittings, enabling stochastic but state-selective switching events that can be used as a single-shot readout of the nuclear projection~\cite{Natterer2017,Natterer2019Ho}. Conceptually, this is the Landau--Zener readout mechanism sketched in Fig.~\ref{fig:fig12}(vi): the nuclear state is inferred from the field value at which a switching event occurs, rather than from a resonant ESR peak. The values of the magnetic field at which those transitions happen are equally spaced and do not depend on the nuclear quadrupole, and are given by Eq.~\eqref{eq:hyperfine_splitting_B}.

\begin{figure}[ht!]
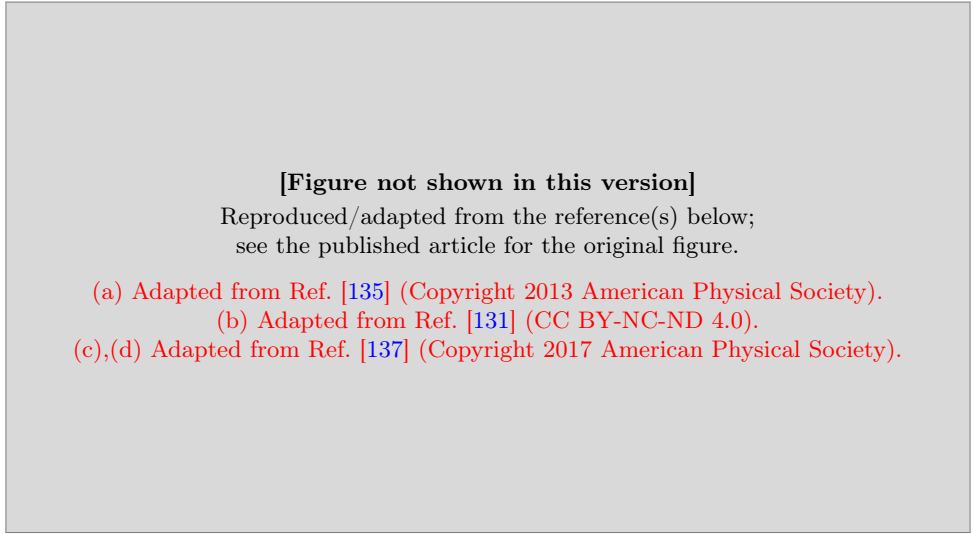

\centering
\reprofig{0.8\columnwidth}{7cm}{\rev{%
(a) Adapted from Ref.~\cite{Thiele2013} (Copyright 2013 American Physical Society).\\
(b) Adapted from Ref.~\cite{Stolte2025SingleShot} (CC BY-NC-ND 4.0).\\
(c),(d) Adapted from Ref.~\cite{Godfrin2017} (Copyright 2017 American Physical Society).}}
\caption{\textbf{Nuclear levels read-out and addressing mechanisms.}
(a) Left: conductance jumps revealing the nuclear spin states (gray) of Tb ($I = 3/2$) and the resulting nuclear-spin trajectory (red). Right: histograms of $\sim$40{,}000 jumps showing four non-overlapping Gaussian-like distributions; shaded bars indicate the time-averaged population $P$ of each nuclear spin state, illustrating the Landau--Zener readout mechanism (vi) in Fig.~\ref{fig:fig12}. Adapted from Ref.~\cite{Thiele2013} {\color{black} with permission. Copyright 2013 American Physical Society.}
(b) For $^{49}$Ti, current histograms measured at different STM tip heights illustrate the readout mechanism (v) in Fig.~\ref{fig:fig12}. The mean current of each trace reflects the total magnetic field; histograms corresponding to the dashed lines in (b) give the time-averaged populations $P$. Adapted from Ref.~\cite{Stolte2025SingleShot} {\color{black} with permission. Copyright 2025 Licensed under CC BY-NC-ND 4.0.}
(c) Rabi oscillations with tunable frequencies from 1.5 to 8~MHz. Colors denote three possible NMR transitions in Tb ($I = 3/2$) (red, green, blue) ((iv) in Fig.~\ref{fig:fig12}).
Middle: Resonance profiles of the three transitions obtained from the maximum Rabi visibility as a function of detuning.
(d) Top: Visibility of the second state versus pulse length and frequency; increasing detuning raises the oscillation frequency and reduces the maximum visibility.
Bottom: Visibility of the second state versus pulse length and power; the Rabi frequency scales linearly with the square root of the pulse power. (c) and (d) Adapted from Ref.~\cite{Godfrin2017} {\color{black} with permission. Copyright 2017 American Physical Society.}
}
\label{fig:fig14}
\end{figure}

Related electronuclear readout strategies have long been exploited in single-molecule magnets, providing a useful benchmark for ESR-STM. In a molecular spin-transistor geometry, a TbPc$_2$ molecule (Tb$^{3+}$ with $J=6$ and nuclear spin $I=3/2$) can be contacted between metallic electrodes such that the tunneling current becomes sensitive to the electronic state $J_z=\pm 6$. It was demonstrated that sweeping the magnetic field across the hyperfine crossings yields discrete conductance jumps at four distinct field values, thereby resolving all four nuclear projections and enabling reconstruction of individual nuclear-spin trajectories (Fig.~\ref{fig:fig14}(a))~\cite{Thiele2013}. In this setting, the nucleus acts as a qudit of dimension $d=2I+1=4$. The experiment established single-shot readout of a nuclear-spin qudit with long lifetimes ($T_1$ of order $10\,\mathrm{s}$) and coherence times $T_2$ of order tens of $\mu\mathrm{s}$, illustrating why nuclear degrees of freedom are attractive memory registers: they couple weakly to the environment yet remain addressable through their hyperfine-coupled electronic partner. In this molecular transistor platform, direct NMR driving can be sufficiently fast to yield a favorable figure of merit $Q=T_2/T_\mathrm{Rabi}\sim 10^4$ for coherent control of the nuclear qudit~\cite{Jankovi2024}.

However, as the Ho/MgO case already suggests, Landau-Zener readout alone does not provide coherent qudit control. Coherent quantum operations require {NMR-level addressing} of transitions within the nuclear manifold, i.e. controlled $\Delta I_z=\pm 1$ rotations and multi-level gate sequences. Indeed, such coherent control of the Tb nuclear spin in TbPc$_2$ using resonant electric RF driving has been demonstrated, with Rabi oscillations observed on selected nuclear transitions (Fig.~\ref{fig:fig14}(d)) and multi-level algorithms implemented using the nuclear qudit~\cite{Godfrin2017}. For ESR-STM, these molecular results provide a clear roadmap: Ho on MgO (and related rare-earth adatoms) already offers robust single-shot nuclear readout via Landau--Zener sweeps, but extending such platforms to coherent quantum-information processing will require integrating local NMR control of the nuclear manifold, thereby promoting the nuclear degree of freedom from a spectroscopic label to an actively driven qudit register. ESR-STM can be useful in this regard by leveraging electron-mediated nuclear driving schemes. Moreover, it is primordial to characterize and optimize (i) the addressing speed and therefore the figure of merit $Q$ and (ii) the nuclear and electronic $T_1$ and $T_2$ as well as the corresponding noise channels in these surface systems, which remains an open challenge involving tomography of the hyperfine states in the time domain.

Figure~\ref{fig:fig14} summarizes representative approaches to nuclear readout and addressing in electron--nuclear STM platforms, spanning Landau-Zener-based trajectories, ESR-detected NMR (ENDOR), and coherent NMR control in molecular devices.

\subsection{Comparison with electron-only control and outlook}

Including nuclear spins fundamentally reshapes the control landscape compared to electron-only ESR-STM. Electron spins provide fast manipulation and high-fidelity electrical readout, but their coherence is intrinsically limited by strong coupling to tunneling electrons, exchange noise, and magnetic-field fluctuations. Nuclear spins, by contrast, are slow and weakly driven, yet benefit from a high degree of isolation that enables relaxation and coherence times orders of magnitude longer, while presenting denser information storage with higher spin states. ESR-STM naturally combines these complementary properties: the electron acts as a fast, addressable interface, while the nucleus serves as a long-lived quantum register, with the hyperfine interaction providing a tunable and site-specific coupling between the two. In this sense, the key objective is not simply extended lifetimes, but the realization of a well-characterized multi-level system whose transition frequencies, selection rules, and coherence properties are stable and reproducible across nominally identical atoms.

From this perspective, nuclear-spin qudits coupled to electrons offer more than just improved memory—they open a route toward intrinsically protected quantum information. Recent theoretical work has shown that qudits can host error-protected logical subspaces in which certain classes of noise (e.g.\ dephasing or amplitude damping) are suppressed by symmetry or encoding, effectively realizing error-corrected qubits within a single physical object~\cite{Lim2025}. In an ESR-STM setting, the combination of a controllable electron spin and a structured nuclear manifold is particularly attractive: fast electron-mediated operations can be used to initialize, read out, and couple logical states, while the nuclear qudit provides a higher-dimensional Hilbert space in which redundancy and error detection can be embedded at the hardware level. Such encodings could relax fidelity requirements on individual gates and measurements, which are known bottlenecks in surface-based spin platforms~\cite{le2025overcoming, Jankovi2024}.
Looking ahead, the central challenge is therefore not only to demonstrate control, but to establish {metrological-grade} qudits: electron–nuclear systems whose $T_1$ and $T_2$ are directly measurable in the time domain, whose noise channels are quantitatively understood, and whose parameters can be engineered reproducibly through choice of atom, adsorption site, and local environment. Achieving this would enable the transition from proof-of-principle demonstrations to architectures in which electron spins act as fast control and coupling resources, while nuclear qudits host protected logical states. In this limit, ESR-STM becomes a platform not merely for observing single-spin quantum phenomena, but for implementing error-resilient quantum degrees of freedom assembled atom by atom on a surface.

\section{Summary}

In this review, we have examined the physical mechanisms underlying ESR-STM signals, with emphasis on how radio-frequency driving, exchange interactions, and many-body effects shape experimentally observed resonances. We first focused on the magneto-electric coupling mechanism and detection in the STM junction. A central conclusion of this discussion is that a proper interpretation of ESR-STM experiments requires a theoretical framework that includes the transport-induced effects in the junction. Whilst effective localized-spin descriptions based on \nameref{subsec:Heisenberg} can capture coherent spin dynamics and provide an intuitive picture of electrically driven ESR through modulation of exchange interactions or anisotropic $g$-factors, they fail to capture current induced effects such as relaxation time $T_2$, current-dependent driving efficiency $\Omega$, and bias-dependent resonance frequency shift \saba{$f_{\rm res}$}. Within their scope, these models reproduce the experimentally observed linear bias dependence of the driving efficiency.

The \nameref{subsec:Kondo} extends this picture by explicitly incorporating the exchange coupling between the impurity spin and conduction electrons in the tip and substrate, thereby enabling a treatment of nonequilibrium transport and higher-order tunneling processes. Within this framework, ESR signals arise from spin-dependent tunneling processes, and both coherent driving and decoherence are governed by exchange-mediated electron transport. \jose{Kondo models that consistently incorporate first-order coherent processes, such as the exchange field, successfully reproduce key experimental trends, including a linear dependence of the driving efficiency on both the RF voltage and the DC current, a linear dependence of the decoherence rate on the DC bias, and a nonlinear bias dependence of the resonance frequency. However, these models lack the spin-transfer-torque mechanism. On the other hand, approaches that incorporate spin-transfer torque lack the exchange field. More importantly, because they remain effective low-energy theories in the infinite-$U$ limit, they cannot fully account for bias-dependent renormalization effects. Kondo models that incorporate neither the exchange field nor spin-transfer torque cannot directly drive the quantum impurity and therefore must rely on an additional driving mechanism, such as the AC magnetic field generated by an RF microwave field, unless charge degrees of freedom are explicitly included. }

The \nameref{subsec:AIM} emerges as a particularly comprehensive framework for describing ESR-STM experiments, as it treats spin and charge fluctuations on equal footing and naturally incorporates nonequilibrium tunneling. Virtual charge fluctuations generate a bias-dependent exchange field that renormalizes the impurity spin levels, while RF modulation of the tunneling amplitudes converts an electric drive into an effective spin torque. Within this model, the driving efficiency shows a linear dependence on both the RF voltage and the DC current, a logarithmic dependence on the DC bias, and the resonance frequency exhibits a logarithmic bias dependence, in good agreement with experimental observations. The Anderson impurity model thereby allows access to crossovers between coherent and transport-dominated spin dynamics that are inaccessible within purely spin-based models.

Finally, in \nameref{subsec:Phenom}, open-quantum systems simulations using a Lindblad or Redfield QME are efficient and have shown to capture relevant physical processes such as Rabi oscillations, decoherence, dephasing well given that they are accurately parameterized. They can serve as the foundation for large-scale simulations, such as higher-spin objects ($S>1/2$) and non-zero nuclear spin $I\neq 0$, where AIM becomes computationally expensive.

The fundamental limitations on the coherence time imposed by the tunneling current and the desire to control systems beyond single spin-$1/2$ QI has motivated the \textit{remote spin} approach where the tunnel impurity acts only as weakly coupled sensor spin. \nameref{sec:multi_control} has been achieved by exploiting exchange interactions between a sensor spin inside the junction and nearby spins located outside this region. Because the STM-induced driving field decays exponentially with distance, direct control of remote spins is inefficient; this limitation is overcome by introducing a single-atom magnet which provides local magnetic field for the spin outside of the tunnel junction. In this configuration, coherent control of remote spins remains possible even when their couplings with the tip are negligible. As a result, remote spins can exhibit longer coherence times and an improved figure of merit compared to sensor spin. This motivates architectures in which a sensor spin inside the junction is used primarily for readout while coherent operations are performed on remote spins. Building on this concept, multi-spin continuous-wave and double-resonance spectroscopy enable selective and simultaneous control of coupled spins, providing direct access to interaction strengths and coherent coupling mechanisms in engineered few-spin systems.

Despite these advances, multi-spin control in ESR-STM remains subject to important limitations that define key directions for future work. One central open challenge concerns the experimental quantification of gate performance. While Rabi oscillations, linewidths, and coherence times provide indirect measures of control quality, a direct determination of gate and state fidelities in ESR-STM experiments has not yet been achieved. In particular, full quantum state or process tomography, which is routinely employed in other solid-state qubit platforms such as superconducting circuits, semiconductor quantum-dot spin qubits, donor spins in silicon, and nitrogen-vacancy centers in diamond~\cite{Matthias2006Tomography,KimDohun2014tomographysemiconductor, HollyStemp2024Tomography, Zhang2023TomographyNV}, has so far remained experimentally inaccessible in STM-based systems. The absence of tomographic protocols limits the ability to rigorously benchmark multi-qubit gates, verify entanglement, and disentangle coherent control errors from decoherence-induced loss of contrast. A further limitation concerns the scalability of multi-spin architectures in ESR-STM. While exchange-coupled spin pairs have enabled demonstrations of coherent multi-spin control, extending these approaches to larger systems faces fundamental and practical challenges. Alternatively, coupled electron–nuclear schemes can leverage hyperfine interactions to enable nuclear-state-resolved ESR spectroscopy, single-shot nuclear readout, dynamic polarization, but protocols for electron-mediated nuclear driving or direct NMR implementations are still required. In the broader context of quantum information processing, scalable operation necessarily involves increasing the number of coherently controlled qubits, as pursued in other qubit platforms. At the same time, \nameref{sec:electron-nuclear} schemes suggest an orthogonal route to scalability, where the accessible Hilbert space is enlarged within each atom.

The STM junction provides atomic-scale precision in an ultra-clean and defect-free environment for coherent spin manipulation and readout. This enables deterministic placement of individual spins, precise control of local exchange interactions, and systematic studies of geometry-dependent coupling mechanisms. While the intrinsically local nature of STM-based control motivates the development of strategies for parallel operation and scalable architectures, the combination of atomic precision and defect-free surroundings establishes the STM as an excellent platform for studying and benchmarking coherent spin dynamics at the single-atom level.

\section{Acknowledgments}
The authors acknowledge support from the Institute for Basic Science under grant IBS-R027-D1. 

We acknowledge fruitful discussions with Andreas Heinrich, Caroline Hommel, Christian Ast, Christopher P. Lutz, Deung-Jang Choi, Dmitriy Borodin, Dominic Ruckert, Fabio Donati, Franklin Cho, Gregory Czap, Hong Thi Bui, Jan Martinek, Joaquín Fernández-Rossier, Kevin Lizárraga, Nicolás Lorente, Patrick Lawes, Paul Greule, Philip Willke, Sander Otte, Sebastian Loth, Sebastian Stepanow, Shinjae Nam, Soo-hyon Phark, Wonjun Jang, Xue Zhang, and Yaowu Liu.





\printbibliography

@article{Role_coh_shakirov_2016,
  title = {Role of coherence in transport through engineered atomic spin devices},
  author = {Shakirov, Alexey M. and Shchadilova, Yulia E. and Rubtsov, Alexey N. and Ribeiro, Pedro},
  journal = {Phys. Rev. B},
  volume = {94},
  issue = {22},
  pages = {224425},
  numpages = {15},
  year = {2016},
  month = {Dec},
  publisher = {American Physical Society},
  doi = {10.1103/PhysRevB.94.224425},
  url = {https://link.aps.org/doi/10.1103/PhysRevB.94.224425}
}

@article{Tu_X_double_barrier_2008,
  title = {Controlling Single-Molecule Negative Differential Resistance in a Double-Barrier Tunnel Junction},
  author = {Tu, X. W. and Mikaelian, G. and Ho, W.},
  journal = {Phys. Rev. Lett.},
  volume = {100},
  issue = {12},
  pages = {126807},
  numpages = {4},
  year = {2008},
  month = {Mar},
  publisher = {American Physical Society},
  doi = {10.1103/PhysRevLett.100.126807},
  url = {https://link.aps.org/doi/10.1103/PhysRevLett.100.126807}
}

@article{caso_model_2014,
	title = {Model for electron spin resonance in {STM} noise},
	volume = {89},
	issn = {1098-0121, 1550-235X},
	url = {https://link.aps.org/doi/10.1103/PhysRevB.89.075412},
	doi = {10.1103/PhysRevB.89.075412},
	number = {7},
	urldate = {2018-07-31},
	journal = {Physical Review B},
	author = {Caso, Alvaro and Horovitz, Baruch and Arrachea, Liliana},
	month = feb,
	year = {2014},
}

@article{Drost2022,
    author = {Drost, Robert and Uhl, Maximilian and Kot, Piotr and Siebrecht, Janis and Schmid, Alexander and Merkt, Jonas and Wünsch, Stefan and Siegel, Michael and Kieler, Oliver and Kleiner, Reinhold and Ast, Christian R.},
    title = {Combining electron spin resonance spectroscopy with scanning tunneling microscopy at high magnetic fields},
    journal = {Review of Scientific Instruments},
    volume = {93},
    number = {4},
    pages = {043705},
    year = {2022},
    month = {04},
    issn = {0034-6748},
    doi = {10.1063/5.0078137},
    url = {https://doi.org/10.1063/5.0078137},
    eprint = {https://pubs.aip.org/aip/rsi/article-pdf/doi/10.1063/5.0078137/16660264/043705_1_online.pdf},
}

@misc{Willke2026,
      title={Atomic-Scale Quantum Control of Single Spin Defects in a Two-Dimensional Semiconductor}, 
      author={Kwan Ho Au-Yeung and Wantong Huang and Johanna Matusche and Paul Greule and Jonas Arnold and Lovis Hardeweg and Máté Stark and Luise Renz and Affan Safeer and Daniel Jansen and Thomas Michely and Jeison Fischer and Wolfgang Wernsdorfer and Christoph Sürgers and Hannu-Pekka Komsa and Johannes Schwenk and Wouter Jolie and Philip Willke},
      year={2026},
      eprint={2602.22301},
      archivePrefix={arXiv},
      primaryClass={cond-mat.mes-hall},
      url={https://arxiv.org/abs/2602.22301}, 
}

@Article{jauho_wingreen_prb_1994,
  title = {Time-dependent transport in interacting and noninteracting resonant-tunneling systems},
  author = {Jauho, Antti-Pekka and Wingreen, Ned S. and Meir, Yigal},
  journal = {Phys. Rev. B},
  volume = {50},
  issue = {8},
  pages = {5528--5544},
  numpages = {0},
  year = {1994},
  month = {Aug},
  publisher = {American Physical Society},
  doi = {10.1103/PhysRevB.50.5528},
  url = {https://link.aps.org/doi/10.1103/PhysRevB.50.5528}
}

@article{Grifoni1998,
title = {Driven quantum tunneling},
journal = {Physics Reports},
volume = {304},
number = {5},
pages = {229-354},
year = {1998},
issn = {0370-1573},
doi = {https://doi.org/10.1016/S0370-1573(98)00022-2},
url = {https://www.sciencedirect.com/science/article/pii/S0370157398000222},
author = {Milena Grifoni and Peter Hänggi}
}

@article{Foden1998,
  title = {Quantum electrodynamic treatment of photon-assisted tunneling},
  author = {Foden, C. L. and Whittaker, D. M.},
  journal = {Phys. Rev. B},
  volume = {58},
  issue = {19},
  pages = {12617--12620},
  numpages = {0},
  year = {1998},
  month = {Nov},
  publisher = {American Physical Society},
  doi = {10.1103/PhysRevB.58.12617},
  url = {https://link.aps.org/doi/10.1103/PhysRevB.58.12617}
}

@article{Yu2023,
doi = {10.1088/1367-2630/ad0a4e},
url = {https://doi.org/10.1088/1367-2630/ad0a4e},
year = {2023},
month = {11},
publisher = {IOP Publishing},
volume = {25},
number = {11},
pages = {113035},
author = {Yu, Jisoo and Urdaniz, Corina and Namgoong, Young and Wolf, Christoph},
title = {On the magnetic bistability of small iron clusters used in scanning tunneling microscopy tip preparation},
journal = {New Journal of Physics}
}

@article{arrachea-moskalets-2006,
  title = {Relation between scattering-matrix and Keldysh formalisms for quantum transport driven by time-periodic fields},
  author = {Arrachea, Liliana and Moskalets, Michael},
  journal = {Phys. Rev. B},
  volume = {74},
  issue = {24},
  pages = {245322},
  numpages = {13},
  year = {2006},
  month = {Dec},
  publisher = {American Physical Society},
  doi = {10.1103/PhysRevB.74.245322},
  url = {https://link.aps.org/doi/10.1103/PhysRevB.74.245322}
}

@article{Galperin_2016,
    author = {Chen, Feng and Ochoa, Maicol A. and Galperin, Michael},
    title = {Nonequilibrium diagrammatic technique for Hubbard Green functions},
    journal = {The Journal of Chemical Physics},
    volume = {146},
    number = {9},
    pages = {092301},
    year = {2016},
    month = {11},
    issn = {0021-9606},
    doi = {10.1063/1.4965825},
    url = {https://doi.org/10.1063/1.4965825},
    eprint = {https://pubs.aip.org/aip/jcp/article-pdf/doi/10.1063/1.4965825/14048116/092301_1_online.pdf},
}

@Misc{TimeESR_github,
title={TimeESR: An STM-ESR code solving a QME in the time domain.},
author={Lorente, Nicol\'as and Reina-G\'alvez, Jose and Wolf, Christoph and Eric D. Switzer},
url={https://github.com/qphensurf/TimeESR},
howpublished = "\url{https://github.com/qphensurf/TimeESR}",
year =         {2022} ,
note= {(Accessed on 2024-12-12)}
}

@article{Kosov_prb_2018,
  title = {Distribution of waiting times between electron cotunneling events},
  author = {Rudge, Samuel L. and Kosov, Daniel S.},
  journal = {Phys. Rev. B},
  volume = {98},
  issue = {24},
  pages = {245402},
  numpages = {13},
  year = {2018},
  month = {Dec},
  publisher = {American Physical Society},
  doi = {10.1103/PhysRevB.98.245402},
  url = {https://link.aps.org/doi/10.1103/PhysRevB.98.245402}
}

@article{Averin_physics_B_1994,
title = {Periodic conductance oscillations in the single-electron tunneling transistor},
journal = {Physica B: Condensed Matter},
volume = {194-196},
pages = {979-980},
year = {1994},
issn = {0921-4526},
doi = {https://doi.org/10.1016/0921-4526(94)90819-2},
url = {https://www.sciencedirect.com/science/article/pii/0921452694908192},
author = {D.V. Averin}
}

@article{Koch_Oppen_2006,
  title = {Theory of the Franck-Condon blockade regime},
  author = {Koch, Jens and von Oppen, Felix and Andreev, A. V.},
  journal = {Phys. Rev. B},
  volume = {74},
  issue = {20},
  pages = {205438},
  numpages = {19},
  year = {2006},
  month = {Nov},
  publisher = {American Physical Society},
  doi = {10.1103/PhysRevB.74.205438},
  url = {https://link.aps.org/doi/10.1103/PhysRevB.74.205438}
}

@article{Koch_Oppen_2004,
  title = {Thermopower of single-molecule devices},
  author = {Koch, Jens and von Oppen, Felix and Oreg, Yuval and Sela, Eran},
  journal = {Phys. Rev. B},
  volume = {70},
  issue = {19},
  pages = {195107},
  numpages = {12},
  year = {2004},
  month = {Nov},
  publisher = {American Physical Society},
  doi = {10.1103/PhysRevB.70.195107},
  url = {https://link.aps.org/doi/10.1103/PhysRevB.70.195107}
}

@article{Turek_2002,
  title = {Cotunneling thermopower of single electron transistors},
  author = {Turek, M. and Matveev, K. A.},
  journal = {Phys. Rev. B},
  volume = {65},
  issue = {11},
  pages = {115332},
  numpages = {9},
  year = {2002},
  month = {Mar},
  publisher = {American Physical Society},
  doi = {10.1103/PhysRevB.65.115332},
  url = {https://link.aps.org/doi/10.1103/PhysRevB.65.115332}
}

@article{Wacker_cotu_2010,
title = {Modeling of cotunneling in quantum dot systems},
journal = {Physica E: Low-dimensional Systems and Nanostructures},
volume = {42},
number = {3},
pages = {595-599},
year = {2010},
note = {Proceedings of the international conference Frontiers of Quantum and Mesoscopic Thermodynamics FQMT '08},
issn = {1386-9477},
doi = {https://doi.org/10.1016/j.physe.2009.06.069},
url = {https://www.sciencedirect.com/science/article/pii/S1386947709002550},
author = {Jonas Nyvold Pedersen and Andreas Wacker}
}

@article{Kosov_prb_2019,
  title = {Nonrenewal statistics in quantum transport from the perspective of first-passage and waiting time distributions},
  author = {Rudge, Samuel L. and Kosov, Daniel S.},
  journal = {Phys. Rev. B},
  volume = {99},
  issue = {11},
  pages = {115426},
  numpages = {14},
  year = {2019},
  month = {Mar},
  publisher = {American Physical Society},
  doi = {10.1103/PhysRevB.99.115426},
  url = {https://link.aps.org/doi/10.1103/PhysRevB.99.115426}
}

@article{Busz_PRB_2025,
  title = {Zero-bias anomaly indicating exchange interaction and spin readout in a canted quantum dot spin valve},
  author = {Busz, Piotr and Tomaszewski, Damian and Martinek, Jan},
  journal = {Phys. Rev. B},
  volume = {111},
  issue = {12},
  pages = {125415},
  numpages = {20},
  year = {2025},
  month = {Mar},
  publisher = {American Physical Society},
  doi = {10.1103/PhysRevB.111.125415},
  url = {https://link.aps.org/doi/10.1103/PhysRevB.111.125415}
}

@article{Weymann_2006_seq_cotu,
  title = {Tunnel magnetoresistance of quantum dots coupled to ferromagnetic leads in the sequential and cotunneling regimes},
  author = {Weymann, Ireneusz and K\"onig, J\"urgen and Martinek, Jan and Barna\ifmmode \acute{s}\else \'{s}\fi{}, J\'ozef and Sch\"on, Gerd},
  journal = {Phys. Rev. B},
  volume = {72},
  issue = {11},
  pages = {115334},
  numpages = {13},
  year = {2005},
  month = {Sep},
  publisher = {American Physical Society},
  doi = {10.1103/PhysRevB.72.115334},
  url = {https://link.aps.org/doi/10.1103/PhysRevB.72.115334}
}

@Article{Braun_Konig_prb_2004,
  title = {Theory of transport through quantum-dot spin valves in the weak-coupling regime},
  author = {Braun, Matthias and K\"onig, J\"urgen and Martinek, Jan},
  journal = {Phys. Rev. B},
  volume = {70},
  issue = {19},
  pages = {195345},
  numpages = {12},
  year = {2004},
  month = {Nov},
  publisher = {American Physical Society},
  doi = {10.1103/PhysRevB.70.195345},
  url = {https://link.aps.org/doi/10.1103/PhysRevB.70.195345}
}

@article{Piotr_Martinek_Hanle_effect_2023,
title = {Hanle effect in transport through single atoms in spin-polarized STM},
journal = {Journal of Magnetism and Magnetic Materials},
volume = {588},
pages = {171465},
year = {2023},
issn = {0304-8853},
doi = {https://doi.org/10.1016/j.jmmm.2023.171465},
url = {https://www.sciencedirect.com/science/article/pii/S0304885323011150},
author = {Piotr Busz and Damian Tomaszewski and Józef Barnaś and Jan Martinek}
}

@article{Binnig1982,
  author = {Binnig, G. and Rohrer, H. and Gerber, Ch. and Weibel, E.},
  title = {Surface Studies by Scanning Tunneling Microscopy},
  journal = {Phys. Rev. Lett.},
  volume = {49},
  pages = {57--61},
  year = {1982},
  doi = {10.1103/PhysRevLett.49.57}
}

@Article{Loth2010,
author={Loth, Sebastian
and von Bergmann, Kirsten
and Ternes, Markus
and Otte, Alexander F.
and Lutz, Christopher P.
and Heinrich, Andreas J.},
title={Controlling the state of quantum spins with electric currents},
journal={Nature Physics},
year={2010},
month={May},
day={01},
volume={6},
number={5},
pages={340-344},
issn={1745-2481},
doi={10.1038/nphys1616},
url={https://doi.org/10.1038/nphys1616}
}

@Article{Xuan2025,
author ="Xuan, Dalong and Wu, Di{'}an and Geng, Xiaobin and Li, Sihao and Wu, Xianglong and Wang, Yu and Zhang, Xue",
title  ="Quantum coherence and relaxation of single spins on surfaces probed by ESR-STM",
journal  ="Nanoscale",
year  ="2025",
volume  ="17",
issue  ="45",
pages  ="26024-26032",
publisher  ="The Royal Society of Chemistry",
doi  ="10.1039/D5NR03773E",
url  ="http://dx.doi.org/10.1039/D5NR03773E"}

@article{Tupkary2022,
  title = {Fundamental limitations in Lindblad descriptions of systems weakly coupled to baths},
  author = {Tupkary, Devashish and Dhar, Abhishek and Kulkarni, Manas and Purkayastha, Archak},
  journal = {Phys. Rev. A},
  volume = {105},
  issue = {3},
  pages = {032208},
  numpages = {14},
  year = {2022},
  month = {Mar},
  publisher = {American Physical Society},
  doi = {10.1103/PhysRevA.105.032208},
  url = {https://link.aps.org/doi/10.1103/PhysRevA.105.032208}
}

@article{Chen2023,
author = {Chen, Yi and Bae, Yujeong and Heinrich, Andreas J.},
title = {Harnessing the Quantum Behavior of Spins on Surfaces},
journal = {Advanced Materials},
volume = {35},
number = {27},
pages = {2107534},
doi = {https://doi.org/10.1002/adma.202107534},
url = {https://advanced.onlinelibrary.wiley.com/doi/abs/10.1002/adma.202107534},
eprint = {https://advanced.onlinelibrary.wiley.com/doi/pdf/10.1002/adma.202107534},
year = {2023}
}

@article{Bastiaans2018,
    author = {Bastiaans, K. M. and Benschop, T. and Chatzopoulos, D. and Cho, D. and Dong, Q. and Jin, Y. and Allan, M. P.},
    title = {Amplifier for scanning tunneling microscopy at MHz frequencies},
    journal = {Review of Scientific Instruments},
    volume = {89},
    number = {9},
    pages = {093709},
    year = {2018},
    month = {09},
    issn = {0034-6748},
    doi = {10.1063/1.5043267},
    url = {https://doi.org/10.1063/1.5043267},
    eprint = {https://pubs.aip.org/aip/rsi/article-pdf/doi/10.1063/1.5043267/16140624/093709_1_online.pdf},
}

@article{Delgado_2011,
  title = {Cotunneling theory of atomic spin inelastic electron tunneling spectroscopy},
  author = {Delgado, F. and Fern\'andez-Rossier, J.},
  journal = {Phys. Rev. B},
  volume = {84},
  issue = {4},
  pages = {045439},
  numpages = {11},
  year = {2011},
  month = {Jul},
  publisher = {American Physical Society},
  doi = {10.1103/PhysRevB.84.045439},
  url = {https://link.aps.org/doi/10.1103/PhysRevB.84.045439}
}

@article{Heinrich2004,
  title = {Single-Atom Spin-Flip Spectroscopy},
  author = {Heinrich, A. J. and Gupta, J. A. and Lutz, C. P. and Eigler, D. M.},
  journal = {Science},
  volume = {306},
  number = {5695},
  pages = {466--469},
  year = {2004},
  publisher = {American Association for the Advancement of Science},
  doi = {10.1126/science.1101077}
}

@article{Gimzewski1988,
  title = {Photon emission of electrons tunneling in a scanning tunneling microscope},
  author = {Gimzewski, J. K. and Reihl, B. and Coombs, J. H. and Schlittler, R. R.},
  journal = {Zeitschrift f{\"u}r Physik B Condensed Matter},
  volume = {72},
  pages = {497--501},
  year = {1988},
  doi = {10.1007/BF01314531}
}

@article{Giessibl1998,
  title = {High-speed force sensor for force microscopy and profilometry utilizing a quartz tuning fork},
  author = {Giessibl, Franz J.},
  journal = {Applied Physics Letters},
  volume = {73},
  number = {26},
  pages = {3956--3958},
  year = {1998},
  doi = {10.1063/1.122944}
}

@article{Giessibl2000,
  title = {Subatomic Features on the Silicon (111)-(7x7) Surface Observed by Atomic Force Microscopy},
  author = {Giessibl, Franz J. and Hembacher, Stefan and Bielefeldt, Hartmut and Mannhart, Jochen},
  journal = {Science},
  volume = {289},
  number = {5478},
  pages = {422--425},
  year = {2000},
  doi = {10.1126/science.289.5478.422}
}

@article{Manassen1989,
  title = {Direct observation of the Larmor precession of individual spins by scanning-tunneling microscopy},
  author = {Manassen, Y. and Hamers, R. J. and Demuth, J. E. and Castellano, A. J.},
  journal = {Phys. Rev. Lett.},
  volume = {62},
  issue = {21},
  pages = {2531--2534},
  year = {1989},
  doi = {10.1103/PhysRevLett.62.2531}
}

@article{Terada2010,
  title = {Real-space imaging of transient carrier dynamics by nanoscale pump--probe microscopy},
  author = {Terada, Yasuhiko and Tayasu, Shoji and Takeuchi, Osamu and Shigekawa, Hidemi},
  journal = {Nature Photonics},
  volume = {4},
  number = {12},
  pages = {869--874},
  year = {2010},
  publisher = {Nature Publishing Group},
  doi = {10.1038/nphoton.2010.239}
}

@article{Cocker2013,
  title = {An ultrafast terahertz scanning tunnelling microscope},
  author = {Cocker, T. L. and Jelic, V. and Gupta, M. and Molesky, S. J. and Burgess, J. A. J. and De Los Reyes, G. and Hegmann, F. A.},
  journal = {Nature Photonics},
  volume = {7},
  number = {8},
  pages = {620--625},
  year = {2013},
  publisher = {Nature Publishing Group},
  doi = {10.1038/nphoton.2013.151}
}

@article{Gimzewski1993,
  title = {Photon emission in scanning tunneling microscopy: Interpretation of photon maps of metallic systems},
  author = {Berndt, Richard and Gimzewski, James K.},
  journal = {Phys. Rev. B},
  volume = {48},
  issue = {7},
  pages = {4746--4754},
  numpages = {0},
  year = {1993},
  month = {Aug},
  publisher = {American Physical Society},
  doi = {10.1103/PhysRevB.48.4746},
  url = {https://link.aps.org/doi/10.1103/PhysRevB.48.4746}
}

@article{Wiesendanger1990,
  author = {Wiesendanger, R. and G\"{u}ntherodt, H.-J. and G\"{u}ntherodt, G. and Gambino, R. J. and Ruf, R.},
  title = {Observation of vacuum tunneling of spin-polarized electrons with the scanning tunneling microscope},
  journal = {Phys. Rev. Lett.},
  volume = {65},
  pages = {247--250},
  year = {1990},
  doi = {10.1103/PhysRevLett.65.247}
}

@article{Birk1995,
  author = {Birk, H. and de Castro, M. J. M. and Yates, S. U. S. and J\"{a}ckel, P. and Riisnes, J. and Sch\"{o}nenberger, C. and Wyder, P.},
  title = {Shot-noise measurements on mesoscopic contacts},
  journal = {Phys. Rev. Lett.},
  volume = {75},
  pages = {1610--1613},
  year = {1995},
  doi = {10.1103/PhysRevLett.75.1610}
}

@article{Stockle2000,
  author = {St\"{o}ckle, R. M. and Suh, Y. D. and Deckert, V. and Zenobi, R.},
  title = {Nanoscale chemical analysis by tip-enhanced Raman spectroscopy},
  journal = {Chem. Phys. Lett.},
  volume = {318},
  pages = {131--136},
  year = {2000},
  doi = {10.1016/S0009-2614(99)01451-7}
}

@article{
Bae2018ScienceAdvance,
author = {Y. Bae  and K. Yang  and P. Willke  and T. Choi  and A. J. Heinrich  and C. P. Lutz },
title = {Enhanced quantum coherence in exchange coupled spins via singlet-triplet transitions},
journal = {Science Advances},
volume = {4},
number = {11},
pages = {eaau4159},
year = {2018},
doi = {10.1126/sciadv.aau4159},
URL = {https://www.science.org/doi/abs/10.1126/sciadv.aau4159}
}

@Article{Heinrich2021,
author={Heinrich, Andreas J.
and Oliver, William D.
and Vandersypen, Lieven M. K.
and Ardavan, Arzhang
and Sessoli, Roberta
and Loss, Daniel
and Jayich, Ania Bleszynski
and Fernandez-Rossier, Joaquin
and Laucht, Arne
and Morello, Andrea},
title={Quantum-coherent nanoscience},
journal={Nature Nanotechnology},
year={2021},
month={Dec},
day={01},
volume={16},
number={12},
pages={1318-1329},
issn={1748-3395},
doi={10.1038/s41565-021-00994-1},
url={https://doi.org/10.1038/s41565-021-00994-1}
}

@article{Baumann_Paul_science_2015,
	title = {Electron paramagnetic resonance of individual atoms on a surface},
	author = {Baumann, Susanne and Paul, William and Choi, Taeyoung and Lutz, Christopher P. and Ardavan, Arzhang and Heinrich, Andreas J.},
	doi = {10.1126/science.aac8703},
	journal = {Science},
	number = {6259},
	pages = {417-420},
	volume = {350},
	year = {2015},
    URL = {https://www.science.org/doi/abs/10.1126/science.aac8703},
}

@article{Lado_Ferron_prb_2017,
    title = {Exchange mechanism for electron paramagnetic resonance of individual adatoms},
    year = {2017},
    journal = {Physical Review B},
    author = {Lado, J. L. and Ferr{\'{o}}n, A. and Fern{\'{a}}ndez-Rossier, J.},
    number = {20},
    month = {11},
    pages = {205420},
    volume = {96},
    publisher = {American Physical Society},
    doi = {10.1103/PhysRevB.96.205420},
    issn = {24699969},
    url = {https://link.aps.org/doi/10.1103/PhysRevB.96.205420}
}

@article{Seifert2020LongitudinalMicroscope,
    title = {{Longitudinal and transverse electron paramagnetic resonance in a scanning tunneling microscope}},
    year = {2020},
    journal = {Science Advances},
    author = {Seifert, Tom S. and Kovarik, Stepan and Juraschek, Dominik M. and Spaldin, Nicola A. and Gambardella, Pietro and Stepanow, Sebastian},
    number = {40},
    month = {10},
    pages = {1--12},
    volume = {6},
    doi = {10.1126/sciadv.abc5511},
    issn = {23752548},
    URL = {https://www.science.org/doi/abs/10.1126/sciadv.abc5511},
}

@article{Ferron2019,
    title = {{ Single spin resonance driven by electric modulation of the  g  -factor anisotropy }},
    year = {2019},
    journal = {Physical Review Research},
    author = {Ferr{\'{o}}n, A. and Rodr{\'{i}}guez, S. A. and G{\'{o}}mez, S. S. and Lado, J. L. and Fern{\'{a}}ndez-Rossier, J.},
    number = {3},
    pages = {1--13},
    volume = {1},
    doi = {10.1103/physrevresearch.1.033185},
    url = {https://link.aps.org/doi/10.1103/PhysRevResearch.1.033185}
}

@article{J_Cuevas_C_Ast_ESR_theory_2024,
  title = {Theory of electron spin resonance in scanning tunneling microscopy},
  author = {Ast, Christian R. and Kot, Piotr and Ismail, Maneesha and de-la-Pe\~na, Sebasti\'an and Fern\'andez-Dom\'{\i}nguez, Antonio I. and Cuevas, Juan Carlos},
  journal = {Phys. Rev. Res.},
  volume = {6},
  issue = {2},
  pages = {023126},
  numpages = {18},
  year = {2024},
  month = {May},
  publisher = {American Physical Society},
  doi = {10.1103/PhysRevResearch.6.023126},
  url = {https://link.aps.org/doi/10.1103/PhysRevResearch.6.023126}
}

@article{J_Reina_Galvez_2023,
	author = {Reina-G\'alvez, Jose and Wolf, Christoph and Lorente, Nicol\'as},
	doi = {10.1103/PhysRevB.107.235404},
	issue = {23},
	journal = {Phys. Rev. B},
	month = {Jun},
	numpages = {8},
	pages = {235404},
	publisher = {American Physical Society},
	title = {Many-body nonequilibrium effects in all-electric electron spin resonance},
	volume = {107},
	year = {2023},
    url = {https://link.aps.org/doi/10.1103/PhysRevB.107.235404}}

@article{J_Reina_Galvez_2021,
	author = {Reina-G\'alvez, Jose and Lorente, Nicol\'as and Delgado, Fernando and Arrachea, Liliana},
	doi = {10.1103/PhysRevB.104.245435},
	issue = {24},
	journal = {Phys. Rev. B},
	month = {Dec},
	numpages = {16},
	pages = {245435},
	publisher = {American Physical Society},
	title = {All-electric electron spin resonance studied by means of Floquet quantum master equations},
	volume = {104},
	year = {2021},
    url = {https://link.aps.org/doi/10.1103/PhysRevB.104.245435}
}

@article{J_Reina_Galvez_2019,
	author = {Reina G\'alvez, J. and Wolf, C. and Delgado, F. and Lorente, N.},
	doi = {10.1103/PhysRevB.100.035411},
	issue = {3},
	journal = {Phys. Rev. B},
	month = {Jul},
	numpages = {10},
	pages = {035411},
	publisher = {American Physical Society},
	title = {Cotunneling mechanism for all-electrical electron spin resonance of single adsorbed atoms},
	volume = {100},
	year = {2019},
    url = {https://link.aps.org/doi/10.1103/PhysRevB.100.035411}
}

@Article{Corina2025,
author={Urdaniz, Corina
and Taherpour, Saba
and Yu, Jisoo
and Reina-Galvez, Jose
and Wolf, Christoph},
title={Transition-Metal Phthalocyanines as Versatile Building Blocks for Molecular Qubits on Surfaces},
journal={The Journal of Physical Chemistry A},
year={2025},
month={Mar},
day={06},
publisher={American Chemical Society},
volume={129},
number={9},
pages={2173-2181},
issn={1089-5639},
doi={10.1021/acs.jpca.4c07627},
url={https://doi.org/10.1021/acs.jpca.4c07627}
}

@article{jose_anh2025,
title = {Efficient driving of a spin qubit using single-atom magnets},
  author = {Reina-G\'alvez, Jose and Le, Hoang-Anh and Bui, Hong Thi and Phark, Soo-hyon and Lorente, Nicol\'as and Wolf, Christoph},
  journal = {Phys. Rev. Res.},
  volume = {7},
  issue = {4},
  pages = {L042055},
  numpages = {7},
  year = {2025},
  month = {Dec},
  publisher = {American Physical Society},
  doi = {10.1103/qpz5-s6fw},
  url = {https://link.aps.org/doi/10.1103/qpz5-s6fw}
}

@article{wolfdelgado2020,
author = {Wolf, Christoph and Delgado, Fernando and Reina, Jos{\'e} and Lorente, Nicolás},
title = {Efficient Ab Initio Multiplet Calculations for Magnetic Adatoms on MgO},
journal = {The Journal of Physical Chemistry A},
volume = {124},
number = {11},
pages = {2318-2327},
year = {2020},
doi = {10.1021/acs.jpca.9b10749},
    note ={PMID: 32098473},

URL = { 
    
        https://doi.org/10.1021/acs.jpca.9b10749
    
    

},
eprint = { 
    
        https://doi.org/10.1021/acs.jpca.9b10749
    
    

}

}

@article{Phark2023electric,
author = {Phark, Soo-hyon and Bui, Hong Thi and Ferrón, Alejandro and Fernández-Rossier, Joaquin and Reina-Gálvez, Jose and Wolf, Christoph and Wang, Yu and Yang, Kai and Heinrich, Andreas J. and Lutz, Christopher P.},
title = {Electric-Field-Driven Spin Resonance by On-Surface Exchange Coupling to a Single-Atom Magnet},
journal = {Advanced Science},
volume = {10},
number = {27},
pages = {2302033},
doi = {https://doi.org/10.1002/advs.202302033},
year = {2023},
url = {https://advanced.onlinelibrary.wiley.com/doi/abs/10.1002/advs.202302033}
}

@article{Ye2024theory,
  title = {Theory of Electron Spin Resonance Spectroscopy in Scanning Tunneling Microscopy},
  author = {Ye, Lyuzhou and Zheng, Xiao and Xu, Xin},
  journal = {Phys. Rev. Lett.},
  volume = {133},
  issue = {17},
  pages = {176201},
  numpages = {6},
  year = {2024},
  month = {Oct},
  publisher = {American Physical Society},
  doi = {10.1103/PhysRevLett.133.176201},
  url = {https://link.aps.org/doi/10.1103/PhysRevLett.133.176201}
}

@article{le2025overcoming,
  title={Overcoming limitations on gate fidelity in noisy static exchange-coupled surface qubits},
  author={Le, Hoang-Anh and Taherpour, Saba and Jankovi{\'c}, Denis and Wolf, Christoph},
  journal={npj Quantum Information},
  year={2026},
  doi={10.1038/s41534-026-01214-1},
  url={https://doi.org/10.1038/s41534-026-01214-1}
}

@article{
KaiYang_science2019,
author = {Kai Yang  and William Paul  and Soo-Hyon Phark  and Philip Willke  and Yujeong Bae  and Taeyoung Choi  and Taner Esat  and Arzhang Ardavan  and Andreas J. Heinrich  and Christopher P. Lutz },
title = {Coherent spin manipulation of individual atoms on a surface},
journal = {Science},
volume = {366},
number = {6464},
pages = {509-512},
year = {2019},
doi = {10.1126/science.aay6779},
URL = {https://www.science.org/doi/abs/10.1126/science.aay6779}
}

@article{Shakirov_spin_torque,
  title = {Spin transfer torque induced paramagnetic resonance},
  author = {Shakirov, Alexey M. and Rubtsov, Alexey N. and Ribeiro, Pedro},
  journal = {Phys. Rev. B},
  volume = {99},
  issue = {5},
  pages = {054434},
  numpages = {7},
  year = {2019},
  month = {Feb},
  publisher = {American Physical Society},
  doi = {10.1103/PhysRevB.99.054434},
  url = {https://link.aps.org/doi/10.1103/PhysRevB.99.054434}
}

@misc{Greg2025magneticresonance,
      title={Magnetic Resonance Imaging of Single Organic Radicals with Sub-Molecular Resolution}, 
      author={Gregory Czap and Christoph Wolf and Jose Reina-Gálvez and Mark H. Sherwood and Christopher P. Lutz},
      year={2025},
      eprint={2504.18043},
      archivePrefix={arXiv},
      primaryClass={cond-mat.mes-hall},
      url={https://arxiv.org/abs/2504.18043}, 
}

@misc{Greule2025spincontrol,
      title={Spin-Electric Control of Individual Molecules on Surfaces}, 
      author={Paul Greule and Wantong Huang and Máté Stark and Kwan Ho Au-Yeung and Johannes Schwenk and Jose Reina-Gálvez and Christoph Sürgers and Wolfgang Wernsdorfer and Christoph Wolf and Philip Willke},
      year={2025},
      eprint={2507.13699},
      archivePrefix={arXiv},
      primaryClass={cond-mat.mes-hall},
      url={https://arxiv.org/abs/2507.13699}, 
}

@article{Wang2023Universal,
  author  = {Wang, Yu and Haze, Masahiro and Bui, Hong T. and Soe, We-hyo and Aubin, Herve and Ardavan, Arzhang and Heinrich, Andreas J. and Phark, Soo-hyon},
  title   = {Universal quantum control of an atomic spin qubit on a surface},
  journal = {npj Quantum Information},
  year    = {2023},
  volume  = {9},
  number  = {1},
  pages   = {48},
  doi     = {10.1038/s41534-023-00737-9},
  url     = {https://doi.org/10.1038/s41534-023-00716-6}
}

@article{Phark2023double,
  author  = {Phark, Soo-hyon and Chen, Yi and Bui, Hong T. and Wang, Yu and Haze, Masahiro and Kim, Jinkyung and Bae, Yujeong and Heinrich, Andreas J. and Wolf, Christoph},
  title   = {Double-Resonance Spectroscopy of Coupled Electron Spins on a Surface},
  journal = {ACS Nano},
  year    = {2023},
  volume  = {17},
  number  = {14},
  pages   = {14144--14151},
  doi     = {10.1021/acsnano.3c04754},
  publisher = {American Chemical Society},
  url     = {https://doi.org/10.1021/acsnano.3c04754}
}

@article{Phark2023multi-qubit,
author = {Yu Wang  and Yi Chen  and Hong T. Bui  and Christoph Wolf  and Masahiro Haze  and Cristina Mier  and Jinkyung Kim  and Deung-Jang Choi  and Christopher P. Lutz  and Yujeong Bae  and Soo-hyon Phark  and Andreas J. Heinrich },
title = {An atomic-scale multi-qubit platform},
journal = {Science},
volume = {382},
number = {6666},
pages = {87-92},
year = {2023},
doi = {10.1126/science.ade5050},
URL = {https://www.science.org/doi/abs/10.1126/science.ade5050},
}

@Article{Guo2026,
author={Guo, Zixuan
and Zhang, Jiajun
and Chen, Yi},
title={Electron spin resonance scanning tunneling microscopy beyond a single spin},
journal={Newton},
year={2026},
month={03},
publisher={Elsevier},
issn={2950-6360},
doi={10.1016/j.newton.2026.100413},
url={https://doi.org/10.1016/j.newton.2026.100413}
}

@article{Willke2021ControlMolecules,
  author  = {Willke, Philip and Bilgeri, Tobias and Zhang, Xue and Wang, Yu and Wolf, Christoph and Aubin, Herve and Heinrich, Andreas and Choi, Taeyoung},
  title   = {Coherent Spin Control of Single Molecules on a Surface},
  journal = {ACS Nano},
  year    = {2021},
  volume  = {15},
  number  = {11},
  pages   = {17959--17965},
  doi     = {10.1021/acsnano.1c06394},
  publisher = {American Chemical Society},
  url     = {https://doi.org/10.1021/acsnano.1c06394 }
}

@article{Broekhoven2024,
  author  = {Broekhoven, Rik and Lee, Curie and Phark, Soo-hyon and Otte, Sander and Wolf, Christoph},
  title   = {Protocol for certifying entanglement in surface spin systems using a scanning tunneling microscope},
  journal = {npj Quantum Information},
  year    = {2024},
  volume  = {10},
  number  = {1},
  pages   = {92},
  doi     = {10.1038/s41534-024-00888-9},
  url     = {https://doi.org/10.1038/s41534-024-00888-9}
}

@article{Kawaguchi2023,
  author  = {Kawaguchi, Ryo and Hashimoto, Katsushi and Kakudate, Toshiyuki and Katoh, Keiichi and Yamashita, Masahiro and Komeda, Tadahiro},
  title   = {Spatially Resolving Electron Spin Resonance of $\pi$-Radical in Single-Molecule Magnet},
  journal = {Nano Letters},
  year    = {2023},
  volume  = {23},
  number  = {1},
  pages   = {213--219},
  doi     = {10.1021/acs.nanolett.2c04049},
  publisher = {American Chemical Society},
  url     = {https://doi.org/10.1021/acs.nanolett.2c04049}
}

@article{Natterer2017,
  author  = {Natterer, Fabian D. and Yang, Kai and Paul, William and Willke, Philip and Choi, Taeyoung and Greber, Thomas and Heinrich, Andreas J. and Lutz, Christopher P.},
  title   = {Reading and writing single-atom magnets},
  journal = {Nature},
  year    = {2017},
  volume  = {543},
  number  = {7644},
  pages   = {226--228},
  doi     = {10.1038/nature21371},
  url     = {https://doi.org/10.1038/nature21371}
}

@article{Choi2017,
  author  = {Choi, Taeyoung and Paul, William and Rolf-Pissarczyk, Steffen and Macdonald, Andrew J. and Natterer, Fabian D. and Yang, Kai and Willke, Philip and Lutz, Christopher P. and Heinrich, Andreas J.},
  title   = {Atomic-scale sensing of the magnetic dipolar field from single atoms},
  journal = {Nature Nanotechnology},
  year    = {2017},
  volume  = {12},
  number  = {5},
  pages   = {420--424},
  doi     = {10.1038/nnano.2017.18},
  url     = {https://doi.org/10.1038/nnano.2017.18}
}

@article{Willke2018Probing,
author = {Philip Willke  and William Paul  and Fabian D. Natterer  and Kai Yang  and Yujeong Bae  and Taeyoung Choi  and Joaquin Fernández-Rossier  and Andreas J. Heinrich  and Christoper P. Lutz },
title = {Probing quantum coherence in single-atom electron spin resonance},
journal = {Science Advances},
volume = {4},
number = {2},
pages = {eaaq1543},
year = {2018},
doi = {10.1126/sciadv.aaq1543},
URL = {https://www.science.org/doi/abs/10.1126/sciadv.aaq1543}
}

@article{Yang2017Engineering,
  title = {Engineering the Eigenstates of Coupled Spin-$1/2$ Atoms on a Surface},
  author = {Yang, Kai and Bae, Yujeong and Paul, William and Natterer, Fabian D. and Willke, Philip and Lado, Jose L. and Ferr\'on, Alejandro and Choi, Taeyoung and Fern\'andez-Rossier, Joaqu\'{\i}n and Heinrich, Andreas J. and Lutz, Christopher P.},
  journal = {Phys. Rev. Lett.},
  volume = {119},
  issue = {22},
  pages = {227206},
  numpages = {5},
  year = {2017},
  month = {Nov},
  publisher = {American Physical Society},
  doi = {10.1103/PhysRevLett.119.227206},
  url = {https://link.aps.org/doi/10.1103/PhysRevLett.119.227206}
}

@article{Weerdenburg2021,
    author = {van Weerdenburg, Werner M. J. and Steinbrecher, Manuel and van Mullekom, Niels P. E. and Gerritsen, Jan W. and von Allwörden, Henning and Natterer, Fabian D. and Khajetoorians, Alexander A.},
    title = {A scanning tunneling microscope capable of electron spin resonance and pump–probe spectroscopy at mK temperature and in vector magnetic field},
    journal = {Review of Scientific Instruments},
    volume = {92},
    number = {3},
    pages = {033906},
    year = {2021},
    month = {03},
    issn = {0034-6748},
    doi = {10.1063/5.0040011},
    url = {https://doi.org/10.1063/5.0040011}
}

@article{Veldman2021cohevolution,
author = {Lukas M. Veldman  and Laëtitia Farinacci  and Rasa Rejali  and Rik Broekhoven  and Jérémie Gobeil  and David Coffey  and Markus Ternes  and Alexander F. Otte },
title = {Free coherent evolution of a coupled atomic spin system initialized by electron scattering},
journal = {Science},
volume = {372},
number = {6545},
pages = {964-968},
year = {2021},
doi = {10.1126/science.abg8223},
URL = {https://www.science.org/doi/abs/10.1126/science.abg8223}
}

@article{reinagalvez2025contrasting,
  title = {Contrasting exchange-field and spin-transfer torque driving mechanisms in all-electric electron spin resonance},
  author = {Reina-G\'alvez, Jose and Nachtigall, Matyas and Lorente, Nicol\'as and Martinek, Jan and Wolf, Christoph},
  journal = {Phys. Rev. B},
  volume = {112},
  issue = {24},
  pages = {245408},
  numpages = {27},
  year = {2025},
  month = {Dec},
  publisher = {American Physical Society},
  doi = {10.1103/nzhr-syhs},
  url = {https://link.aps.org/doi/10.1103/nzhr-syhs}
}

@article{Paul2017,
author={Paul, William
and Yang, Kai
and Baumann, Susanne
and Romming, Niklas
and Choi, Taeyoung
and Lutz, Christopher P.
and Heinrich, Andreas J.},
title={Control of the millisecond spin lifetime of an electrically probed atom},
journal={Nature Physics},
year={2017},
month={Apr},
day={01},
volume={13},
number={4},
pages={403-407},
issn={1745-2481},
doi={10.1038/nphys3965},
url={https://doi.org/10.1038/nphys3965}
}

@article{Willke2019Tuning,
  author  = {Willke, Philip and Singha, Aparajita and Zhang, Xue and Esat, Taner and Lutz, Christopher P. and Heinrich, Andreas J. and Choi, Taeyoung},
  title   = {Tuning Single-Atom Electron Spin Resonance in a Vector Magnetic Field},
  journal = {Nano Letters},
  year    = {2019},
  volume  = {19},
  number  = {11},
  pages   = {8201--8206},
  doi     = {10.1021/acs.nanolett.9b03559},
  publisher = {American Chemical Society},
  url     = {https://doi.org/10.1021/acs.nanolett.9b03559}
}

@article{Willke2019MRI,
  author  = {Willke, Philip and Yang, Kai and Bae, Yujeong and Heinrich, Andreas J. and Lutz, Christopher P.},
  title   = {Magnetic Resonance Imaging of Single Atoms on a Surface},
  journal = {Nature Physics},
  year    = {2019},
  volume  = {15},
  number  = {10},
  pages   = {1005--1010},
  doi     = {10.1038/s41567-019-0573-x},
  url     = {https://doi.org/10.1038/s41567-019-0573-x}
}

@article{Yang2019Tuning,
  title = {Tuning the Exchange Bias on a Single Atom from 1 mT to 10 T},
  author = {Yang, Kai and Paul, William and Natterer, Fabian D. and Lado, Jose L. and Bae, Yujeong and Willke, Philip and Choi, Taeyoung and Ferr\'on, Alejandro and Fern\'andez-Rossier, Joaqu\'{\i}n and Heinrich, Andreas J. and Lutz, Christopher P.},
  journal = {Phys. Rev. Lett.},
  volume = {122},
  issue = {22},
  pages = {227203},
  numpages = {6},
  year = {2019},
  month = {Jun},
  publisher = {American Physical Society},
  doi = {10.1103/PhysRevLett.122.227203},
  url = {https://link.aps.org/doi/10.1103/PhysRevLett.122.227203}
}

@article{Steinbrecher2021,
  title = {Quantifying the interplay between fine structure and geometry of an individual molecule on a surface},
  author = {Steinbrecher, Manuel and van Weerdenburg, Werner M. J. and Walraven, Etienne F. and van Mullekom, Niels P. E. and Gerritsen, Jan W. and Natterer, Fabian D. and Badrtdinov, Danis I. and Rudenko, Alexander N. and Mazurenko, Vladimir V. and Katsnelson, Mikhail I. and van der Avoird, Ad and Groenenboom, Gerrit C. and Khajetoorians, Alexander A.},
  journal = {Phys. Rev. B},
  volume = {103},
  issue = {15},
  pages = {155405},
  numpages = {13},
  year = {2021},
  month = {Apr},
  publisher = {American Physical Society},
  doi = {10.1103/PhysRevB.103.155405},
  url = {https://link.aps.org/doi/10.1103/PhysRevB.103.155405}
}

@article{Kovarik2022,
  author  = {Kovarik, Stepan and Robles, Roberto and Schlitz, Richard and Seifert, Tom Sebastian and Lorente, Nicolas and Gambardella, Pietro and Stepanow, Sebastian},
  title   = {Electron Paramagnetic Resonance of Alkali Metal Atoms and Dimers on Ultrathin MgO},
  journal = {Nano Letters},
  year    = {2022},
  volume  = {22},
  number  = {10},
  pages   = {4176--4181},
  doi     = {10.1021/acs.nanolett.2c00980},
  publisher = {American Chemical Society},
  url     = {https://doi.org/10.1021/acs.nanolett.2c00980}
}

@article{Chen2025FePc,
  title = {Microscopic Mechanism of Coexisting Electron Spin Resonance and Kondo Resonance in a Single Iron Phthalocyanine Molecule},
  author = {Chen, Qi and Du, Hongjian and Meng, Xinyong and Wang, Jufeng and Liao, Jingzun and Li, Bin and Hu, Wei and Fan, Qitang and Tan, Shijing and Ma, Chuanxu and Yang, Jinlong and Hou, J. G. and Wang, Bing},
  journal = {Phys. Rev. Lett.},
  volume = {135},
  issue = {8},
  pages = {086403},
  numpages = {8},
  year = {2025},
  month = {Aug},
  publisher = {American Physical Society},
  doi = {10.1103/cgq3-dyxb},
  url = {https://link.aps.org/doi/10.1103/cgq3-dyxb}
}

@article{Taner2024,
  author  = {Esat, Taner and Borodin, Dmitriy and Oh, Jeongmin and Heinrich, Andreas J. and Tautz, F. Stefan and Bae, Yujeong and Temirov, Ruslan},
  title   = {A Quantum Sensor for Atomic-Scale Electric and Magnetic Fields},
  journal = {Nature Nanotechnology},
  year    = {2024},
  volume  = {19},
  number  = {10},
  pages   = {1466--1471},
  doi     = {10.1038/s41565-024-01724-z},
  url     = {https://doi.org/10.1038/s41565-024-01724-z}
}

@article{Hong2024stark,
  author  = {Bui, Hong T. and Wolf, Christoph and Wang, Yu and Haze, Masahiro and Ardavan, Arzhang and Heinrich, Andreas J. and Phark, Soo-hyon},
  title   = {All-Electrical Driving and Probing of Dressed States in a Single Spin},
  journal = {ACS Nano},
  year    = {2024},
  volume  = {18},
  number  = {19},
  pages   = {12187--12193},
  doi     = {10.1021/acsnano.4c00196},
  publisher = {American Chemical Society},
  url     = {https://doi.org/10.1021/acsnano.4c00196}
}

@article{Otte2009Kondo,
  title = {Spin Excitations of a Kondo-Screened Atom Coupled to a Second Magnetic Atom},
  author = {Otte, A. F. and Ternes, M. and Loth, S. and Lutz, C. P. and Hirjibehedin, C. F. and Heinrich, A. J.},
  journal = {Phys. Rev. Lett.},
  volume = {103},
  issue = {10},
  pages = {107203},
  numpages = {4},
  year = {2009},
  month = {Sep},
  publisher = {American Physical Society},
  doi = {10.1103/PhysRevLett.103.107203},
  url = {https://link.aps.org/doi/10.1103/PhysRevLett.103.107203}
}

@article{Otte2008MagneticAnisotropy,
  author  = {Otte, Alexander F. and Ternes, Markus and von Bergmann, Kirsten and Loth, Sebastian and Brune, Harald and Lutz, Christopher P. and Hirjibehedin, Cyrus F. and Heinrich, Andreas J.},
  title   = {The Role of Magnetic Anisotropy in the Kondo Effect},
  journal = {Nature Physics},
  year    = {2008},
  volume  = {4},
  number  = {11},
  pages   = {847--850},
  doi     = {10.1038/nphys1072},
  url     = {https://doi.org/10.1038/nphys1072}
}

@article{Ternes2015Spinexcitations,
doi = {10.1088/1367-2630/17/6/063016},
url = {https://doi.org/10.1088/1367-2630/17/6/063016},
year = {2015},
month = {June},
publisher = {IOP Publishing},
volume = {17},
number = {6},
pages = {063016},
author = {Ternes, Markus},
title = {Spin excitations and correlations in scanning tunneling spectroscopy},
journal = {New Journal of Physics}
}

@article{Kot2023ElectricControl,
  author  = {Kot, Piotr and Ismail, Maneesha and Drost, Robert and Siebrecht, Janis and Huang, Haonan and Ast, Christian R.},
  title   = {Electric Control of Spin Transitions at the Atomic Scale},
  journal = {Nature Communications},
  year    = {2023},
  volume  = {14},
  number  = {1},
  pages   = {6612},
  doi     = {10.1038/s41467-023-42287-2},
  url     = {https://doi.org/10.1038/s41467-023-42287-2}
}

@article{Huang2025Ferrimagnets,
  author  = {Huang, Wantong and Stark, Máté and Greule, Paul and Au-Yeung, Kwan Ho and Sostina, Daria and Reina Gálvez, José and Sürgers, Christoph and Wernsdorfer, Wolfgang and Wolf, Christoph and Willke, Philip},
  title   = {Quantum Spin-Engineering in On-Surface Molecular Ferrimagnets},
  journal = {Nature Communications},
  year    = {2025},
  volume  = {16},
  number  = {1},
  pages   = {5208},
  doi     = {10.1038/s41467-025-60409-w},
  url     = {https://doi.org/10.1038/s41467-025-60409-w}
}

@misc{zhang2025controllingexchangefieldsurface,
      title={Controlling the Exchange Field of Surface Spin Impurities via DC Voltages}, 
      author={Xue Zhang and Jose Reina-Gálvez and Di'an Wu and Jan Martinek and Andreas J. Heinrich and Taeyoung Choi and Christoph Wolf},
      year={2025},
      eprint={2412.03866},
      archivePrefix={arXiv},
      primaryClass={cond-mat.mes-hall},
      url={https://arxiv.org/abs/2412.03866}, 
}

@book{Abragam1970,
  author    = {Abragam, A. and Bleaney, B.},
  title     = {Electron Paramagnetic Resonance of Transition Ions},
  publisher = {Oxford University Press},
  series    = {Oxford Classic Texts in the Physical Sciences},
  year      = {1970}
}

@book{Wiesendanger1994SPM,
  author    = {Wiesendanger, Roland},
  title     = {Scanning Probe Microscopy and Spectroscopy: Methods and Applications},
  publisher = {Cambridge University Press},
  year      = {1994}
}

@Article{Eric2025entanglement,
author ="Switzer, Eric D. and Reina-Gálvez, Jose and Giedke, Géza and Rahman, Talat S. and Wolf, Christoph and Choi, Deung-Jang and Lorente, Nicolás",
title  ="Unraveling spin entanglement using quantum gates with scanning tunneling microscopy-driven electron spin resonance",
journal  ="Nanoscale Adv.",
year  ="2025",
pages  ="8048-8057",
publisher  ="RSC",
doi  ="10.1039/D5NA00421G",
url  ="http://dx.doi.org/10.1039/D5NA00421G",
}

@article{Evert2025nuclearspin,
  author    = {Stolte, Evert W. and Lee, Jinwon and Vennema, Hester G. and Broekhoven, Rik and Teng, Esther and Katan, Allard J. and Veldman, Lukas M. and Willke, Philip and Otte, Sander},
  title     = {Single-shot readout of the nuclear spin of an on-surface atom},
  journal   = {Nature Communications},
  year      = {2025},
  volume    = {16},
  number    = {1},
  pages     = {7785},
  doi       = {10.1038/s41467-025-63232-5},
  url       = {https://doi.org/10.1038/s41467-025-63232-5}
}

@article{Lambe1968VibrationSpectra,
  title = {Molecular Vibration Spectra by Inelastic Electron Tunneling},
  author = {Lambe, J. and Jaklevic, R. C.},
  journal = {Phys. Rev.},
  volume = {165},
  issue = {3},
  pages = {821--832},
  numpages = {0},
  year = {1968},
  month = {Jan},
  publisher = {American Physical Society},
  doi = {10.1103/PhysRev.165.821},
  url = {https://link.aps.org/doi/10.1103/PhysRev.165.821}
}

@article{Song2010Highresolution,
  author    = {Song, Young Jae and Otte, Alexander F. and Kuk, Young and Hu, Yike and Torrance, David B. and First, Phillip N. and de Heer, Walt A. and Min, Hongki and Adam, Shaffique and Stiles, Mark D. and MacDonald, Allan H. and Stroscio, Joseph A.},
  title     = {High-resolution tunnelling spectroscopy of a graphene quartet},
  journal   = {Nature},
  year      = {2010},
  volume    = {467},
  number    = {7312},
  pages     = {185--189},
  doi       = {10.1038/nature09330},
  url       = {https://doi.org/10.1038/nature09330}
}

@article{Ast2016quantumlimit,
  author    = {Ast, Christian R. and Jäck, Berthold and Senkpiel, Jacob and Eltschka, Matthias and Etzkorn, Markus and Ankerhold, Joachim and Kern, Klaus},
  title     = {Sensing the quantum limit in scanning tunnelling spectroscopy},
  journal   = {Nature Communications},
  year      = {2016},
  volume    = {7},
  number    = {1},
  pages     = {13009},
  doi       = {10.1038/ncomms13009},
  url       = {https://doi.org/10.1038/ncomms13009}
}

@article{Schrieffer1966,
  title = {Relation between the Anderson and Kondo Hamiltonians},
  author = {Schrieffer, J. R. and Wolff, P. A.},
  journal = {Phys. Rev.},
  volume = {149},
  issue = {2},
  pages = {491--492},
  numpages = {0},
  year = {1966},
  month = {Sep},
  publisher = {American Physical Society},
  doi = {10.1103/PhysRev.149.491},
  url = {https://link.aps.org/doi/10.1103/PhysRev.149.491}
}

@article{Natterer2019,
    author = {Natterer, Fabian D. and Patthey, François and Bilgeri, Tobias and Forrester, Patrick R. and Weiss, Nicolas and Brune, Harald},
    title = {Upgrade of a low-temperature scanning tunneling microscope for electron-spin resonance},
    journal = {Review of Scientific Instruments},
    volume = {90},
    number = {1},
    pages = {013706},
    year = {2019},
    month = {01},
    issn = {0034-6748},
    doi = {10.1063/1.5065384},
    url = {https://doi.org/10.1063/1.5065384},
    eprint = {https://pubs.aip.org/aip/rsi/article-pdf/doi/10.1063/1.5065384/14757622/013706_1_online.pdf},
}

@article{Hwang2022,
    author = {Hwang, Jiyoon and Krylov, Denis and Elbertse, Robbie and Yoon, Sangwon and Ahn, Taehong and Oh, Jeongmin and Fang, Lei and Jang, Won-jun and Cho, Franklin H. and Heinrich, Andreas J. and Bae, Yujeong},
    title = {Development of a scanning tunneling microscope for variable temperature electron spin resonance},
    journal = {Review of Scientific Instruments},
    volume = {93},
    number = {9},
    pages = {093703},
    year = {2022},
    month = {09},
    issn = {0034-6748},
    doi = {10.1063/5.0096081},
    url = {https://doi.org/10.1063/5.0096081},
    eprint = {https://pubs.aip.org/aip/rsi/article-pdf/doi/10.1063/5.0096081/16603984/093703_1_online.pdf},
}

@article{Stefano2024driving4f,
  author    = {Reale, Stefano and Hwang, Jiyoon and Oh, Jeongmin and Brune, Harald and Heinrich, Andreas J. and Donati, Fabio and Bae, Yujeong},
  title     = {Electrically driven spin resonance of 4f electrons in a single atom on a surface},
  journal   = {Nature Communications},
  year      = {2024},
  volume    = {15},
  number    = {1},
  pages     = {5289},
  doi       = {10.1038/s41467-024-49447-y},
  url       = {https://doi.org/10.1038/s41467-024-49447-y}
}

@article{Zhang2022FePc,
  author    = {Zhang, Xue and Wolf, Christoph and Wang, Yu and Aubin, Hervé and Bilgeri, Tobias and Willke, Philip and Heinrich, Andreas J. and Choi, Taeyoung},
  title     = {Electron spin resonance of single iron phthalocyanine molecules and role of their non-localized spins in magnetic interactions},
  journal   = {Nature Chemistry},
  year      = {2022},
  volume    = {14},
  number    = {1},
  pages     = {59--65},
  doi       = {10.1038/s41557-021-00827-7},
  url       = {https://doi.org/10.1038/s41557-021-00827-7}
}

@article{JK2021vectormagneticfield,
  title = {Spin resonance amplitude and frequency of a single atom on a surface in a vector magnetic field},
  author = {Kim, Jinkyung and Jang, Won-jun and Bui, Thi Hong and Choi, Deung-Jang and Wolf, Christoph and Delgado, Fernando and Chen, Yi and Krylov, Denis and Lee, Soonhyeong and Yoon, Sangwon and Lutz, Christopher P. and Heinrich, Andreas J. and Bae, Yujeong},
  journal = {Phys. Rev. B},
  volume = {104},
  issue = {17},
  pages = {174408},
  numpages = {9},
  year = {2021},
  month = {Nov},
  publisher = {American Physical Society},
  doi = {10.1103/PhysRevB.104.174408},
  url = {https://link.aps.org/doi/10.1103/PhysRevB.104.174408}
}

@article{Stepan2024pentacene,
author = {Stepan Kovarik  and Richard Schlitz  and Aishwarya Vishwakarma  and Dominic Ruckert  and Pietro Gambardella  and Sebastian Stepanow },
title = {Spin torque–driven electron paramagnetic resonance of a single spin in a pentacene molecule},
journal = {Science},
volume = {384},
number = {6702},
pages = {1368-1373},
year = {2024},
doi = {10.1126/science.adh4753},
URL = {https://www.science.org/doi/abs/10.1126/science.adh4753}
}

@article{phark2025Ti3ML,
author  = {Phark, Soo-hyon and
     Bui, Hong Thi and
     Seo, We-hyo and
     Liu, Yaowu and
     Sheina, Valeria and
     Lee, Curie and
     Wolf, Christoph and
     Heinrich, Andreas J. and
     Robles, Roberto and
     Lorente, Nicol{\'a}s},
title   = {Spin-state engineering of single titanium adsorbates on ultrathin magnesium oxide},
journal = {Nature Communications},
year    = {2026},
doi     = {10.1038/s41467-026-68314-6},
url     = {https://doi.org/10.1038/s41467-026-68314-6},
issn    = {2041-1723}
}

@article{Willke2018Hyperfine,
author = {Philip Willke  and Yujeong Bae  and Kai Yang  and Jose L. Lado  and Alejandro Ferrón  and Taeyoung Choi  and Arzhang Ardavan  and Joaquín Fernández-Rossier  and Andreas J. Heinrich  and Christopher P. Lutz },
title = {Hyperfine interaction of individual atoms on a surface},
journal = {Science},
volume = {362},
number = {6412},
pages = {336-339},
year = {2018},
doi = {10.1126/science.aat7047},
URL = {https://www.science.org/doi/abs/10.1126/science.aat7047}
}

@article{JK2022Hyperfine,
  author    = {Kim, Jinkyung and Noh, Kyungju and Chen, Yi and Donati, Fabio and Heinrich, Andreas J. and Wolf, Christoph and Bae, Yujeong},
  title     = {Anisotropic Hyperfine Interaction of Surface-Adsorbed Single Atoms},
  journal   = {Nano Letters},
  year      = {2022},
  volume    = {22},
  number    = {23},
  pages     = {9766--9772},
  doi       = {10.1021/acs.nanolett.2c02782},
  url       = {https://doi.org/10.1021/acs.nanolett.2c02782}
}

@article{FernandoNico2021,
title = {A theoretical review on the single-impurity electron spin resonance on surfaces},
journal = {Progress in Surface Science},
volume = {96},
number = {2},
pages = {100625},
year = {2021},
issn = {0079-6816},
doi = {https://doi.org/10.1016/j.progsurf.2021.100625},
url = {https://www.sciencedirect.com/science/article/pii/S0079681621000137},
author = {Fernando Delgado and Nicolás Lorente}
}

@article{Yang2018,
  title = {Electrically controlled nuclear polarization of individual atoms},
  volume = {13},
  ISSN = {1748-3395},
  url = {http://dx.doi.org/10.1038/s41565-018-0296-7},
  DOI = {10.1038/s41565-018-0296-7},
  number = {12},
  journal = {Nature Nanotechnology},
  publisher = {Springer Science and Business Media LLC},
  author = {Yang,  Kai and Willke,  Philip and Bae,  Yujeong and Ferrón,  Alejandro and Lado,  Jose L. and Ardavan,  Arzhang and Fernández-Rossier,  Joaquín and Heinrich,  Andreas J. and Lutz,  Christopher P.},
  year = {2018},
  month = nov,
  pages = {1120–1125}
}

@article{Manassen2018ENDOR,
  title = {Fingerprints of single nuclear spin energy levels using STM – ENDOR},
  volume = {289},
  ISSN = {1090-7807},
  url = {http://dx.doi.org/10.1016/j.jmr.2018.02.005},
  DOI = {10.1016/j.jmr.2018.02.005},
  journal = {Journal of Magnetic Resonance},
  publisher = {Elsevier BV},
  author = {Manassen,  Yishay and Averbukh,  Michael and Jbara,  Moamen and Siebenhofer,  Bernhard and Shnirman,  Alexander and Horovitz,  Baruch},
  year = {2018},
  month = apr,
  pages = {107–112}
}

@article{Veldman2024Coherent,
  title = {Coherent spin dynamics between electron and nucleus within a single atom},
  volume = {15},
  ISSN = {2041-1723},
  url = {http://dx.doi.org/10.1038/s41467-024-52270-0},
  DOI = {10.1038/s41467-024-52270-0},
  number = {1},
  journal = {Nature Communications},
  publisher = {Springer Science and Business Media LLC},
  author = {Veldman,  Lukas M. and Stolte,  Evert W. and Canavan,  Mark P. and Broekhoven,  Rik and Willke,  Philip and Farinacci,  Laëtitia and Otte,  Sander},
  year = {2024},
  month = sep 
}

@article{Natterer2019Ho,
  title = {Quantum state manipulation of single atom magnets using the hyperfine interaction},
  volume = {100},
  ISSN = {2469-9969},
  url = {http://dx.doi.org/10.1103/PhysRevB.100.180405},
  DOI = {10.1103/physrevb.100.180405},
  number = {18},
  journal = {Physical Review B},
  publisher = {American Physical Society (APS)},
  author = {Forrester,  Patrick Robert and Patthey,  Fran\c{c}ois and Fernandes,  Edgar and Sblendorio,  Dante Phillipe and Brune,  Harald and Natterer,  Fabian Donat},
  year = {2019},
  month = nov 
}

@article{Stolte2025SingleShot,
  title = {Single-shot readout of the nuclear spin of an on-surface atom},
  volume = {16},
  ISSN = {2041-1723},
  url = {http://dx.doi.org/10.1038/s41467-025-63232-5},
  DOI = {10.1038/s41467-025-63232-5},
  number = {1},
  journal = {Nature Communications},
  publisher = {Springer Science and Business Media LLC},
  author = {Stolte,  Evert W. and Lee,  Jinwon and Vennema,  Hester G. and Broekhoven,  Rik and Teng,  Esther and Katan,  Allard J. and Veldman,  Lukas M. and Willke,  Philip and Otte,  Sander},
  year = {2025},
  month = aug 
}

@Article{Goerz2019krotov,
	title={{Krotov: A Python implementation of Krotov's method for quantum optimal control}},
	author={Michael H. Goerz and Daniel Basilewitsch and Fernando Gago-Encinas and Matthias G. Krauss and Karl P. Horn and Daniel M. Reich and Christiane P. Koch},
	journal={SciPost Phys.},
	volume={7},
	pages={080},
	year={2019},
	publisher={SciPost},
	doi={10.21468/SciPostPhys.7.6.080},
	url={https://scipost.org/10.21468/SciPostPhys.7.6.080},
}

@article{BarGill2013NV,
  author  = {Bar-Gill, Nir and Pham, Linh M. and Jarmola, Andrey and Budker, Dmitry and Walsworth, Ronald L.},
  title   = {Solid-state electronic spin coherence time approaching one second},
  journal = {Nature Communications},
  year    = {2013},
  volume  = {4},
  number  = {1},
  pages   = {1743},
  doi     = {10.1038/ncomms2771},
  url     = {https://doi.org/10.1038/ncomms2771}
}

@article{Krantz2019superconducting,
    author = {Krantz, P. and Kjaergaard, M. and Yan, F. and Orlando, T. P. and Gustavsson, S. and Oliver, W. D.},
    title = {A quantum engineer's guide to superconducting qubits},
    journal = {Applied Physics Reviews},
    volume = {6},
    number = {2},
    pages = {021318},
    year = {2019},
    month = {06},
    issn = {1931-9401},
    doi = {10.1063/1.5089550},
    url = {https://doi.org/10.1063/1.5089550}
}

@article{Bruzewicz2019Trappedion,
    author = {Bruzewicz, Colin D. and Chiaverini, John and McConnell, Robert and Sage, Jeremy M.},
    title = {Trapped-ion quantum computing: Progress and challenges},
    journal = {Applied Physics Reviews},
    volume = {6},
    number = {2},
    pages = {021314},
    year = {2019},
    month = {05},
    issn = {1931-9401},
    doi = {10.1063/1.5088164},
    url = {https://doi.org/10.1063/1.5088164}
}

@article{Jirovec2021Ge,
  author  = {Jirovec, Daniel and Hofmann, Andrea and Ballabio, Andrea and Mutter, Philipp M. and Tavani, Giulio and Botifoll, Marc and Crippa, Alessandro and Kukucka, Josip and Sagi, Oliver and Martins, Frederico and Saez-Mollejo, Jaime and Prieto, Ivan and Borovkov, Maksim and Arbiol, Jordi and Chrastina, Daniel and Isella, Giovanni and Katsaros, Georgios},
  title   = {A singlet-triplet hole spin qubit in planar Ge},
  journal = {Nature Materials},
  year    = {2021},
  volume  = {20},
  number  = {8},
  pages   = {1106--1112},
  doi     = {10.1038/s41563-021-01022-2},
  url     = {https://doi.org/10.1038/s41563-021-01022-2}
}

@article{Hall2016,
  author  = {Hall, L. T. and Kehayias, P. and Simpson, D. A. and Jarmola, A. and Stacey, A. and Budker, D. and Hollenberg, L. C. L.},
  title   = {Detection of nanoscale electron spin resonance spectra demonstrated using nitrogen-vacancy centre probes in diamond},
  journal = {Nature Communications},
  year    = {2016},
  volume  = {7},
  number  = {1},
  pages   = {10211},
  doi     = {10.1038/ncomms10211},
  url     = {https://doi.org/10.1038/ncomms10211}
}

@article{Cao2025,
    author = {Cao, Jiaan and Ye, Lyuzhou and Xu, Rui-Xue and Zheng, Xiao},
    title = {Influence of radio-frequency voltage on electron spin resonance spectroscopy in scanning tunneling microscopy†},
    journal = {Chinese Journal of Chemical Physics},
    volume = {38},
    number = {4},
    pages = {375-381},
    year = {2025},
    month = {08},
    issn = {1674-0068},
    doi = {10.1063/1674-0068/cjcp2409127},
    url = {https://doi.org/10.1063/1674-0068/cjcp2409127}
}

@book{BreuerPetruccione2002,
  author    = {Breuer, Heinz-Peter and Petruccione, Francesco},
  title     = {The Theory of Open Quantum Systems},
  publisher = {Oxford University Press},
  address   = {Oxford},
  year      = {2002},
  isbn      = {978-0-19-852063-4}
}

@article{Baumann2015PRL,
  title = {Origin of Perpendicular Magnetic Anisotropy and Large Orbital Moment in Fe Atoms on MgO},
  author = {Baumann, S. and Donati, F. and Stepanow, S. and Rusponi, S. and Paul, W. and Gangopadhyay, S. and Rau, I. G. and Pacchioni, G. E. and Gragnaniello, L. and Pivetta, M. and Dreiser, J. and Piamonteze, C. and Lutz, C. P. and Macfarlane, R. M. and Jones, B. A. and Gambardella, P. and Heinrich, A. J. and Brune, H.},
  journal = {Phys. Rev. Lett.},
  volume = {115},
  issue = {23},
  pages = {237202},
  numpages = {6},
  year = {2015},
  month = {Dec},
  publisher = {American Physical Society},
  doi = {10.1103/PhysRevLett.115.237202},
  url = {https://link.aps.org/doi/10.1103/PhysRevLett.115.237202}
}

@article{Shavit2019,
  title = {Generalized open quantum system approach for the electron paramagnetic resonance of magnetic atoms},
  author = {Shavit, Gal and Horovitz, Baruch and Goldstein, Moshe},
  journal = {Phys. Rev. B},
  volume = {99},
  issue = {19},
  pages = {195433},
  numpages = {13},
  year = {2019},
  month = {May},
  publisher = {American Physical Society},
  doi = {10.1103/PhysRevB.99.195433},
  url = {https://link.aps.org/doi/10.1103/PhysRevB.99.195433}
}

@article{Tien1963,
  title = {Multiphoton Process Observed in the Interaction of Microwave Fields with the Tunneling between Superconductor Films},
  author = {Tien, P. K. and Gordon, J. P.},
  journal = {Phys. Rev.},
  volume = {129},
  issue = {2},
  pages = {647--651},
  numpages = {0},
  year = {1963},
  month = {Jan},
  publisher = {American Physical Society},
  doi = {10.1103/PhysRev.129.647},
  url = {https://link.aps.org/doi/10.1103/PhysRev.129.647}
}

@article{Godfrin2017,
  title = {Operating Quantum States in Single Magnetic Molecules: Implementation of Grover’s Quantum Algorithm},
  volume = {119},
  ISSN = {1079-7114},
  url = {http://dx.doi.org/10.1103/PhysRevLett.119.187702},
  DOI = {10.1103/physrevlett.119.187702},
  number = {18},
  journal = {Physical Review Letters},
  publisher = {American Physical Society (APS)},
  author = {Godfrin,  C. and Ferhat,  A. and Ballou,  R. and Klyatskaya,  S. and Ruben,  M. and Wernsdorfer,  W. and Balestro,  F.},
  year = {2017},
  month = nov 
}

@misc{Otte2025ENDOR,
  doi = {10.48550/ARXIV.2512.11652},
  url = {https://arxiv.org/abs/2512.11652},
  author = {Vennema,  Hester G. and Mier,  Cristina and Stolte,  Evert W. and Edens,  Leonard and Lee,  Jinwon and Otte,  Sander},
  title = {Nuclear magnetic resonance on a single atom with a local probe},
  publisher = {arXiv},
  year = {2025},
  copyright = {arXiv.org perpetual,  non-exclusive license}
}

@article{Thiele2013,
  title = {Electrical Readout of Individual Nuclear Spin Trajectories in a Single-Molecule Magnet Spin Transistor},
  volume = {111},
  ISSN = {1079-7114},
  url = {http://dx.doi.org/10.1103/PhysRevLett.111.037203},
  DOI = {10.1103/physrevlett.111.037203},
  number = {3},
  journal = {Physical Review Letters},
  publisher = {American Physical Society (APS)},
  author = {Thiele,  S. and Vincent,  R. and Holzmann,  M. and Klyatskaya,  S. and Ruben,  M. and Balestro,  F. and Wernsdorfer,  W.},
  year = {2013},
  month = jul 
}

@article{KimDohun2014tomographysemiconductor,
  author  = {Kim, Dohun and Shi, Zhan and Simmons, C. B. and Ward, D. R. and Prance, J. R. and Koh, Teck Seng and Gamble, John King and Savage, D. E. and Lagally, M. G. and Friesen, Mark and Coppersmith, S. N. and Eriksson, Mark A.},
  title   = {Quantum control and process tomography of a semiconductor quantum dot hybrid qubit},
  journal = {Nature},
  volume  = {511},
  number  = {7507},
  pages   = {70--74},
  year    = {2014},
  doi     = {10.1038/nature13407},
  issn    = {1476-4687}
}

@article{HollyStemp2024Tomography,
  author  = {Stemp, Holly G. and Asaad, Serwan and van Blankenstein, Mark R. and Vaartjes, Arjen and Johnson, Mark A. I. and M{\k a}dzik, Mateusz T. and Heskes, Amber J. A. and Firgau, Hannes R. and Su, Rocky Y. and Yang, Chih Hwan and Laucht, Arne and Ostrove, Corey I. and Rudinger, Kenneth M. and Young, Kevin and Blume-Kohout, Robin and Hudson, Fay E. and Dzurak, Andrew S. and Itoh, Kohei M. and Jakob, Alexander M. and Johnson, Brett C. and Jamieson, David N. and Morello, Andrea},
  title   = {Tomography of entangling two-qubit logic operations in exchange-coupled donor electron spin qubits},
  journal = {Nature Communications},
  volume  = {15},
  number  = {1},
  pages   = {8415},
  year    = {2024},
  doi     = {10.1038/s41467-024-52795-4},
  issn    = {2041-1723}
}

@article{Zhang2023TomographyNV,
  title = {Fast Quantum State Tomography in the Nitrogen Vacancy Center of Diamond},
  author = {Zhang, Jingfu and Hegde, Swathi S. and Suter, Dieter},
  journal = {Phys. Rev. Lett.},
  volume = {130},
  issue = {9},
  pages = {090801},
  numpages = {6},
  year = {2023},
  month = {Feb},
  publisher = {American Physical Society},
  doi = {10.1103/PhysRevLett.130.090801},
  url = {https://link.aps.org/doi/10.1103/PhysRevLett.130.090801}
}

@article{Matthias2006Tomography,
author = {Matthias Steffen  and M. Ansmann  and Radoslaw C. Bialczak  and N. Katz  and Erik Lucero  and R. McDermott  and Matthew Neeley  and E. M. Weig  and A. N. Cleland  and John M. Martinis },
title = {Measurement of the Entanglement of Two Superconducting Qubits via State Tomography},
journal = {Science},
volume = {313},
number = {5792},
pages = {1423-1425},
year = {2006},
doi = {10.1126/science.1130886},
URL = {https://www.science.org/doi/abs/10.1126/science.1130886}
}

@BOOK{Schweiger2001Principles,
  title     = "Principles of pulse electron paramagnetic resonance",
  author    = "Schweiger, Arthur and Jeschke, Gunnar",
  publisher = "Oxford University Press",
  month     =  oct,
  year      =  2001,
  address   = "London, England",
  language  = "en"
}

@article{Lim2025,
  title = {Designing quantum error correction codes for practical spin qudit systems},
  volume = {112},
  ISSN = {2469-9934},
  url = {http://dx.doi.org/10.1103/7q6l-d4qh},
  DOI = {10.1103/7q6l-d4qh},
  number = {2},
  journal = {Physical Review A},
  publisher = {American Physical Society (APS)},
  author = {Lim,  Sumin and Ardavan,  Arzhang},
  year = {2025},
  month = aug 
}

@article{Jankovi2024,
  title = {Noisy qudit vs multiple qubits: conditions on gate efficiency for enhancing fidelity},
  volume = {10},
  ISSN = {2056-6387},
  url = {http://dx.doi.org/10.1038/s41534-024-00829-6},
  DOI = {10.1038/s41534-024-00829-6},
  number = {1},
  journal = {npj Quantum Information},
  publisher = {Springer Science and Business Media LLC},
  author = {Janković,  Denis and Hartmann,  Jean-Gabriel and Ruben,  Mario and Hervieux,  Paul-Antoine},
  year = {2024},
  month = jun 
}

@article{Farinacci2022,
  title = {Experimental Determination of a Single Atom Ground State Orbital through Hyperfine Anisotropy},
  volume = {22},
  ISSN = {1530-6992},
  url = {http://dx.doi.org/10.1021/acs.nanolett.2c02783},
  DOI = {10.1021/acs.nanolett.2c02783},
  number = {21},
  journal = {Nano Letters},
  publisher = {American Chemical Society (ACS)},
  author = {Farinacci,  Laëtitia and Veldman,  Lukas M. and Willke,  Philip and Otte,  Sander},
  year = {2022},
  month = oct,
  pages = {8470–8474}
}

@article{DJ2025review,
title = {Electron spin resonance with scanning tunneling microscopy: a tool for an on-surface quantum platform of identical qubits},
journal = {Nanoscale Advances},
volume = {7},
number = {15},
pages = {4551-4558},
year = {2025},
issn = {2516-0230},
doi = {https://doi.org/10.1039/d5na00316d},
url = {https://www.sciencedirect.com/science/article/pii/S2516023025002588},
author = {Deung-Jang Choi and Soo-hyon Phark and Andreas J. Heinrich and Nicolás Lorente}
}

@article{Bi2023review,
title = {Recent progress in probing atomic and molecular quantum coherence with scanning tunneling microscopy},
journal = {Progress in Surface Science},
volume = {98},
number = {1},
pages = {100696},
year = {2023},
issn = {0079-6816},
doi = {https://doi.org/10.1016/j.progsurf.2022.100696},
url = {https://www.sciencedirect.com/science/article/pii/S0079681622000442},
author = {Liya Bi and Kangkai Liang and Gregory Czap and Hao Wang and Kai Yang and Shaowei Li}
}

@article{HAZAN2023107377,
title = {ESR-STM on diamagnetic molecule: C60 on graphene},
journal = {Journal of Magnetic Resonance},
volume = {348},
pages = {107377},
year = {2023},
issn = {1090-7807},
doi = {https://doi.org/10.1016/j.jmr.2023.107377},
url = {https://www.sciencedirect.com/science/article/pii/S1090780723000125},
author = {Zion Hazan and Michael Averbukh and Yishay Manassen}
}

@article{Horovitz2021,
  title = {Spin entanglement via scanning tunneling microscope current},
  author = {Horovitz, Baruch and Henkel, Carsten},
  journal = {Phys. Rev. B},
  volume = {104},
  issue = {8},
  pages = {L081405},
  numpages = {5},
  year = {2021},
  month = {Aug},
  publisher = {American Physical Society},
  doi = {10.1103/PhysRevB.104.L081405},
  url = {https://link.aps.org/doi/10.1103/PhysRevB.104.L081405}
}

@article{HaoWang2024topo,
  author    = {Hao Wang and Peng Fan and Jing Chen and Lili Jiang and Hong-Jun Gao and Jose L. Lado and Kai Yang},
  title     = {Construction of topological quantum magnets from atomic spins on surfaces},
  journal   = {Nature Nanotechnology},
  year      = {2024},
  volume    = {19},
  number    = {12},
  pages     = {1782--1788},
  doi       = {10.1038/s41565-024-01775-2},
  url       = {https://doi.org/10.1038/s41565-024-01775-2},
  issn      = {1748-3395}
  }

@article{Yang2021RVB,
  author    = {Kai Yang and Soo-Hyon Phark and Yujeong Bae and Taner Esat and Philip Willke and Arzhang Ardavan and Andreas J. Heinrich and Christopher P. Lutz},
  title     = {Probing resonating valence bond states in artificial quantum magnets},
  journal   = {Nature Communications},
  year      = {2021},
  volume    = {12},
  number    = {1},
  pages     = {993},
  doi       = {10.1038/s41467-021-21274-5},
  url       = {https://doi.org/10.1038/s41467-021-21274-5}
}

@article{HaoWang2025LZSM,
  author    = {Hao Wang and Jing Chen and Peng Fan and Yelko del Castillo and Alejandro Ferr{\'o}n and Lili Jiang and Zilong Wu and Shijie Li and Hong-Jun Gao and Heng Fan and Joaqu{\'i}n Fern{\'a}ndez-Rossier and Kai Yang},
  title     = {Electrically tunable quantum interference of atomic spins on surfaces},
  journal   = {Nature Communications},
  year      = {2025},
  volume    = {16},
  number    = {1},
  pages     = {8988},
  doi       = {10.1038/s41467-025-64022-9},
  url       = {https://doi.org/10.1038/s41467-025-64022-9}
}


\end{document}